\documentclass{aifrontiers}
\usepackage{aifrontiers}

\usepackage[utf8]{inputenc}
\usepackage{booktabs}
\usepackage{graphicx}
\graphicspath{{figures/}}
\usepackage{multirow}
\usepackage{longtable}
\usepackage{enumitem}
\usepackage{pifont}
\usepackage{colortbl}
\usepackage{float}
\tcbuselibrary{most}
\newtcolorbox{diagnosticbox}{
  colback=gray!5, colframe=gray!60, boxrule=0.4pt,
  left=6pt, right=6pt, top=4pt, bottom=4pt,
  fontupper=\small, sharp corners,
  borderline west={2pt}{0pt}{black!50}
}
\usepackage{makecell}
\usepackage{wrapfig}
\usepackage{tikz}
\usetikzlibrary{arrows.meta, positioning, fit, backgrounds, shapes.geometric, calc, shadows.blur}
\newcommand{\circled}[1]{\tikz[baseline=(char.base)]{\node[circle, fill=black!60, text=white, font=\scriptsize\bfseries, inner sep=1pt, minimum size=0.35cm] (char) {#1};}}

\colorlet{pipGreenDk}{green!60!black}
\colorlet{pipGreen}{green!70!black}
\colorlet{pipGreenTeal}{green!60!teal}
\colorlet{pipBlue}{blue!70}
\colorlet{pipBlueCyan}{blue!60!cyan}
\colorlet{pipOrange}{orange!80!black}
\colorlet{pipRed}{red!70!black}
\colorlet{pipTeal}{teal!80!black}
\colorlet{pipViolet}{violet!70}
\colorlet{pipVioletLt}{violet!60}
\colorlet{pipVioletDk}{violet!80!black}
\tikzset{
    stagebox/.style={draw=#1!50!black, rounded corners=8pt, inner sep=10pt,
                     line width=1pt, fill=#1!4},
    stagebox/.default=gray,
    snum/.style={circle, fill=#1!70!black, text=white, font=\large\bfseries,
                 minimum size=0.6cm, inner sep=0pt},
    snum/.default=blue,
    stitle/.style={font=\Large\bfseries\sffamily, color=#1!70!black},
    stitle/.default=blue,
    mybox/.style={draw=#1!50!black, rounded corners=4pt, minimum height=0.8cm,
                align=center, font=\footnotesize, fill=#1!8, thick},
    mybox/.default=gray,
    iconbox/.style={draw=#1!40!black, rounded corners=3pt, minimum size=0.7cm,
                    align=center, font=\scriptsize, fill=#1!6},
    grplabel/.style={font=\footnotesize\bfseries\sffamily, color=#1!60!black},
    grpbox/.style={draw=#1!30!black, rounded corners=4pt, fill=#1!4,
                   dashed, line width=0.6pt, inner sep=6pt},
}

\definecolor{darkblue}{rgb}{0, 0, 0.5}
\hypersetup{colorlinks=true, citecolor=darkblue, linkcolor=darkblue, urlcolor=darkblue}

\newtcolorbox{findingbox}{
  colback=gray!4, colframe=gray!40, boxrule=0.3pt,
  left=6pt, right=6pt, top=5pt, bottom=5pt,
  fontupper=\small, sharp corners,
  borderline west={2pt}{0pt}{black!60}
}
\newcommand{\finding}[1]{\begin{findingbox}\textbf{Finding.}~#1\end{findingbox}}

\renewcommand{\AIFLogo}{%
  \raisebox{-0.3ex}{\includegraphics[height=1.4em]{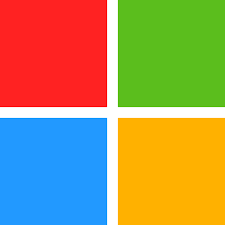}}%
  \hspace{0.6em}%
  {\sffamily\bfseries\large Microsoft}%
}

\title{ADK Arena: Evaluating Agent Development Kits via LLM-as-a-Developer}
\shorttitle{ADK Arena}
\author{
    Jintao Huang$^{\spadesuit}$\thanks{Work done during internship at Microsoft CoreAI. Correspondence: Jintao Huang \texttt{huang.5692@osu.edu}, Xiaomin Li \texttt{xiaominli@microsoft.com}.}, Xiaomin Li$^{\heartsuit}$, Gaurav Mittal$^{\heartsuit}$, Yu Hu$^{\heartsuit}$\\
    {\normalsize $^{\spadesuit}$The Ohio State University \quad $^{\heartsuit}$Microsoft}
}
\date{2026}

\begin{document}

\begin{abstract}
The rapid proliferation of Agent Development Kits (ADKs), SDK-level frameworks for building LLM-powered autonomous agents, has outpaced any empirical understanding of how framework choice affects agent performance. We propose \textbf{LLM-as-a-Developer}, a methodology that replaces human developers with an LLM coding agent that learns each framework's API from documentation, writes agent code, and iteratively repairs it through a validate-and-feedback loop until tests pass. By holding the developer constant and varying only the framework, generation effort becomes a quantitative proxy for API usability and the resulting agents provide a controlled measure of framework effectiveness. We implement this in \textbf{ADK Arena}, a fully automated pipeline with per-framework Docker isolation, a three-level validation pipeline, and benchmark adapters for SWE-bench, $\tau^2$-bench, Terminal-Bench, and MCP-Atlas. Evaluating all 51 popular Python ADK frameworks (204 agent--benchmark pairs), we find that: (1)~generation succeeds for 57\% of runs, and its cost varies 5.6$\times$ across frameworks (\$0.6 to \$3.4 per agent), a quantitative proxy for API complexity, though cost alone does not predict success; (2)~no single framework dominates: the best single-benchmark ADK agents resolve up to 80\% of tasks and can even \emph{beat} general-purpose frontier coding agents at a fraction of the cost, yet the median framework resolves only 32\%; (3)~across information-source ablations, genuine framework usage stays within a narrow 28--40\% band (highest with raw source access and still 33\% with no reference material at all), indicating that documentation, source code, and parametric knowledge are largely substitutable rather than any one being a hard bottleneck\footnote{We release the full ADK Arena pipeline at \url{https://github.com/jintao-h/ADK-Arena}.}.
\end{abstract}

\maketitle
\section{Introduction}
\label{sec:intro}

A developer building an LLM-powered agent in 2025 faces dozens of \emph{Agent Development Kits} (ADKs), SDK-level frameworks that provide reusable abstractions for tool calling~\citep{schick2024toolformer,patil2023gorilla}, agentic loops~\citep{yao2023react,shinn2023reflexion}, multi-agent orchestration~\citep{wu2024autogen,li2023camel,chen2024agentverse}, and other capabilities for autonomous agents~\citep{xi2025rise,wang2024survey,sumers2024cognitive}. Fueled by advances in large language models~\citep{brown2020language,openai2023gpt4,anthropic2024claude,touvron2023llama}, every major AI vendor has shipped an official ADK within 18 months, alongside dozens of open-source alternatives. Our landscape survey identifies 51 popular ADK frameworks distributed as Python packages (\S\ref{sec:collection}). Yet despite this growth, developers have no empirical basis for choosing among them: \emph{which framework produces the best agents, and at what development cost?}
\looseness=-1

The barrier is methodological. Traditional evaluation requires experts to manually implement benchmark workloads against each framework's API, an $O(N \times M)$ effort that introduces experimenter bias and cannot scale~\citep{techempower,tpch,mlperf}. Developer surveys~\citep{wang2025developer,liu2026multiagent,hasan2025testing} reveal what developers \emph{say} about frameworks, but not how the frameworks \emph{perform}. Agent benchmarks~\citep{jimenez2024swebench,tau2bench,liu2023agentbench,mialon2023gaia} compare \textit{models}, holding the framework constant. No prior study has attempted to compare frameworks at ecosystem scale.
\looseness=-1

\begin{figure*}[t]
\centering
\includegraphics[width=\textwidth]{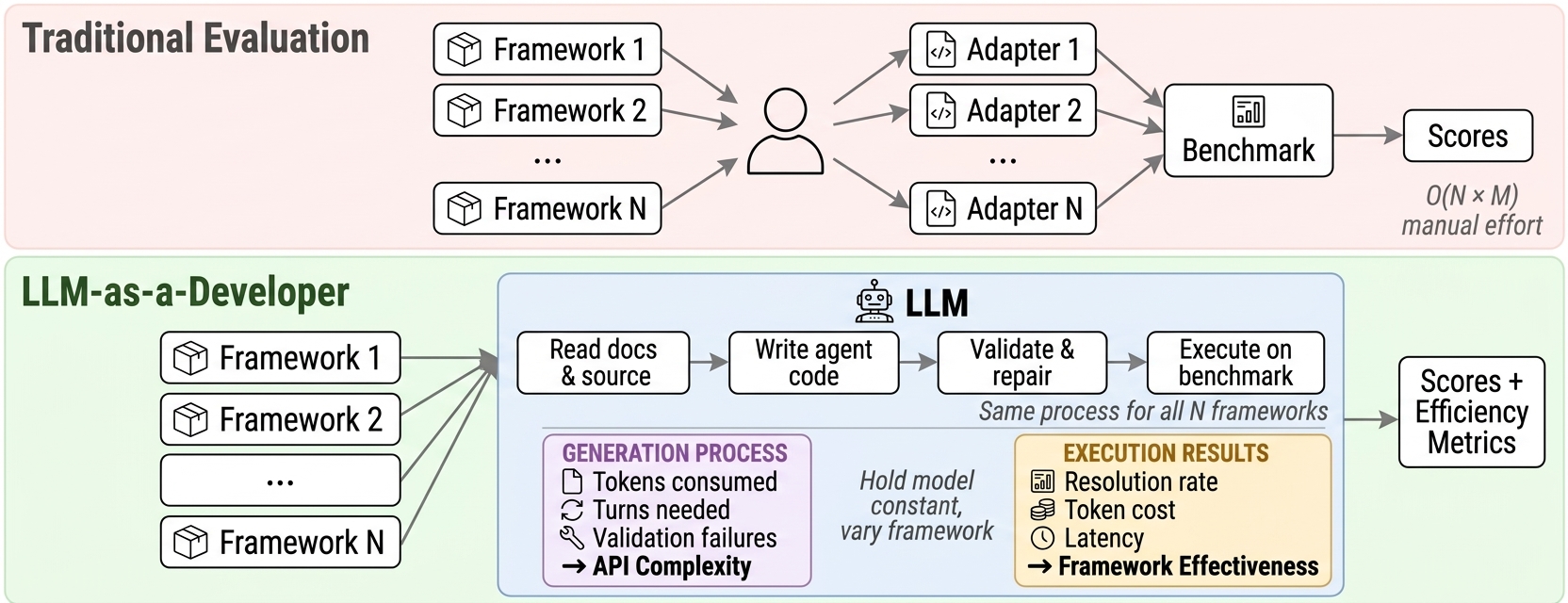}
\caption{LLM-as-a-Developer. }
\label{fig:llm-as-dev}
\end{figure*}

We propose \textbf{LLM-as-a-Developer} (\autoref{fig:llm-as-dev}), inspired by LLM-as-a-Judge~\citep{zheng2023judging}: instead of replacing human \emph{evaluators}, we replace human \emph{developers}. A single LLM-as-a-Developer agent learns each framework from its documentation, writes task-solving agent code, and iteratively repairs it through a validate-and-feedback loop, mirroring the explore--write--test--fix cycle a human developer follows when adopting a new SDK. By holding the developer constant and varying only the framework, two complementary signals emerge: the \emph{generation process} (tokens consumed, turns needed, failure rate) quantifies API usability, while the \emph{execution results} (resolution rate, cost, latency) measure framework effectiveness.
\looseness=-1

We realize this methodology in \textbf{ADK Arena}, a fully automated pipeline that takes a framework's GitHub repository as input and produces benchmark scores as output, with no manually written agent code. The LLM developer operates inside isolated Docker environments with access to framework documentation and source code, iterating through a three-level validation pipeline that catches errors ranging from import failures to runtime crashes on real benchmark tasks. We evaluate across four established benchmarks spanning software engineering (SWE-bench), conversational tool use ($\tau^2$-bench), multi-tool orchestration (MCP-Atlas), and terminal interaction (Terminal-Bench).
\looseness=-1

Applying ADK Arena to all 51 frameworks (204 agent-benchmark pairs) reveals that generation cost varies 5.6$\times$ across frameworks (\$0.6 to \$3.4 per agent), directly reflecting API complexity: well-designed APIs such as LangGraph and OpenAI Agents are among the cheapest to target, while large or poorly documented ones cost several times more to reach a working agent. No single framework dominates: the best single-benchmark ADK agents resolve up to 80\% of tasks and can even beat general-purpose frontier coding agents at a fraction of the cost, yet the median framework resolves only 32\%; strikingly, the developer model that \emph{writes} the agent matters more than the one that runs it: Opus-authored agents resolve roughly twice as many tasks as GPT-authored ones on the same backbone. Ablation experiments show that documentation, raw source, and parametric knowledge are largely interchangeable: genuine framework usage stays within a 28 to 40\% band across conditions (highest with raw source access), and even with no reference material 33\% of agents still pass, so no single information source is a hard bottleneck.
\looseness=-1

\paragraph{Contributions.} We make the following contributions:
\begin{enumerate}[nosep,leftmargin=*]
  \item \textbf{LLM-as-a-Developer} (\S\ref{sec:llm-as-dev}), a methodology that uses an LLM-as-a-Developer agent as a controlled proxy developer, yielding generation effort as a quantitative measure of API usability and task performance as a measure of framework effectiveness.
  \item \textbf{ADK Arena} (\S\ref{sec:methodology}), a fully automated system with per-framework Docker isolation, a three-level Validate-and-Repair pipeline, token-level telemetry, and adapters for four benchmarks.
  \item \textbf{Large-scale empirical results} (\S\ref{sec:results}) from 51 frameworks (204 agent-benchmark pairs), including the first ecosystem-wide comparison of ADK frameworks against frontier coding agents, and information-source ablations that isolate the contributions of documentation, source code, and parametric knowledge.
\end{enumerate}

\section{Methodology: LLM-as-a-Developer}
\label{sec:llm-as-dev}

\subsection{Limitations of Traditional Framework Evaluation}
\label{sec:limitations}

Software framework evaluation has a long tradition: TPC benchmarks compare database engines~\citep{tpch}, MLPerf compares deep-learning frameworks~\citep{mlperf}, and TechEmpower compares web frameworks~\citep{techempower}. These expert-driven approaches work well when frameworks are mature and few, but applying them to the rapidly growing ADK ecosystem~\citep{xi2025rise,wang2024survey,hong2024metagpt,wu2024autogen,qian2024chatdev} exposes three limitations:
\begin{itemize}[nosep,leftmargin=*]
  \item \textbf{Experimenter bias.} Hand-written benchmark code reflects the author's familiarity with each framework, a serious confound when APIs evolve weekly. MAFBench~\citep{orogat2026mafbench}, the closest prior work, covers only 7 frameworks with hand-written micro-tasks.
  \item \textbf{Scalability.} $N$ frameworks $\times$ $M$ benchmarks requires $O(N \times M)$ expert implementations. We target 51 frameworks across 4 benchmarks (204 pairs).
  \item \textbf{No unified performance metric.} Developer surveys~\citep{wang2025developer,liu2026multiagent,hasan2025testing} capture opinions but not runtime behavior; agent benchmarks~\citep{jimenez2024swebench,tau2bench,liu2023agentbench,mialon2023gaia} compare models, not frameworks.
\end{itemize}

\subsection{The Developer Analogy}
\label{sec:developer-analogy}

We treat an LLM as a developer who learns each framework from documentation, writes code, and iteratively debugs until it works, following the same explore--write--test--fix cycle a human follows when adopting a new SDK. This builds on three ideas:
\begin{itemize}[nosep,leftmargin=*]
  \item \textbf{From LLM-as-a-Judge to LLM-as-a-Developer.} LLM-as-a-Judge~\citep{zheng2023judging} replaces human evaluators for scalable assessment. We apply the same principle to a different bottleneck (replacing human \textit{developers}) to achieve scalability, reproducibility, and cost efficiency across dozens of frameworks.
  \item \textbf{Grounded in coding capability.} LLM coding has progressed from code completion~\citep{chen2021evaluating,li2023starcoder,roziere2024code} to tool-augmented agents~\citep{schick2024toolformer,patil2023gorilla,qin2024toolllm} that resolve $>$70\% of real GitHub issues~\citep{jimenez2024swebench,yang2024sweagent,wang2024executable}. The analogy is strongest for the ``first-time developer'' experience, making it well-suited for measuring onboarding difficulty.
  \item \textbf{Controlled-variable protocol.} Identical prompts, tools, and token budgets for every framework; information restricted to each framework's own repository~\citep{mlperf,tpch}. Two signals emerge: \emph{generation effort} (tokens, LLM calls, failures) measures API complexity; \emph{execution performance} (resolution rate, cost) measures framework effectiveness.
\end{itemize}

\subsection{Assumptions and Scope}
\label{sec:design-principles}

The methodology rests on three key assumptions that directly address the limitations identified in \S\ref{sec:limitations}:

\begin{itemize}[nosep,leftmargin=*]
  \item \textbf{Controlled and scalable.} The LLM starts from the same baseline for every framework, eliminating the experimenter-familiarity confound~\citep{tichy1998should,sim2003using}. The pipeline requires only a framework's repository as input, scaling to 204 pairs without per-framework engineering, mirroring the scalability gains of LLM-as-a-Judge~\citep{zheng2023judging} and LLM-based test generation~\citep{lemieux2023codamosa,xia2024fuzz4all}.
  \item \textbf{Usability as a measurable signal.} Generation metrics (tokens consumed, LLM calls needed, validation failures) serve as quantitative proxies for API complexity and documentation quality, relating to cognitive dimensions of API usability~\citep{clarke2004measuring,stylos2007usability,rama2015api}. Unlike developer surveys~\citep{wang2025developer} or issue mining~\citep{hasan2025testing}, these signals are objective and reproducible.
  \item \textbf{LLM capability as a constant.} We assume the LLM is a sufficiently capable developer held constant across frameworks, analogous to controlled SE experiments that fix expertise while varying tools~\citep{ko2004six,murphy2006java}. We mitigate the training-data confound through a three-condition ablation (\S\ref{sec:ablation}) that disentangles prior familiarity from documentation quality. Prior work has extensively validated that LLM judgments align with human judgments across diverse evaluation tasks~\citep{zheng2023judging,chiang2024chatbot,li2024crowdsourced}, supporting the use of LLM-based proxies for developer behavior.
\end{itemize}

\section{ADK Arena}
\label{sec:methodology}

\begin{figure*}[t]
\centering
\includegraphics[width=\textwidth]{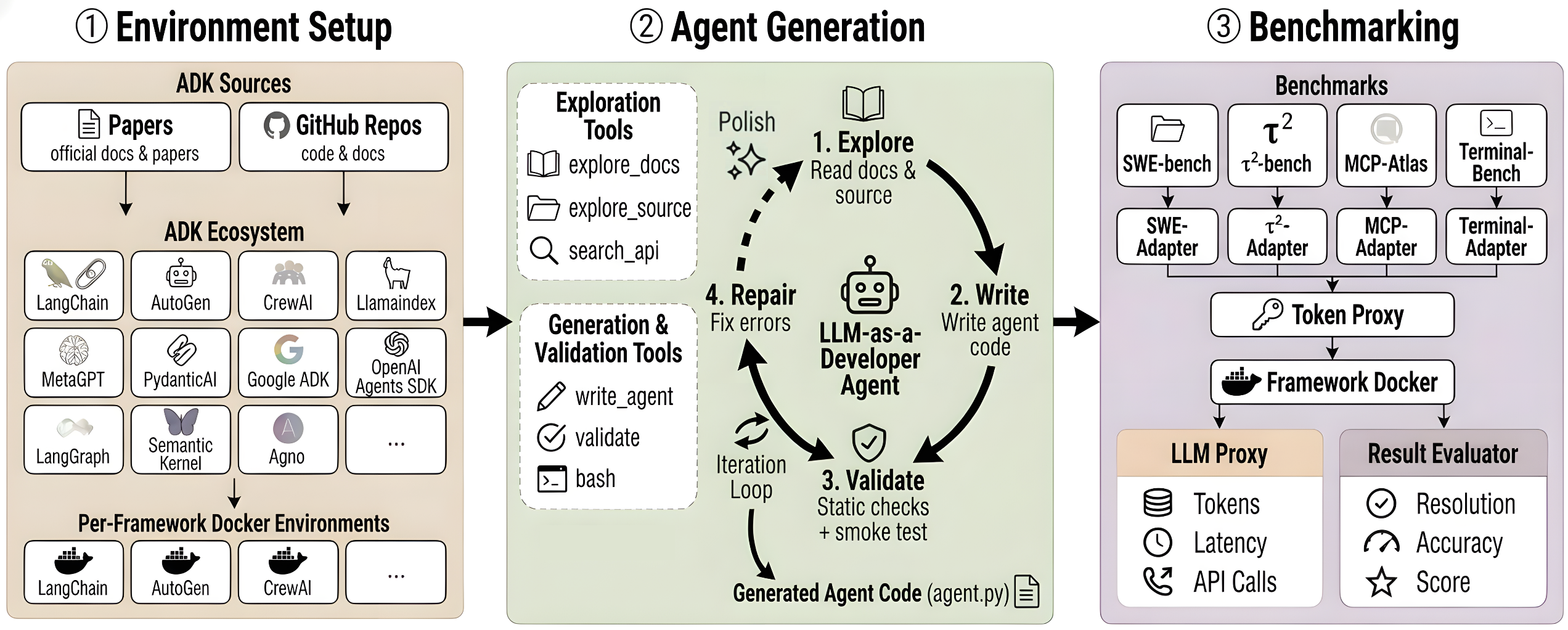}
\caption{\textbf{The ADK Arena pipeline.}
\textit{\protect\circled{1}~Environment Setup}: collect repos, build Docker images.
\textit{\protect\circled{2}~Agent Generation}: explore docs, write code, validate and repair.
\textit{\protect\circled{3}~Benchmarking}: execute in containers, score output.}
\label{fig:pipeline}
\end{figure*}

We realize the LLM-as-a-Developer methodology (\S\ref{sec:llm-as-dev}) as a fully automated pipeline (\autoref{fig:pipeline}).

\subsection{Environment Setup}
\label{sec:collection}

We first curate a comprehensive set of ADKs and prepare isolated execution environments for each.

\paragraph{Collection.}
We collect 51 Python ADK frameworks from three sources: academic venues (\texttt{ICLR}, \texttt{NeurIPS}, \texttt{ICML}, \texttt{ACL}, \texttt{EMNLP}, \texttt{NAACL}, \texttt{AAAI}, \texttt{IJCAI}, \texttt{AAMAS}, \texttt{KDD}, \texttt{WWW}, \texttt{COLM}), GitHub topic/keyword search (${\geq}1{,}000$ stars), and curated awesome-lists. After filtering out agent applications, no-code platforms, and non-Python projects, the final set spans both research and industry releases (Appendix~\ref{sec:ecosystem}).

\paragraph{Per-framework environment.}
\label{sec:stage1}
Each framework receives a dedicated Docker image built from a shared base with the framework's packages pre-installed. Benchmark-specific images extend this base with task-specific tooling (e.g., \texttt{git} and test runners for SWE-bench). All LLM traffic is routed through a local proxy, ensuring controlled and reproducible environments. Inside each container, the LLM developer has access to two information sources:
\begin{itemize}[nosep,leftmargin=*]
  \item \textbf{Curated documentation.} We collect each framework's documentation from its \texttt{docs/} directory into a unified format. The LLM developer reads it via the \texttt{explore\_docs} tool that returns 20K characters per call, allowing incremental exploration of large API references.
  \item \textbf{Source code.} The framework's full repository is mounted read-only, enabling the LLM to inspect implementation details, discover undocumented APIs, and verify function signatures. Accessible via \texttt{explore\_source} and \texttt{search\_api} (regex grep) tools (\autoref{tab:dev-tools}).
\end{itemize}

\subsection{Agent Generation}
\label{sec:stage2}

We build an \emph{LLM-as-a-Developer} agent using OpenAI Codex (GPT-5.4) that automates the entire learning-and-coding workflow a human developer would follow. Given only a framework's documentation and source code, it explores the API, writes a task-solving agent, and iteratively repairs it until validation passes.

\paragraph{Agent interface.}
Each generated agent is a single Python file (\texttt{agent.py}) exposing one entry point: \texttt{solve(prompt:~str, workdir:~str) -> str}. The function receives a natural-language task description and a working directory, then orchestrates the framework's agents and tools to produce a solution. This uniform interface decouples generation from evaluation. Any benchmark adapter can invoke the agent identically regardless of which framework it wraps.

\paragraph{Generation loop.}
The LLM-as-a-Developer drives the generation loop with a 5-minute budget. Each iteration follows: \emph{explore} documentation/source, \emph{write} agent code, \emph{validate} against a three-level pipeline, and \emph{repair} based on diagnostic feedback. The loop terminates when all three validation levels pass or the time budget expires.

\paragraph{Iteration loop.}
\label{par:validate-and-repair}
Within each generation loop iteration, an inner \emph{iteration loop} repeatedly validates the current code and revises it based on structured diagnostic feedback, enabling targeted single-edit repairs rather than blind regeneration. Only once validation passes all three levels does the agent proceed to the next round of exploration and polishing. The validation pipeline has increasing fidelity:
\begin{enumerate}[nosep,leftmargin=*]
  \item \textbf{Step~1: Static analysis}: compile check (\texttt{py\_compile}), import verification, framework usage check (reject raw API fallbacks), and 40+ AST/regex patterns detecting common anti-patterns (Appendix~\ref{sec:validation-patterns}).
  \item \textbf{Step~2: Real LLM smoke test}: execute \texttt{solve()} with a real LLM through a token-recording proxy. Verifies that the agent can successfully call the LLM, handle the response format, and return a valid result.
  \item \textbf{Step~3: Real benchmark task}: run one task from the target benchmark's official task set under realistic conditions. We do not run the task to completion; instead, a benchmark-specific early-exit check monitors the proxy for expected behavior (e.g., minimum number of LLM calls with tool use for MCP-Atlas). If the check is satisfied, validation passes immediately. Otherwise, after a short timeout the task is marked as failed. Passing Step~3 guarantees the agent can execute on the full benchmark without crashing.
\end{enumerate}
\noindent On failure at any level, the validator pattern-matches the error against 70+ known signatures and emits a structured diagnostic hint containing a root-cause diagnosis and a concrete fix. For example, a \texttt{ConnectionRefusedError} produces:

\begin{diagnosticbox}
\textbf{DIAGNOSIS}: Agent is connecting to a hardcoded URL (e.g., \texttt{https://api.openai.com}) instead of the proxy endpoint. The ADK execution environment routes all LLM traffic through a local proxy at \texttt{ADK\_BASE\_URL}.\\[4pt]
\textbf{FIX}: Replace all hardcoded URLs with \texttt{os.environ["ADK\_BASE\_URL"]}.
\end{diagnosticbox}

\paragraph{Development tools.}
To support the generation and iteration loops, we provide the LLM developer with six CLI tools inside the per-framework Docker container (\autoref{tab:dev-tools}). The generated agent must use the framework's native API (raw OpenAI/httpx fallbacks are rejected by validation).

\begin{table}[t]
\caption{Development tools available to the LLM developer during agent generation.}
\label{tab:dev-tools}
\centering
\resizebox{\linewidth}{!}{%
\small
\begin{tabular}{@{}ll@{\hspace{16pt}}ll@{}}
\toprule
\multicolumn{2}{c}{\textbf{Exploration}} & \multicolumn{2}{c}{\textbf{Development}} \\
\cmidrule(r){1-2} \cmidrule(l){3-4}
\texttt{explore\_docs}   & Browse curated API documentation  & \texttt{write\_agent} & Write \texttt{agent.py} with auto compatibility fixes \\
\texttt{explore\_source} & Read framework \texttt{.py} source and directory structure & \texttt{validate}     & Run 3-level Validate-and-Repair pipeline \\
\texttt{search\_api}     & Regex grep over source for API usage patterns     & \texttt{bash}         & Run shell commands (\texttt{python -c}, version checks) \\
\bottomrule
\end{tabular}%
}
\end{table}

\subsection{Benchmarking}
\label{sec:stage3}

We evaluate on four established benchmarks (50 tasks each): SWE-bench Verified~\citep{jimenez2024swebench}, $\tau^2$-bench~\citep{tau2bench}, MCP-Atlas~\citep{mcpatlas2025}, and Terminal-Bench~\citep{terminalbench2025}. Execution uses GPT-5.4 Nano via the token proxy.

\paragraph{Benchmark adapters.}
The four benchmarks differ substantially in input format, execution environment, and scoring methodology. Each \emph{adapter} encapsulates these differences behind a uniform interface: given an agent's \texttt{solve()} function, a task prompt, and a working directory, the adapter handles environment setup, execution orchestration, and evaluation scoring. This allows a single generated agent to be evaluated identically across all four benchmarks without benchmark-specific code.

\paragraph{Unified LLM proxy.}
ADK frameworks use diverse HTTP stacks (OpenAI SDK, \texttt{httpx}, \texttt{requests}) and target different provider APIs (OpenAI, Anthropic, Google). To evaluate all 51 frameworks against the same backbone LLM regardless of which provider API each natively targets, we route all LLM requests through a transparent \emph{token proxy}:
\begin{itemize}[nosep,leftmargin=*]
  \item \textbf{Protocol translation}: full bidirectional conversion between Anthropic and OpenAI message formats, so each framework sees its expected wire protocol regardless of the backend LLM.
  \item \textbf{Metric capture}: extracting per-call token counts (input, output, cached) from both streaming and non-streaming responses, and logging latency, request/response content, tool-call sequences, and per-call breakdowns (system prompt, tool schemas, history) for post-hoc analysis.
\end{itemize}

\section{Evaluation}
\label{sec:results}

\begin{table*}[t]
\caption{Generation effort for 51 ADK frameworks (\textsc{Normal} condition), ordered by PyPI
monthly downloads. Values averaged across benchmarks. Each cell shows GPT-5.4\,/\,Opus-4.6 results.
\textbf{Pass}: number of benchmarks (out of 4) for which the generated agent passes all three
validation levels, which means it can execute the benchmark but does not necessarily finish the task successfully.}
\label{tab:gen_effort}
\centering
\resizebox{\linewidth}{!}{%
\renewcommand{\arraystretch}{1.15}
\begin{tabular}{@{\hskip 2pt}l ccccc @{\hspace{6pt}\vrule width 0.4pt\hspace{6pt}} l ccccc@{\hskip 2pt}}
\toprule
\textbf{Framework} & \textbf{Pass} & \textbf{In Tok} & \textbf{Out Tok} & \textbf{Cache Tok} & \textbf{Cost (\$)} &
\textbf{Framework} & \textbf{Pass} & \textbf{In Tok} & \textbf{Out Tok} & \textbf{Cache Tok} & \textbf{Cost (\$)} \\
\midrule
    \raisebox{-0.4ex}{\includegraphics[height=1.1em]{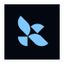}}\,LangChain & 3\,/\,3 & 1,077K\,/\,1,119K & 14K\,/\,12K & 851K\,/\,1,057K & 1.24\,/\,1.14 & \raisebox{-0.4ex}{\includegraphics[height=1.1em]{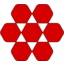}}\,Swarms & 2\,/\,1 & 1,206K\,/\,1,308K & 12K\,/\,9K & 696K\,/\,1,260K & 1.83\,/\,1.08 \\
    \raisebox{-0.4ex}{\includegraphics[height=1.1em]{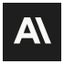}}\,Anthropic SDK & 4\,/\,1 & 1,704K\,/\,1,572K & 14K\,/\,13K & 1,073K\,/\,1,512K & 2.38\,/\,1.37 & \raisebox{-0.4ex}{\includegraphics[height=1.1em]{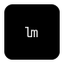}}\,MCP Agent & 3\,/\,3 & 2,444K\,/\,1,565K & 17K\,/\,11K & 1,538K\,/\,1,506K & 3.39\,/\,1.32 \\
    \raisebox{-0.4ex}{\includegraphics[height=1.1em]{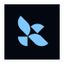}}\,LangGraph & 2\,/\,4 & 1,008K\,/\,693K & 10K\,/\,15K & 556K\,/\,630K & 1.58\,/\,1.00 & \raisebox{-0.4ex}{\includegraphics[height=1.1em]{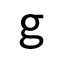}}\,Guidance & 3\,/\,3 & 1,904K\,/\,1,847K & 12K\,/\,11K & 1,104K\,/\,1,774K & 2.81\,/\,1.53 \\
    \raisebox{-0.4ex}{\includegraphics[height=1.1em]{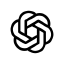}}\,OpenAI Agents & 3\,/\,3 & 2,013K\,/\,1,085K & 12K\,/\,13K & 1,128K\,/\,1,022K & 3.04\,/\,1.15 & \raisebox{-0.4ex}{\includegraphics[height=1.1em]{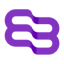}}\,AtomicAgents & 3\,/\,2 & 2,003K\,/\,1,115K & 13K\,/\,11K & 1,066K\,/\,1,063K & 3.14\,/\,1.08 \\
\addlinespace[0.3em]
    \raisebox{-0.4ex}{\includegraphics[height=1.1em]{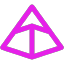}}\,PydanticAI & 3\,/\,3 & 1,469K\,/\,891K & 9K\,/\,11K & 808K\,/\,846K & 2.24\,/\,0.93 & \raisebox{-0.4ex}{\includegraphics[height=1.1em]{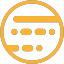}}\,Griptape & 3\,/\,0 & 1,706K\,/\,1,824K & 13K\,/\,11K & 939K\,/\,1,767K & 2.63\,/\,1.45 \\
    \raisebox{-0.4ex}{\includegraphics[height=1.1em]{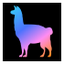}}\,LlamaIndex & 1\,/\,1 & 1,726K\,/\,1,095K & 12K\,/\,11K & 1,127K\,/\,1,050K & 2.32\,/\,1.02 & \raisebox{-0.4ex}{\includegraphics[height=1.1em]{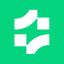}}\,Upsonic & 2\,/\,1 & 1,321K\,/\,1,449K & 9K\,/\,7K & 849K\,/\,1,397K & 1.79\,/\,1.14 \\
    \raisebox{-0.4ex}{\includegraphics[height=1.1em]{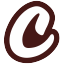}}\,CrewAI & 3\,/\,4 & 1,433K\,/\,1,155K & 10K\,/\,12K & 817K\,/\,1,040K & 2.16\,/\,1.39 & \raisebox{-0.4ex}{\includegraphics[height=1.1em]{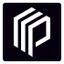}}\,ControlFlow & 0\,/\,0 & 1,972K\,/\,1,107K & 8K\,/\,8K & 1,214K\,/\,1,065K & 2.73\,/\,0.94 \\
    \raisebox{-0.4ex}{\includegraphics[height=1.1em]{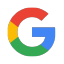}}\,Google ADK & 2\,/\,3 & 1,777K\,/\,1,436K & 18K\,/\,11K & 1,091K\,/\,1,383K & 2.58\,/\,1.24 & \raisebox{-0.4ex}{\includegraphics[height=1.1em]{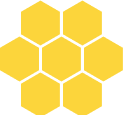}}\,AgencySwarm & 3\,/\,3 & 1,796K\,/\,1,156K & 12K\,/\,8K & 1,078K\,/\,1,051K & 2.59\,/\,1.26 \\
\addlinespace[0.3em]
    \raisebox{-0.4ex}{\includegraphics[height=1.1em]{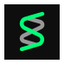}}\,Strands & 2\,/\,2 & 1,217K\,/\,1,064K & 9K\,/\,8K & 815K\,/\,1,019K & 1.60\,/\,0.93 & \raisebox{-0.4ex}{\includegraphics[height=1.1em]{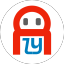}}\,Agently & 2\,/\,2 & 1,648K\,/\,1,288K & 9K\,/\,15K & 1,018K\,/\,1,233K & 2.30\,/\,1.27 \\
    \raisebox{-0.4ex}{\includegraphics[height=1.1em]{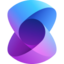}}\,SemanticKernel & 2\,/\,2 & 1,397K\,/\,1,180K & 8K\,/\,12K & 843K\,/\,1,138K & 1.99\,/\,1.07 & \raisebox{-0.4ex}{\includegraphics[height=1.1em]{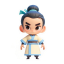}}\,Lagent & 1\,/\,2 & 1,456K\,/\,1,345K & 9K\,/\,20K & 816K\,/\,1,284K & 2.20\,/\,1.44 \\
    \raisebox{-0.4ex}{\includegraphics[height=1.1em]{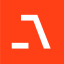}}\,Agno & 4\,/\,3 & 1,953K\,/\,1,452K & 17K\,/\,14K & 1,204K\,/\,1,365K & 2.79\,/\,1.46 & \raisebox{-0.4ex}{\includegraphics[height=1.1em]{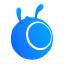}}\,AgentUniverse & 1\,/\,4 & 1,999K\,/\,1,482K & 16K\,/\,15K & 1,255K\,/\,1,405K & 2.81\,/\,1.47 \\
    \raisebox{-0.4ex}{\includegraphics[height=1.1em]{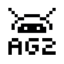}}\,AG2 & 4\,/\,4 & 1,679K\,/\,1,078K & 13K\,/\,14K & 839K\,/\,1,016K & 2.75\,/\,1.15 & \raisebox{-0.4ex}{\includegraphics[height=1.1em]{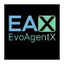}}\,EvoAgentX & 1\,/\,2 & 1,558K\,/\,1,665K & 9K\,/\,14K & 929K\,/\,1,597K & 2.24\,/\,1.49 \\
\addlinespace[0.3em]
    \raisebox{-0.4ex}{\includegraphics[height=1.1em]{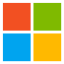}}\,AutoGen & 3\,/\,4 & 1,812K\,/\,1,008K & 12K\,/\,14K & 1,123K\,/\,953K & 2.55\,/\,1.10 & \raisebox{-0.4ex}{\includegraphics[height=1.1em]{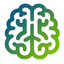}}\,Nerve & 3\,/\,3 & 2,271K\,/\,1,484K & 12K\,/\,11K & 1,517K\,/\,1,409K & 2.95\,/\,1.37 \\
    \raisebox{-0.4ex}{\includegraphics[height=1.1em]{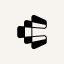}}\,Composio & 2\,/\,4 & 2,171K\,/\,1,481K & 13K\,/\,14K & 1,301K\,/\,1,424K & 3.12\,/\,1.34 & \raisebox{-0.4ex}{\includegraphics[height=1.1em]{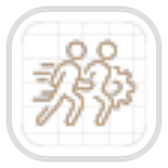}}\,AgentSquad & 3\,/\,4 & 2,048K\,/\,1,217K & 16K\,/\,14K & 1,139K\,/\,1,100K & 3.15\,/\,1.49 \\
    \raisebox{-0.4ex}{\includegraphics[height=1.1em]{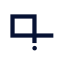}}\,Haystack & 4\,/\,4 & 1,447K\,/\,1,239K & 13K\,/\,14K & 935K\,/\,1,156K & 1.99\,/\,1.29 & \raisebox{-0.4ex}{\includegraphics[height=1.1em]{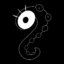}}\,MotleyCrew & 3\,/\,2 & 1,735K\,/\,1,356K & 14K\,/\,10K & 1,057K\,/\,1,307K & 2.49\,/\,1.14 \\
    \raisebox{-0.4ex}{\includegraphics[height=1.1em]{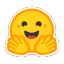}}\,SmoLAgents & 2\,/\,3 & 1,014K\,/\,1,057K & 9K\,/\,12K & 596K\,/\,1,007K & 1.51\,/\,1.04 & \raisebox{-0.4ex}{\includegraphics[height=1.1em]{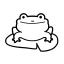}}\,TaskflowAI & 1\,/\,1 & 1,781K\,/\,446K & 11K\,/\,7K & 985K\,/\,399K & 2.71\,/\,0.61 \\
\addlinespace[0.3em]
    \raisebox{-0.4ex}{\includegraphics[height=1.1em]{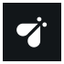}}\,AgentFramework & 3\,/\,4 & 2,040K\,/\,1,447K & 13K\,/\,13K & 995K\,/\,1,361K & 3.36\,/\,1.44 & \raisebox{-0.4ex}{\includegraphics[height=1.1em]{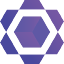}}\,CouncilAI & 1\,/\,2 & 1,814K\,/\,2,618K & 14K\,/\,21K & 1,050K\,/\,2,366K & 2.70\,/\,2.97 \\
    \raisebox{-0.4ex}{\includegraphics[height=1.1em]{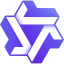}}\,Qwen Agent & 2\,/\,0 & 1,265K\,/\,920K & 10K\,/\,10K & 695K\,/\,878K & 1.95\,/\,0.90 & \raisebox{-0.4ex}{\includegraphics[height=1.1em]{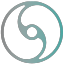}}\,AgentFlow & 1\,/\,1 & 1,567K\,/\,1,926K & 10K\,/\,11K & 931K\,/\,1,850K & 2.27\,/\,1.59 \\
    \raisebox{-0.4ex}{\includegraphics[height=1.1em]{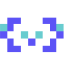}}\,AgentScope & 2\,/\,2 & 1,371K\,/\,2,211K & 10K\,/\,12K & 864K\,/\,2,142K & 1.91\,/\,1.71 & \raisebox{-0.4ex}{\includegraphics[height=1.1em]{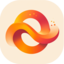}}\,AutoAgent & 2\,/\,3 & 1,807K\,/\,763K & 10K\,/\,8K & 1,075K\,/\,715K & 2.60\,/\,0.79 \\
    \raisebox{-0.4ex}{\includegraphics[height=1.1em]{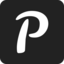}}\,PraisonAI & 3\,/\,3 & 2,138K\,/\,1,447K & 14K\,/\,13K & 1,314K\,/\,1,387K & 3.02\,/\,1.31 & \raisebox{-0.4ex}{\includegraphics[height=1.1em]{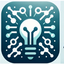}}\,AgentLite & 3\,/\,3 & 1,565K\,/\,760K & 10K\,/\,43K & 939K\,/\,706K & 2.26\,/\,1.71 \\
\addlinespace[0.3em]
    \raisebox{-0.4ex}{\includegraphics[height=1.1em]{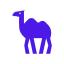}}\,CAMEL & 3\,/\,3 & 1,550K\,/\,1,083K & 14K\,/\,11K & 878K\,/\,1,030K & 2.36\,/\,1.06 & \raisebox{-0.4ex}{\includegraphics[height=1.1em]{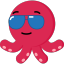}}\,Octotools & 3\,/\,2 & 1,783K\,/\,936K & 13K\,/\,30K & 1,109K\,/\,832K & 2.51\,/\,1.69 \\
    \raisebox{-0.4ex}{\includegraphics[height=1.1em]{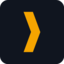}}\,FastAgent & 0\,/\,4 & 2,385K\,/\,1,786K & 10K\,/\,12K & 1,583K\,/\,1,721K & 3.09\,/\,1.48 & \raisebox{-0.4ex}{\includegraphics[height=1.1em]{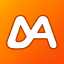}}\,GPTSwarm & 2\,/\,2 & 2,039K\,/\,1,552K & 11K\,/\,12K & 1,335K\,/\,1,497K & 2.71\,/\,1.33 \\
    \raisebox{-0.4ex}{\includegraphics[height=1.1em]{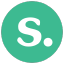}}\,Solace & 1\,/\,1 & 979K\,/\,995K & 10K\,/\,7K & 565K\,/\,946K & 1.49\,/\,0.90 & \raisebox{-0.4ex}{\includegraphics[height=1.1em]{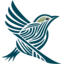}}\,Taskweaver & 1\,/\,1 & 2,091K\,/\,730K & 13K\,/\,8K & 1,354K\,/\,682K & 2.82\,/\,0.79 \\
    \raisebox{-0.4ex}{\includegraphics[height=1.1em]{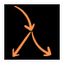}}\,Langroid & 0\,/\,0 & 512K\,/\,171K & 4K\,/\,1K & 334K\,/\,142K & 0.69\,/\,0.24 & \raisebox{-0.4ex}{\includegraphics[height=1.1em]{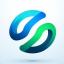}}\,AutoAgents & 1\,/\,2 & 2,072K\,/\,1,165K & 12K\,/\,36K & 1,269K\,/\,1,106K & 2.92\,/\,1.75 \\
\addlinespace[0.3em]
    \raisebox{-0.4ex}{\includegraphics[height=1.1em]{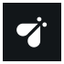}}\,BeeAI & 2\,/\,2 & 2,169K\,/\,1,346K & 17K\,/\,37K & 1,416K\,/\,1,288K & 2.94\,/\,1.85 & \raisebox{-0.4ex}{\includegraphics[height=1.1em]{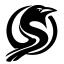}}\,OpenSage & 3\,/\,1 & 1,873K\,/\,991K & 11K\,/\,9K & 1,140K\,/\,938K & 2.65\,/\,0.95 \\
    \raisebox{-0.4ex}{\includegraphics[height=1.1em]{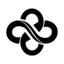}}\,MetaGPT & 3\,/\,1 & 1,921K\,/\,1,063K & 11K\,/\,7K & 1,202K\,/\,1,010K & 2.66\,/\,0.95 &  &  &  &  &  &  \\
\bottomrule
\end{tabular}%
}
\end{table*}

We evaluate 51 frameworks along two dimensions: generation performance (\S\ref{sec:gen_results}), measuring how efficiently the LLM developer learns and produces working agent code, and execution performance (\S\ref{sec:agent_quality}), measuring how well the resulting agents solve real tasks. We then compare against frontier coding agents (\S\ref{sec:reference}) and stress-test the pipeline by ablating information sources (\S\ref{sec:ablation}).

\subsection{Generation Performance}
\label{sec:gen_results}

\autoref{tab:gen_effort} reports generation effort for all 51 frameworks across four benchmarks. Of the 408 generated agents (two per framework--benchmark pair, one from each developer model), 232 (57\%) pass all three validation levels, showing that fully automated agent generation is viable for over half of the ecosystem. Per-agent generation cost (averaged across benchmarks) ranges from \$0.6 to \$3.4, a 5.6$\times$ spread.

Cost tracks API usability rather than raw spend. The most robust frameworks pass all four benchmarks under \emph{both} developer models, \textit{Haystack} (\$2.0\,/\,\$1.3 per agent for GPT\,/\,Opus) and \textit{AG2} (\$2.8\,/\,\$1.2), exposing clean, learnable APIs. The cheapest runs are \emph{not} the best: \textit{Langroid} (\$0.2--0.7), \textit{Solace} (\$0.9--1.5), and \textit{Qwen Agent} produce few or no passing agents, because with no productive path into the API the developer gives up early and each attempt stays short. The highest-cost frameworks (\textit{MCP Agent}, \textit{AgentFramework}, \textit{AtomicAgents}, all near \$3.4 on the GPT side) carry the largest API surfaces, yet high spend does not doom them: MCP Agent still passes 3/4. Cost thus signals API complexity, not a clean success/failure split.

\finding{The pipeline produces a validated agent for 57\% of runs (232/408). Per-agent generation cost varies 5.6$\times$ across frameworks and serves as a quantitative proxy for API complexity, though cost alone does not predict success.}

\subsection{Execution Performance}
\label{sec:agent_quality}

\begingroup
\scriptsize
\setlength{\LTcapwidth}{\textwidth}
\setlength{\tabcolsep}{3pt}
\renewcommand{\arraystretch}{1.15}
\begin{longtable}{@{}l ccccc@{}}
\caption{Full execution performance for all 51 ADK frameworks (normal condition). Each cell packs the four benchmarks in order SWE-bench\,/\,$\tau^2$-bench\,/\,MCP-Atlas\,/\,TerminalBench. First row: agent generated by GPT-5.4 (Codex). Second row: by Opus-4.6 (Claude Code). Both execute on GPT-5.4 Nano. ``--'' = agent failed to build/execute or no data recorded.}\label{tab:exec_full}\\
\toprule
\textbf{Framework} & \textbf{Resolve (\%)} & \textbf{In Tok} & \textbf{Out Tok} & \textbf{Cache Tok} & \textbf{Cost (\$)} \\
\midrule
\endfirsthead
\multicolumn{6}{c}{\tablename\ \thetable{} -- continued from previous page}\\
\toprule
\textbf{Framework} & \textbf{Resolve (\%)} & \textbf{In Tok} & \textbf{Out Tok} & \textbf{Cache Tok} & \textbf{Cost (\$)} \\
\midrule
\endhead
\midrule
\multicolumn{6}{r}{\itshape continued on next page}\\
\endfoot
\bottomrule
\endlastfoot
\nopagebreak
    \multirow{2}{*}{\raisebox{-0.4ex}{\includegraphics[height=1.1em]{figures/logos/langchain.png}}\,LangChain} & 44{\tiny/}4{\tiny/}58{\tiny/}40 & 910.5K{\tiny/}1.4M{\tiny/}54.0K{\tiny/}471.8K & 11.4K{\tiny/}15.1K{\tiny/}2.5K{\tiny/}17.0K & 580.8K{\tiny/}317.7K{\tiny/}17.4K{\tiny/}376.8K & 0.13{\tiny/}0.26{\tiny/}0.01{\tiny/}0.07 \\*
     & 30{\tiny/}76{\tiny/}--{\tiny/}48 & 436.9K{\tiny/}55.0K{\tiny/}--{\tiny/}374.4K & 10.0K{\tiny/}9.3K{\tiny/}--{\tiny/}19.9K & 269.3K{\tiny/}35.6K{\tiny/}--{\tiny/}259.5K & 0.07{\tiny/}0.01{\tiny/}--{\tiny/}0.06 \\
\addlinespace[0.25em]
\nopagebreak
    \multirow{2}{*}{\raisebox{-0.4ex}{\includegraphics[height=1.1em]{figures/logos/anthropic-sdk.png}}\,Anthropic SDK} & 0{\tiny/}66{\tiny/}0{\tiny/}0 & 973{\tiny/}76.7K{\tiny/}67{\tiny/}1.9K & 1.2K{\tiny/}9.2K{\tiny/}252{\tiny/}1.7K & 148{\tiny/}51.8K{\tiny/}0{\tiny/}0 & 0.00{\tiny/}0.02{\tiny/}0.00{\tiny/}0.00 \\*
     & --{\tiny/}60{\tiny/}--{\tiny/}-- & --{\tiny/}63.5K{\tiny/}--{\tiny/}-- & --{\tiny/}10.6K{\tiny/}--{\tiny/}-- & --{\tiny/}40.1K{\tiny/}--{\tiny/}-- & --{\tiny/}0.02{\tiny/}--{\tiny/}-- \\
\addlinespace[0.25em]
\nopagebreak
    \multirow{2}{*}{\raisebox{-0.4ex}{\includegraphics[height=1.1em]{figures/logos/langgraph.png}}\,LangGraph} & 0{\tiny/}0{\tiny/}--{\tiny/}16 & 4.0K{\tiny/}0{\tiny/}--{\tiny/}104.4K & 2.4K{\tiny/}0{\tiny/}--{\tiny/}6.0K & 0{\tiny/}0{\tiny/}--{\tiny/}79.1K & 0.00{\tiny/}0.00{\tiny/}--{\tiny/}0.02 \\*
     & 26{\tiny/}66{\tiny/}46{\tiny/}36 & 310.3K{\tiny/}80.1K{\tiny/}45.7K{\tiny/}552.1K & 8.2K{\tiny/}10.4K{\tiny/}2.6K{\tiny/}19.0K & 177.1K{\tiny/}48.1K{\tiny/}20.1K{\tiny/}397.8K & 0.05{\tiny/}0.02{\tiny/}0.01{\tiny/}0.09 \\
\addlinespace[0.25em]
\nopagebreak
    \multirow{2}{*}{\raisebox{-0.4ex}{\includegraphics[height=1.1em]{figures/logos/openai-agents.png}}\,OpenAI Agents} & 6{\tiny/}74{\tiny/}--{\tiny/}10 & 59.2K{\tiny/}69.5K{\tiny/}--{\tiny/}30.5K & 2.5K{\tiny/}10.2K{\tiny/}--{\tiny/}9.6K & 20.2K{\tiny/}44.6K{\tiny/}--{\tiny/}12.2K & 0.01{\tiny/}0.02{\tiny/}--{\tiny/}0.01 \\*
     & 44{\tiny/}70{\tiny/}--{\tiny/}54 & 1.6M{\tiny/}57.9K{\tiny/}--{\tiny/}1.1M & 34.6K{\tiny/}9.5K{\tiny/}--{\tiny/}42.4K & 1.3M{\tiny/}39.2K{\tiny/}--{\tiny/}850.9K & 0.23{\tiny/}0.02{\tiny/}--{\tiny/}0.17 \\
\addlinespace[0.25em]
\nopagebreak
    \multirow{2}{*}{\raisebox{-0.4ex}{\includegraphics[height=1.1em]{figures/logos/pydantic-ai.png}}\,PydanticAI} & 0{\tiny/}42{\tiny/}--{\tiny/}2 & 6.2K{\tiny/}41.2K{\tiny/}--{\tiny/}2.8K & 10.7K{\tiny/}10.0K{\tiny/}--{\tiny/}2.4K & 2.4K{\tiny/}23.9K{\tiny/}--{\tiny/}594 & 0.01{\tiny/}0.01{\tiny/}--{\tiny/}0.00 \\*
     & 30{\tiny/}66{\tiny/}--{\tiny/}14 & 920.1K{\tiny/}50.4K{\tiny/}--{\tiny/}143.4K & 24.5K{\tiny/}8.1K{\tiny/}--{\tiny/}8.4K & 502.6K{\tiny/}30.5K{\tiny/}--{\tiny/}96.5K & 0.15{\tiny/}0.01{\tiny/}--{\tiny/}0.03 \\
\addlinespace[0.25em]
\nopagebreak
    \multirow{2}{*}{\raisebox{-0.4ex}{\includegraphics[height=1.1em]{figures/logos/llamaindex.png}}\,LlamaIndex} & --{\tiny/}0{\tiny/}0{\tiny/}-- & --{\tiny/}1.8K{\tiny/}0{\tiny/}-- & --{\tiny/}822{\tiny/}0{\tiny/}-- & --{\tiny/}0{\tiny/}0{\tiny/}-- & --{\tiny/}0.00{\tiny/}0.00{\tiny/}-- \\*
     & --{\tiny/}68{\tiny/}--{\tiny/}-- & --{\tiny/}61.1K{\tiny/}--{\tiny/}-- & --{\tiny/}9.4K{\tiny/}--{\tiny/}-- & --{\tiny/}39.7K{\tiny/}--{\tiny/}-- & --{\tiny/}0.02{\tiny/}--{\tiny/}-- \\
\addlinespace[0.25em]
\nopagebreak
    \multirow{2}{*}{\raisebox{-0.4ex}{\includegraphics[height=1.1em]{figures/logos/crewai.png}}\,CrewAI} & 30{\tiny/}44{\tiny/}--{\tiny/}0 & 760.5K{\tiny/}42.4K{\tiny/}--{\tiny/}202.5K & 20.5K{\tiny/}12.0K{\tiny/}--{\tiny/}17.5K & 555.7K{\tiny/}22.1K{\tiny/}--{\tiny/}129.1K & 0.11{\tiny/}0.02{\tiny/}--{\tiny/}0.04 \\*
     & 46{\tiny/}0{\tiny/}30{\tiny/}56 & 2.7M{\tiny/}1.8K{\tiny/}23.1K{\tiny/}803.4K & 17.3K{\tiny/}799{\tiny/}2.2K{\tiny/}27.5K & 2.4M{\tiny/}0{\tiny/}12.2K{\tiny/}577.6K & 0.32{\tiny/}0.00{\tiny/}0.01{\tiny/}0.12 \\
\addlinespace[0.25em]
\nopagebreak
    \multirow{2}{*}{\raisebox{-0.4ex}{\includegraphics[height=1.1em]{figures/logos/google-adk.png}}\,Google ADK} & 44{\tiny/}--{\tiny/}--{\tiny/}0 & 1.6M{\tiny/}--{\tiny/}--{\tiny/}723 & 37.6K{\tiny/}--{\tiny/}--{\tiny/}8.1K & 1.1M{\tiny/}--{\tiny/}--{\tiny/}0 & 0.23{\tiny/}--{\tiny/}--{\tiny/}0.01 \\*
     & --{\tiny/}80{\tiny/}42{\tiny/}52 & --{\tiny/}70.9K{\tiny/}12.9K{\tiny/}480.4K & --{\tiny/}10.9K{\tiny/}1.1K{\tiny/}15.4K & --{\tiny/}47.9K{\tiny/}3.2K{\tiny/}367.4K & --{\tiny/}0.02{\tiny/}0.00{\tiny/}0.07 \\
\addlinespace[0.25em]
\nopagebreak
    \multirow{2}{*}{\raisebox{-0.4ex}{\includegraphics[height=1.1em]{figures/logos/strands-agents.png}}\,Strands} & --{\tiny/}74{\tiny/}--{\tiny/}10 & --{\tiny/}55.5K{\tiny/}--{\tiny/}13.6K & --{\tiny/}9.0K{\tiny/}--{\tiny/}3.0K & --{\tiny/}36.4K{\tiny/}--{\tiny/}3.6K & --{\tiny/}0.01{\tiny/}--{\tiny/}0.00 \\*
     & 16{\tiny/}34{\tiny/}--{\tiny/}24 & 787.5K{\tiny/}76.0K{\tiny/}--{\tiny/}518.6K & 9.1K{\tiny/}8.9K{\tiny/}--{\tiny/}19.2K & 501.7K{\tiny/}52.4K{\tiny/}--{\tiny/}407.0K & 0.11{\tiny/}0.02{\tiny/}--{\tiny/}0.08 \\
\addlinespace[0.25em]
\nopagebreak
    \multirow{2}{*}{\raisebox{-0.4ex}{\includegraphics[height=1.1em]{figures/logos/semantic-kernel.png}}\,SemanticKernel} & --{\tiny/}40{\tiny/}2{\tiny/}-- & --{\tiny/}64.5K{\tiny/}8.0K{\tiny/}-- & --{\tiny/}9.8K{\tiny/}3.8K{\tiny/}-- & --{\tiny/}33.9K{\tiny/}3.5K{\tiny/}-- & --{\tiny/}0.02{\tiny/}0.00{\tiny/}-- \\*
     & --{\tiny/}0{\tiny/}--{\tiny/}6 & --{\tiny/}25.3K{\tiny/}--{\tiny/}16.2K & --{\tiny/}5.5K{\tiny/}--{\tiny/}6.0K & --{\tiny/}18.6K{\tiny/}--{\tiny/}1.1K & --{\tiny/}0.01{\tiny/}--{\tiny/}0.01 \\
\addlinespace[0.25em]
\nopagebreak
    \multirow{2}{*}{\raisebox{-0.4ex}{\includegraphics[height=1.1em]{figures/logos/agno.png}}\,Agno} & 34{\tiny/}0{\tiny/}68{\tiny/}0 & 1.4M{\tiny/}12.3K{\tiny/}38.3K{\tiny/}5.1K & 13.0K{\tiny/}3.3K{\tiny/}3.1K{\tiny/}1.1K & 1.1M{\tiny/}7.2K{\tiny/}11.5K{\tiny/}0 & 0.18{\tiny/}0.00{\tiny/}0.01{\tiny/}0.00 \\*
     & 52{\tiny/}60{\tiny/}--{\tiny/}54 & 1.3M{\tiny/}48.7K{\tiny/}--{\tiny/}763.2K & 11.9K{\tiny/}8.3K{\tiny/}--{\tiny/}20.8K & 960.5K{\tiny/}31.7K{\tiny/}--{\tiny/}558.8K & 0.18{\tiny/}0.01{\tiny/}--{\tiny/}0.11 \\
\addlinespace[0.25em]
\nopagebreak
    \multirow{2}{*}{\raisebox{-0.4ex}{\includegraphics[height=1.1em]{figures/logos/ag2.png}}\,AG2} & 0{\tiny/}60{\tiny/}0{\tiny/}6 & 541.3K{\tiny/}48.2K{\tiny/}17.3K{\tiny/}563.3K & 16.4K{\tiny/}7.3K{\tiny/}572{\tiny/}14.5K & 508.0K{\tiny/}34.5K{\tiny/}15.0K{\tiny/}496.6K & 0.07{\tiny/}0.01{\tiny/}0.00{\tiny/}0.07 \\*
     & 48{\tiny/}54{\tiny/}62{\tiny/}60 & 1.2M{\tiny/}46.8K{\tiny/}60.2K{\tiny/}0 & 17.1K{\tiny/}7.1K{\tiny/}2.9K{\tiny/}0 & 892.3K{\tiny/}33.5K{\tiny/}20.8K{\tiny/}0 & 0.17{\tiny/}0.01{\tiny/}0.01{\tiny/}0.00 \\
\addlinespace[0.25em]
\nopagebreak
    \multirow{2}{*}{\raisebox{-0.4ex}{\includegraphics[height=1.1em]{figures/logos/microsoft.png}}\,AutoGen} & 34{\tiny/}14{\tiny/}--{\tiny/}0 & 1.7M{\tiny/}18.5K{\tiny/}--{\tiny/}131.0K & 18.6K{\tiny/}5.4K{\tiny/}--{\tiny/}14.5K & 1.3M{\tiny/}12.2K{\tiny/}--{\tiny/}88.2K & 0.22{\tiny/}0.01{\tiny/}--{\tiny/}0.03 \\*
     & 40{\tiny/}70{\tiny/}46{\tiny/}48 & 762.4K{\tiny/}55.9K{\tiny/}45.7K{\tiny/}456.1K & 7.4K{\tiny/}7.9K{\tiny/}2.3K{\tiny/}15.7K & 530.1K{\tiny/}35.8K{\tiny/}18.5K{\tiny/}304.8K & 0.11{\tiny/}0.01{\tiny/}0.01{\tiny/}0.07 \\
\addlinespace[0.25em]
\nopagebreak
    \multirow{2}{*}{\raisebox{-0.4ex}{\includegraphics[height=1.1em]{figures/logos/composio.png}}\,Composio} & 12{\tiny/}--{\tiny/}--{\tiny/}4 & 59.9K{\tiny/}--{\tiny/}--{\tiny/}305.7K & 2.6K{\tiny/}--{\tiny/}--{\tiny/}9.4K & 35.4K{\tiny/}--{\tiny/}--{\tiny/}275.2K & 0.01{\tiny/}--{\tiny/}--{\tiny/}0.04 \\*
     & 58{\tiny/}70{\tiny/}40{\tiny/}46 & 1.0M{\tiny/}72.3K{\tiny/}22.6K{\tiny/}514.2K & 12.9K{\tiny/}10.4K{\tiny/}1.7K{\tiny/}17.3K & 671.3K{\tiny/}42.4K{\tiny/}3.4K{\tiny/}352.7K & 0.15{\tiny/}0.02{\tiny/}0.01{\tiny/}0.08 \\
\addlinespace[0.25em]
\nopagebreak
    \multirow{2}{*}{\raisebox{-0.4ex}{\includegraphics[height=1.1em]{figures/logos/haystack.png}}\,Haystack} & 0{\tiny/}66{\tiny/}8{\tiny/}0 & 41.5K{\tiny/}43.3K{\tiny/}1.0K{\tiny/}3.4K & 14.6K{\tiny/}7.2K{\tiny/}728{\tiny/}4.3K & 14.0K{\tiny/}26.6K{\tiny/}0{\tiny/}26 & 0.02{\tiny/}0.01{\tiny/}0.00{\tiny/}0.00 \\*
     & 36{\tiny/}74{\tiny/}44{\tiny/}52 & 702.3K{\tiny/}65.9K{\tiny/}76.1K{\tiny/}641.4K & 11.9K{\tiny/}8.7K{\tiny/}2.3K{\tiny/}22.1K & 467.1K{\tiny/}47.4K{\tiny/}22.2K{\tiny/}447.9K & 0.10{\tiny/}0.02{\tiny/}0.01{\tiny/}0.10 \\
\addlinespace[0.25em]
\nopagebreak
    \multirow{2}{*}{\raisebox{-0.4ex}{\includegraphics[height=1.1em]{figures/logos/smolagents.png}}\,SmoLAgents} & 4{\tiny/}66{\tiny/}--{\tiny/}-- & 32.8K{\tiny/}55.1K{\tiny/}--{\tiny/}-- & 9.7K{\tiny/}7.6K{\tiny/}--{\tiny/}-- & 16.3K{\tiny/}39.2K{\tiny/}--{\tiny/}-- & 0.01{\tiny/}0.01{\tiny/}--{\tiny/}-- \\*
     & 28{\tiny/}0{\tiny/}--{\tiny/}32 & 911.0K{\tiny/}1.8K{\tiny/}--{\tiny/}766.1K & 13.6K{\tiny/}814{\tiny/}--{\tiny/}35.7K & 590.3K{\tiny/}0{\tiny/}--{\tiny/}626.7K & 0.13{\tiny/}0.00{\tiny/}--{\tiny/}0.12 \\
\addlinespace[0.25em]
\nopagebreak
    \multirow{2}{*}{\raisebox{-0.4ex}{\includegraphics[height=1.1em]{figures/logos/agent-framework.png}}\,AgentFramework} & --{\tiny/}2{\tiny/}68{\tiny/}14 & --{\tiny/}5.8K{\tiny/}108.2K{\tiny/}244.5K & --{\tiny/}1.8K{\tiny/}3.9K{\tiny/}9.4K & --{\tiny/}3.6K{\tiny/}45.3K{\tiny/}0 & --{\tiny/}0.00{\tiny/}0.02{\tiny/}0.06 \\*
     & 80{\tiny/}38{\tiny/}32{\tiny/}40 & 2.0M{\tiny/}87.4K{\tiny/}36.4K{\tiny/}356.3K & 13.7K{\tiny/}5.9K{\tiny/}1.3K{\tiny/}24.5K & 0{\tiny/}65.2K{\tiny/}0{\tiny/}213.4K & 0.40{\tiny/}0.02{\tiny/}0.01{\tiny/}0.07 \\
\addlinespace[0.25em]
\nopagebreak
    \multirow{2}{*}{\raisebox{-0.4ex}{\includegraphics[height=1.1em]{figures/logos/qwen-agent.png}}\,Qwen Agent} & 0{\tiny/}44{\tiny/}--{\tiny/}-- & 591{\tiny/}40.2K{\tiny/}--{\tiny/}-- & 85{\tiny/}9.4K{\tiny/}--{\tiny/}-- & 0{\tiny/}20.2K{\tiny/}--{\tiny/}-- & 0.00{\tiny/}0.01{\tiny/}--{\tiny/}-- \\*
     & --{\tiny/}--{\tiny/}--{\tiny/}-- & --{\tiny/}--{\tiny/}--{\tiny/}-- & --{\tiny/}--{\tiny/}--{\tiny/}-- & --{\tiny/}--{\tiny/}--{\tiny/}-- & --{\tiny/}--{\tiny/}--{\tiny/}-- \\
\addlinespace[0.25em]
\nopagebreak
    \multirow{2}{*}{\raisebox{-0.4ex}{\includegraphics[height=1.1em]{figures/logos/agentscope.png}}\,AgentScope} & 0{\tiny/}--{\tiny/}--{\tiny/}0 & 3.1K{\tiny/}--{\tiny/}--{\tiny/}17.3K & 3.0K{\tiny/}--{\tiny/}--{\tiny/}10.2K & 92{\tiny/}--{\tiny/}--{\tiny/}2.2K & 0.00{\tiny/}--{\tiny/}--{\tiny/}0.01 \\*
     & 56{\tiny/}0{\tiny/}--{\tiny/}-- & 643.0K{\tiny/}14.2K{\tiny/}--{\tiny/}-- & 17.1K{\tiny/}4.1K{\tiny/}--{\tiny/}-- & 537.2K{\tiny/}11.1K{\tiny/}--{\tiny/}-- & 0.09{\tiny/}0.01{\tiny/}--{\tiny/}-- \\
\addlinespace[0.25em]
\nopagebreak
    \multirow{2}{*}{\raisebox{-0.4ex}{\includegraphics[height=1.1em]{figures/logos/praisonai.png}}\,PraisonAI} & 0{\tiny/}14{\tiny/}--{\tiny/}32 & 2.3K{\tiny/}58.5K{\tiny/}--{\tiny/}274.4K & 8.5K{\tiny/}4.7K{\tiny/}--{\tiny/}13.8K & 328{\tiny/}39.3K{\tiny/}--{\tiny/}202.3K & 0.01{\tiny/}0.01{\tiny/}--{\tiny/}0.05 \\*
     & 16{\tiny/}76{\tiny/}--{\tiny/}2 & 98.0K{\tiny/}64.9K{\tiny/}--{\tiny/}11.6K & 3.5K{\tiny/}9.7K{\tiny/}--{\tiny/}1.7K & 16.6K{\tiny/}40.2K{\tiny/}--{\tiny/}5.3K & 0.02{\tiny/}0.02{\tiny/}--{\tiny/}0.00 \\
\addlinespace[0.25em]
\nopagebreak
    \multirow{2}{*}{\raisebox{-0.4ex}{\includegraphics[height=1.1em]{figures/logos/camel.png}}\,CAMEL} & 50{\tiny/}22{\tiny/}--{\tiny/}4 & 1.1M{\tiny/}45.4K{\tiny/}--{\tiny/}144.6K & 17.0K{\tiny/}5.8K{\tiny/}--{\tiny/}8.2K & 843.9K{\tiny/}30.8K{\tiny/}--{\tiny/}86.3K & 0.15{\tiny/}0.01{\tiny/}--{\tiny/}0.03 \\*
     & 52{\tiny/}64{\tiny/}--{\tiny/}60 & 1.1M{\tiny/}50.0K{\tiny/}--{\tiny/}417.3K & 15.1K{\tiny/}6.7K{\tiny/}--{\tiny/}20.5K & 731.1K{\tiny/}29.9K{\tiny/}--{\tiny/}262.1K & 0.15{\tiny/}0.01{\tiny/}--{\tiny/}0.07 \\
\addlinespace[0.25em]
\nopagebreak
    \multirow{2}{*}{\raisebox{-0.4ex}{\includegraphics[height=1.1em]{figures/logos/fast-agent.png}}\,FastAgent} & --{\tiny/}0{\tiny/}--{\tiny/}-- & --{\tiny/}0{\tiny/}--{\tiny/}-- & --{\tiny/}0{\tiny/}--{\tiny/}-- & --{\tiny/}0{\tiny/}--{\tiny/}-- & --{\tiny/}0.00{\tiny/}--{\tiny/}-- \\*
     & 0{\tiny/}72{\tiny/}2{\tiny/}40 & 469.8K{\tiny/}61.9K{\tiny/}388{\tiny/}245.3K & 7.4K{\tiny/}10.6K{\tiny/}63{\tiny/}5.2K & 335.3K{\tiny/}40.8K{\tiny/}0{\tiny/}223.5K & 0.07{\tiny/}0.02{\tiny/}0.00{\tiny/}0.03 \\
\addlinespace[0.25em]
\nopagebreak
    \multirow{2}{*}{\raisebox{-0.4ex}{\includegraphics[height=1.1em]{figures/logos/solace-agent-mesh.png}}\,Solace} & --{\tiny/}68{\tiny/}--{\tiny/}-- & --{\tiny/}56.3K{\tiny/}--{\tiny/}-- & --{\tiny/}9.2K{\tiny/}--{\tiny/}-- & --{\tiny/}36.9K{\tiny/}--{\tiny/}-- & --{\tiny/}0.01{\tiny/}--{\tiny/}-- \\*
     & --{\tiny/}--{\tiny/}--{\tiny/}32 & --{\tiny/}--{\tiny/}--{\tiny/}1.1M & --{\tiny/}--{\tiny/}--{\tiny/}39.6K & --{\tiny/}--{\tiny/}--{\tiny/}874.5K & --{\tiny/}--{\tiny/}--{\tiny/}0.17 \\
\addlinespace[0.25em]
\nopagebreak
    \multirow{2}{*}{\raisebox{-0.4ex}{\includegraphics[height=1.1em]{figures/logos/langroid.png}}\,Langroid} & --{\tiny/}--{\tiny/}--{\tiny/}-- & --{\tiny/}--{\tiny/}--{\tiny/}-- & --{\tiny/}--{\tiny/}--{\tiny/}-- & --{\tiny/}--{\tiny/}--{\tiny/}-- & --{\tiny/}--{\tiny/}--{\tiny/}-- \\*
     & --{\tiny/}--{\tiny/}--{\tiny/}-- & --{\tiny/}--{\tiny/}--{\tiny/}-- & --{\tiny/}--{\tiny/}--{\tiny/}-- & --{\tiny/}--{\tiny/}--{\tiny/}-- & --{\tiny/}--{\tiny/}--{\tiny/}-- \\
\addlinespace[0.25em]
\nopagebreak
    \multirow{2}{*}{\raisebox{-0.4ex}{\includegraphics[height=1.1em]{figures/logos/beeai.png}}\,BeeAI} & 4{\tiny/}--{\tiny/}--{\tiny/}0 & 5.2K{\tiny/}--{\tiny/}--{\tiny/}4.0K & 17.4K{\tiny/}--{\tiny/}--{\tiny/}23.3K & 26{\tiny/}--{\tiny/}--{\tiny/}1.1K & 0.01{\tiny/}--{\tiny/}--{\tiny/}0.02 \\*
     & --{\tiny/}0{\tiny/}2{\tiny/}-- & --{\tiny/}1.8K{\tiny/}5.5K{\tiny/}-- & --{\tiny/}811{\tiny/}1.4K{\tiny/}-- & --{\tiny/}0{\tiny/}338{\tiny/}-- & --{\tiny/}0.00{\tiny/}0.00{\tiny/}-- \\
\addlinespace[0.25em]
\nopagebreak
    \multirow{2}{*}{\raisebox{-0.4ex}{\includegraphics[height=1.1em]{figures/logos/metagpt.png}}\,MetaGPT} & 0{\tiny/}68{\tiny/}--{\tiny/}0 & 7.5K{\tiny/}46.0K{\tiny/}--{\tiny/}3.2K & 6.9K{\tiny/}7.2K{\tiny/}--{\tiny/}3.8K & 620{\tiny/}28.1K{\tiny/}--{\tiny/}599 & 0.01{\tiny/}0.01{\tiny/}--{\tiny/}0.00 \\*
     & --{\tiny/}0{\tiny/}--{\tiny/}-- & --{\tiny/}24.6K{\tiny/}--{\tiny/}-- & --{\tiny/}5.3K{\tiny/}--{\tiny/}-- & --{\tiny/}18.0K{\tiny/}--{\tiny/}-- & --{\tiny/}0.01{\tiny/}--{\tiny/}-- \\
\addlinespace[0.25em]
\nopagebreak
    \multirow{2}{*}{\raisebox{-0.4ex}{\includegraphics[height=1.1em]{figures/logos/swarms.png}}\,Swarms} & 0{\tiny/}34{\tiny/}--{\tiny/}-- & 22.5K{\tiny/}56.8K{\tiny/}--{\tiny/}-- & 6.9K{\tiny/}8.1K{\tiny/}--{\tiny/}-- & 4.0K{\tiny/}42.2K{\tiny/}--{\tiny/}-- & 0.01{\tiny/}0.01{\tiny/}--{\tiny/}-- \\*
     & 0{\tiny/}64{\tiny/}--{\tiny/}28 & 14.8K{\tiny/}55.8K{\tiny/}--{\tiny/}0 & 635{\tiny/}8.6K{\tiny/}--{\tiny/}0 & 3.7K{\tiny/}38.2K{\tiny/}--{\tiny/}0 & 0.00{\tiny/}0.01{\tiny/}--{\tiny/}0.00 \\
\addlinespace[0.25em]
\nopagebreak
    \multirow{2}{*}{\raisebox{-0.4ex}{\includegraphics[height=1.1em]{figures/logos/mcp-agent.png}}\,MCP Agent} & 10{\tiny/}68{\tiny/}--{\tiny/}0 & 46.0K{\tiny/}64.8K{\tiny/}--{\tiny/}3.8K & 3.8K{\tiny/}9.0K{\tiny/}--{\tiny/}15.7K & 11.8K{\tiny/}45.6K{\tiny/}--{\tiny/}788 & 0.01{\tiny/}0.02{\tiny/}--{\tiny/}0.01 \\*
     & 18{\tiny/}62{\tiny/}--{\tiny/}42 & 742.6K{\tiny/}50.9K{\tiny/}--{\tiny/}575.1K & 17.1K{\tiny/}7.8K{\tiny/}--{\tiny/}15.3K & 383.5K{\tiny/}34.0K{\tiny/}--{\tiny/}425.1K & 0.12{\tiny/}0.01{\tiny/}--{\tiny/}0.08 \\
\addlinespace[0.25em]
\nopagebreak
    \multirow{2}{*}{\raisebox{-0.4ex}{\includegraphics[height=1.1em]{figures/logos/guidance.png}}\,Guidance} & 0{\tiny/}72{\tiny/}--{\tiny/}0 & 4.4K{\tiny/}55.1K{\tiny/}--{\tiny/}2.1K & 385{\tiny/}8.3K{\tiny/}--{\tiny/}616 & 0{\tiny/}33.4K{\tiny/}--{\tiny/}0 & 0.00{\tiny/}0.01{\tiny/}--{\tiny/}0.00 \\*
     & 38{\tiny/}68{\tiny/}--{\tiny/}52 & 1.2M{\tiny/}55.2K{\tiny/}--{\tiny/}854.2K & 12.2K{\tiny/}8.7K{\tiny/}--{\tiny/}28.7K & 818.1K{\tiny/}38.8K{\tiny/}--{\tiny/}623.8K & 0.17{\tiny/}0.01{\tiny/}--{\tiny/}0.13 \\
\addlinespace[0.25em]
\nopagebreak
    \multirow{2}{*}{\raisebox{-0.4ex}{\includegraphics[height=1.1em]{figures/logos/atomic-agents.png}}\,AtomicAgents} & 0{\tiny/}76{\tiny/}--{\tiny/}0 & 10.5K{\tiny/}53.8K{\tiny/}--{\tiny/}151.4K & 3.1K{\tiny/}8.1K{\tiny/}--{\tiny/}14.5K & 1.6K{\tiny/}34.3K{\tiny/}--{\tiny/}100.8K & 0.00{\tiny/}0.01{\tiny/}--{\tiny/}0.03 \\*
     & 64{\tiny/}--{\tiny/}--{\tiny/}48 & 676.7K{\tiny/}--{\tiny/}--{\tiny/}333.6K & 23.2K{\tiny/}--{\tiny/}--{\tiny/}25.1K & 586.9K{\tiny/}--{\tiny/}--{\tiny/}221.8K & 0.10{\tiny/}--{\tiny/}--{\tiny/}0.06 \\
\addlinespace[0.25em]
\nopagebreak
    \multirow{2}{*}{\raisebox{-0.4ex}{\includegraphics[height=1.1em]{figures/logos/griptape.png}}\,Griptape} & --{\tiny/}66{\tiny/}2{\tiny/}0 & --{\tiny/}62.2K{\tiny/}1.5K{\tiny/}89.8K & --{\tiny/}10.0K{\tiny/}299{\tiny/}14.0K & --{\tiny/}41.4K{\tiny/}23{\tiny/}58.8K & --{\tiny/}0.02{\tiny/}0.00{\tiny/}0.02 \\*
     & --{\tiny/}--{\tiny/}--{\tiny/}-- & --{\tiny/}--{\tiny/}--{\tiny/}-- & --{\tiny/}--{\tiny/}--{\tiny/}-- & --{\tiny/}--{\tiny/}--{\tiny/}-- & --{\tiny/}--{\tiny/}--{\tiny/}-- \\
\addlinespace[0.25em]
\nopagebreak
    \multirow{2}{*}{\raisebox{-0.4ex}{\includegraphics[height=1.1em]{figures/logos/upsonic.png}}\,Upsonic} & 24{\tiny/}0{\tiny/}--{\tiny/}-- & 1.6M{\tiny/}14.4K{\tiny/}--{\tiny/}-- & 18.2K{\tiny/}3.6K{\tiny/}--{\tiny/}-- & 1.1M{\tiny/}11.0K{\tiny/}--{\tiny/}-- & 0.22{\tiny/}0.00{\tiny/}--{\tiny/}-- \\*
     & 32{\tiny/}--{\tiny/}--{\tiny/}-- & 1.0M{\tiny/}--{\tiny/}--{\tiny/}-- & 14.6K{\tiny/}--{\tiny/}--{\tiny/}-- & 756.3K{\tiny/}--{\tiny/}--{\tiny/}-- & 0.14{\tiny/}--{\tiny/}--{\tiny/}-- \\
\addlinespace[0.25em]
\nopagebreak
    \multirow{2}{*}{\raisebox{-0.4ex}{\includegraphics[height=1.1em]{figures/logos/controlflow.png}}\,ControlFlow} & --{\tiny/}--{\tiny/}--{\tiny/}-- & --{\tiny/}--{\tiny/}--{\tiny/}-- & --{\tiny/}--{\tiny/}--{\tiny/}-- & --{\tiny/}--{\tiny/}--{\tiny/}-- & --{\tiny/}--{\tiny/}--{\tiny/}-- \\*
     & --{\tiny/}--{\tiny/}--{\tiny/}-- & --{\tiny/}--{\tiny/}--{\tiny/}-- & --{\tiny/}--{\tiny/}--{\tiny/}-- & --{\tiny/}--{\tiny/}--{\tiny/}-- & --{\tiny/}--{\tiny/}--{\tiny/}-- \\
\addlinespace[0.25em]
\nopagebreak
    \multirow{2}{*}{\raisebox{-0.4ex}{\includegraphics[height=1.1em]{figures/logos/agency-swarm.png}}\,AgencySwarm} & 42{\tiny/}46{\tiny/}--{\tiny/}0 & 306.4K{\tiny/}16.1K{\tiny/}--{\tiny/}484.3K & 3.5K{\tiny/}1.9K{\tiny/}--{\tiny/}12.0K & 0{\tiny/}1.1K{\tiny/}--{\tiny/}0 & 0.06{\tiny/}0.00{\tiny/}--{\tiny/}0.11 \\*
     & 52{\tiny/}--{\tiny/}34{\tiny/}16 & 210.8K{\tiny/}--{\tiny/}18.3K{\tiny/}80.9K & 722{\tiny/}--{\tiny/}268{\tiny/}1.5K & 165.5K{\tiny/}--{\tiny/}5.4K{\tiny/}64.1K & 0.03{\tiny/}--{\tiny/}0.00{\tiny/}0.01 \\
\addlinespace[0.25em]
\nopagebreak
    \multirow{2}{*}{\raisebox{-0.4ex}{\includegraphics[height=1.1em]{figures/logos/agently.png}}\,Agently} & 0{\tiny/}--{\tiny/}0{\tiny/}-- & 4.3K{\tiny/}--{\tiny/}2.8K{\tiny/}-- & 14.8K{\tiny/}--{\tiny/}3.8K{\tiny/}-- & 701{\tiny/}--{\tiny/}0{\tiny/}-- & 0.01{\tiny/}--{\tiny/}0.00{\tiny/}-- \\*
     & --{\tiny/}--{\tiny/}0{\tiny/}0 & --{\tiny/}--{\tiny/}4.8K{\tiny/}1.4K & --{\tiny/}--{\tiny/}2.4K{\tiny/}6.8K & --{\tiny/}--{\tiny/}0{\tiny/}0 & --{\tiny/}--{\tiny/}0.00{\tiny/}0.01 \\
\addlinespace[0.25em]
\nopagebreak
    \multirow{2}{*}{\raisebox{-0.4ex}{\includegraphics[height=1.1em]{figures/logos/lagent.png}}\,Lagent} & --{\tiny/}0{\tiny/}--{\tiny/}0 & --{\tiny/}0{\tiny/}--{\tiny/}1.7K & --{\tiny/}0{\tiny/}--{\tiny/}1.5K & --{\tiny/}0{\tiny/}--{\tiny/}72 & --{\tiny/}0.00{\tiny/}--{\tiny/}0.00 \\*
     & --{\tiny/}0{\tiny/}--{\tiny/}0 & --{\tiny/}27.2K{\tiny/}--{\tiny/}3.0K & --{\tiny/}6.1K{\tiny/}--{\tiny/}8.1K & --{\tiny/}19.0K{\tiny/}--{\tiny/}445 & --{\tiny/}0.01{\tiny/}--{\tiny/}0.01 \\
\addlinespace[0.25em]
\nopagebreak
    \multirow{2}{*}{\raisebox{-0.4ex}{\includegraphics[height=1.1em]{figures/logos/agentuniverse.png}}\,AgentUniverse} & --{\tiny/}--{\tiny/}18{\tiny/}-- & --{\tiny/}--{\tiny/}1.0K{\tiny/}-- & --{\tiny/}--{\tiny/}967{\tiny/}-- & --{\tiny/}--{\tiny/}0{\tiny/}-- & --{\tiny/}--{\tiny/}0.00{\tiny/}-- \\*
     & 58{\tiny/}66{\tiny/}38{\tiny/}56 & 1.8M{\tiny/}64.2K{\tiny/}22.1K{\tiny/}962.7K & 32.0K{\tiny/}10.3K{\tiny/}2.2K{\tiny/}46.6K & 1.3M{\tiny/}40.5K{\tiny/}7.6K{\tiny/}717.1K & 0.26{\tiny/}0.02{\tiny/}0.01{\tiny/}0.16 \\
\addlinespace[0.25em]
\nopagebreak
    \multirow{2}{*}{\raisebox{-0.4ex}{\includegraphics[height=1.1em]{figures/logos/evoagentx.png}}\,EvoAgentX} & 0{\tiny/}0{\tiny/}--{\tiny/}-- & 1.3K{\tiny/}0{\tiny/}--{\tiny/}-- & 403{\tiny/}0{\tiny/}--{\tiny/}-- & 41{\tiny/}0{\tiny/}--{\tiny/}-- & 0.00{\tiny/}0.00{\tiny/}--{\tiny/}-- \\*
     & --{\tiny/}68{\tiny/}--{\tiny/}14 & --{\tiny/}52.6K{\tiny/}--{\tiny/}75.3K & --{\tiny/}9.3K{\tiny/}--{\tiny/}9.1K & --{\tiny/}30.5K{\tiny/}--{\tiny/}62.1K & --{\tiny/}0.01{\tiny/}--{\tiny/}0.02 \\
\addlinespace[0.25em]
\nopagebreak
    \multirow{2}{*}{\raisebox{-0.4ex}{\includegraphics[height=1.1em]{figures/logos/nerve.png}}\,Nerve} & 26{\tiny/}2{\tiny/}--{\tiny/}0 & 418.3K{\tiny/}6.1K{\tiny/}--{\tiny/}230.5K & 10.4K{\tiny/}1.5K{\tiny/}--{\tiny/}9.6K & 271.4K{\tiny/}4.4K{\tiny/}--{\tiny/}177.3K & 0.06{\tiny/}0.00{\tiny/}--{\tiny/}0.04 \\*
     & 10{\tiny/}68{\tiny/}--{\tiny/}24 & 933.7K{\tiny/}52.1K{\tiny/}--{\tiny/}300.6K & 11.2K{\tiny/}7.9K{\tiny/}--{\tiny/}18.6K & 612.0K{\tiny/}30.4K{\tiny/}--{\tiny/}201.8K & 0.13{\tiny/}0.01{\tiny/}--{\tiny/}0.05 \\
\addlinespace[0.25em]
\nopagebreak
    \multirow{2}{*}{\raisebox{-0.4ex}{\includegraphics[height=1.1em]{figures/logos/agent-squad.png}}\,AgentSquad} & 0{\tiny/}0{\tiny/}--{\tiny/}4 & 7.6K{\tiny/}1.8K{\tiny/}--{\tiny/}147.6K & 2.6K{\tiny/}800{\tiny/}--{\tiny/}8.0K & 2.6K{\tiny/}0{\tiny/}--{\tiny/}106.0K & 0.00{\tiny/}0.00{\tiny/}--{\tiny/}0.03 \\*
     & 34{\tiny/}52{\tiny/}48{\tiny/}48 & 383.0K{\tiny/}51.3K{\tiny/}54.0K{\tiny/}686.7K & 10.2K{\tiny/}6.6K{\tiny/}3.3K{\tiny/}25.8K & 234.3K{\tiny/}35.0K{\tiny/}19.4K{\tiny/}483.7K & 0.06{\tiny/}0.01{\tiny/}0.01{\tiny/}0.11 \\
\addlinespace[0.25em]
\nopagebreak
    \multirow{2}{*}{\raisebox{-0.4ex}{\includegraphics[height=1.1em]{figures/logos/motleycrew.png}}\,MotleyCrew} & 38{\tiny/}68{\tiny/}--{\tiny/}0 & 509.8K{\tiny/}85.0K{\tiny/}--{\tiny/}27.2K & 22.5K{\tiny/}10.0K{\tiny/}--{\tiny/}3.4K & 338.0K{\tiny/}60.0K{\tiny/}--{\tiny/}9.7K & 0.09{\tiny/}0.02{\tiny/}--{\tiny/}0.01 \\*
     & 0{\tiny/}--{\tiny/}--{\tiny/}40 & 242.2K{\tiny/}--{\tiny/}--{\tiny/}190.5K & 9.0K{\tiny/}--{\tiny/}--{\tiny/}14.8K & 164.5K{\tiny/}--{\tiny/}--{\tiny/}105.8K & 0.04{\tiny/}--{\tiny/}--{\tiny/}0.04 \\
\addlinespace[0.25em]
\nopagebreak
    \multirow{2}{*}{\raisebox{-0.4ex}{\includegraphics[height=1.1em]{figures/logos/taskflowai.png}}\,TaskflowAI} & --{\tiny/}66{\tiny/}--{\tiny/}-- & --{\tiny/}54.3K{\tiny/}--{\tiny/}-- & --{\tiny/}8.6K{\tiny/}--{\tiny/}-- & --{\tiny/}34.3K{\tiny/}--{\tiny/}-- & --{\tiny/}0.01{\tiny/}--{\tiny/}-- \\*
     & --{\tiny/}76{\tiny/}--{\tiny/}-- & --{\tiny/}77.3K{\tiny/}--{\tiny/}-- & --{\tiny/}10.4K{\tiny/}--{\tiny/}-- & --{\tiny/}52.9K{\tiny/}--{\tiny/}-- & --{\tiny/}0.02{\tiny/}--{\tiny/}-- \\
\addlinespace[0.25em]
\nopagebreak
    \multirow{2}{*}{\raisebox{-0.4ex}{\includegraphics[height=1.1em]{figures/logos/council-ai.png}}\,CouncilAI} & --{\tiny/}--{\tiny/}--{\tiny/}0 & --{\tiny/}--{\tiny/}--{\tiny/}1.6K & --{\tiny/}--{\tiny/}--{\tiny/}568 & --{\tiny/}--{\tiny/}--{\tiny/}0 & --{\tiny/}--{\tiny/}--{\tiny/}0.00 \\*
     & --{\tiny/}72{\tiny/}--{\tiny/}46 & --{\tiny/}61.0K{\tiny/}--{\tiny/}822.5K & --{\tiny/}10.0K{\tiny/}--{\tiny/}22.5K & --{\tiny/}38.7K{\tiny/}--{\tiny/}640.5K & --{\tiny/}0.02{\tiny/}--{\tiny/}0.12 \\
\addlinespace[0.25em]
\nopagebreak
    \multirow{2}{*}{\raisebox{-0.4ex}{\includegraphics[height=1.1em]{figures/logos/agentflow.png}}\,AgentFlow} & 0{\tiny/}--{\tiny/}--{\tiny/}-- & 0{\tiny/}--{\tiny/}--{\tiny/}-- & 0{\tiny/}--{\tiny/}--{\tiny/}-- & 0{\tiny/}--{\tiny/}--{\tiny/}-- & 0.00{\tiny/}--{\tiny/}--{\tiny/}-- \\*
     & --{\tiny/}68{\tiny/}--{\tiny/}-- & --{\tiny/}52.2K{\tiny/}--{\tiny/}-- & --{\tiny/}9.0K{\tiny/}--{\tiny/}-- & --{\tiny/}32.7K{\tiny/}--{\tiny/}-- & --{\tiny/}0.01{\tiny/}--{\tiny/}-- \\
\addlinespace[0.25em]
\nopagebreak
    \multirow{2}{*}{\raisebox{-0.4ex}{\includegraphics[height=1.1em]{figures/logos/autoagent.png}}\,AutoAgent} & 0{\tiny/}44{\tiny/}--{\tiny/}-- & 7.4K{\tiny/}69.4K{\tiny/}--{\tiny/}-- & 430{\tiny/}9.5K{\tiny/}--{\tiny/}-- & 980{\tiny/}51.7K{\tiny/}--{\tiny/}-- & 0.00{\tiny/}0.02{\tiny/}--{\tiny/}-- \\*
     & --{\tiny/}72{\tiny/}42{\tiny/}44 & --{\tiny/}56.8K{\tiny/}21.8K{\tiny/}671.8K & --{\tiny/}9.1K{\tiny/}1.5K{\tiny/}34.5K & --{\tiny/}34.5K{\tiny/}4.7K{\tiny/}449.1K & --{\tiny/}0.02{\tiny/}0.01{\tiny/}0.12 \\
\addlinespace[0.25em]
\nopagebreak
    \multirow{2}{*}{\raisebox{-0.4ex}{\includegraphics[height=1.1em]{figures/logos/agentlite.png}}\,AgentLite} & 0{\tiny/}22{\tiny/}--{\tiny/}0 & 2.2K{\tiny/}72.4K{\tiny/}--{\tiny/}383 & 10.8K{\tiny/}6.6K{\tiny/}--{\tiny/}360 & 0{\tiny/}45.8K{\tiny/}--{\tiny/}0 & 0.01{\tiny/}0.02{\tiny/}--{\tiny/}0.00 \\*
     & 0{\tiny/}66{\tiny/}14{\tiny/}-- & 43.7K{\tiny/}59.5K{\tiny/}5.5K{\tiny/}-- & 8.8K{\tiny/}9.3K{\tiny/}4.6K{\tiny/}-- & 33.5K{\tiny/}42.9K{\tiny/}1.1K{\tiny/}-- & 0.01{\tiny/}0.02{\tiny/}0.00{\tiny/}-- \\
\addlinespace[0.25em]
\nopagebreak
    \multirow{2}{*}{\raisebox{-0.4ex}{\includegraphics[height=1.1em]{figures/logos/octotools.png}}\,Octotools} & 0{\tiny/}60{\tiny/}--{\tiny/}8 & 0{\tiny/}82.4K{\tiny/}--{\tiny/}75.2K & 0{\tiny/}10.5K{\tiny/}--{\tiny/}16.9K & 0{\tiny/}58.6K{\tiny/}--{\tiny/}52.7K & 0.00{\tiny/}0.02{\tiny/}--{\tiny/}0.02 \\*
     & 0{\tiny/}62{\tiny/}--{\tiny/}-- & 3.9K{\tiny/}53.8K{\tiny/}--{\tiny/}-- & 587{\tiny/}9.0K{\tiny/}--{\tiny/}-- & 2.5K{\tiny/}30.2K{\tiny/}--{\tiny/}-- & 0.00{\tiny/}0.01{\tiny/}--{\tiny/}-- \\
\addlinespace[0.25em]
\nopagebreak
    \multirow{2}{*}{\raisebox{-0.4ex}{\includegraphics[height=1.1em]{figures/logos/gptswarm.png}}\,GPTSwarm} & 0{\tiny/}--{\tiny/}64{\tiny/}-- & 0{\tiny/}--{\tiny/}84.4K{\tiny/}-- & 0{\tiny/}--{\tiny/}3.1K{\tiny/}-- & 0{\tiny/}--{\tiny/}24.1K{\tiny/}-- & 0.00{\tiny/}--{\tiny/}0.02{\tiny/}-- \\*
     & 54{\tiny/}--{\tiny/}58{\tiny/}-- & 770.8K{\tiny/}--{\tiny/}29.8K{\tiny/}-- & 11.8K{\tiny/}--{\tiny/}2.9K{\tiny/}-- & 514.4K{\tiny/}--{\tiny/}9.5K{\tiny/}-- & 0.11{\tiny/}--{\tiny/}0.01{\tiny/}-- \\
\addlinespace[0.25em]
\nopagebreak
    \multirow{2}{*}{\raisebox{-0.4ex}{\includegraphics[height=1.1em]{figures/logos/taskweaver.png}}\,Taskweaver} & --{\tiny/}32{\tiny/}--{\tiny/}-- & --{\tiny/}49.9K{\tiny/}--{\tiny/}-- & --{\tiny/}7.4K{\tiny/}--{\tiny/}-- & --{\tiny/}37.9K{\tiny/}--{\tiny/}-- & --{\tiny/}0.01{\tiny/}--{\tiny/}-- \\*
     & --{\tiny/}68{\tiny/}--{\tiny/}-- & --{\tiny/}56.3K{\tiny/}--{\tiny/}-- & --{\tiny/}8.9K{\tiny/}--{\tiny/}-- & --{\tiny/}39.4K{\tiny/}--{\tiny/}-- & --{\tiny/}0.01{\tiny/}--{\tiny/}-- \\
\addlinespace[0.25em]
\nopagebreak
    \multirow{2}{*}{\raisebox{-0.4ex}{\includegraphics[height=1.1em]{figures/logos/autoagents.png}}\,AutoAgents} & --{\tiny/}0{\tiny/}--{\tiny/}-- & --{\tiny/}0{\tiny/}--{\tiny/}-- & --{\tiny/}0{\tiny/}--{\tiny/}-- & --{\tiny/}0{\tiny/}--{\tiny/}-- & --{\tiny/}0.00{\tiny/}--{\tiny/}-- \\*
     & 0{\tiny/}66{\tiny/}--{\tiny/}-- & 93.3K{\tiny/}53.3K{\tiny/}--{\tiny/}-- & 22.0K{\tiny/}8.5K{\tiny/}--{\tiny/}-- & 11.8K{\tiny/}34.1K{\tiny/}--{\tiny/}-- & 0.04{\tiny/}0.01{\tiny/}--{\tiny/}-- \\
\addlinespace[0.25em]
\nopagebreak
    \multirow{2}{*}{\raisebox{-0.4ex}{\includegraphics[height=1.1em]{figures/logos/opensage.png}}\,OpenSage} & 0{\tiny/}68{\tiny/}--{\tiny/}40 & 386.3K{\tiny/}70.5K{\tiny/}--{\tiny/}326.7K & 16.9K{\tiny/}10.0K{\tiny/}--{\tiny/}11.9K & 216.0K{\tiny/}48.6K{\tiny/}--{\tiny/}249.9K & 0.07{\tiny/}0.02{\tiny/}--{\tiny/}0.05 \\*
     & 18{\tiny/}--{\tiny/}--{\tiny/}28 & 1.1M{\tiny/}--{\tiny/}--{\tiny/}275.8K & 10.8K{\tiny/}--{\tiny/}--{\tiny/}11.4K & 764.0K{\tiny/}--{\tiny/}--{\tiny/}191.1K & 0.15{\tiny/}--{\tiny/}--{\tiny/}0.05 \\
\end{longtable}
\endgroup

\autoref{tab:exec_full} reports the resolution rate, token usage, and cost for all 51 frameworks across the four benchmarks under the ``normal'' information condition (framework documentation + repository access). For each framework we evaluate two generated agents, one written by GPT-5.4 (Codex), one by Opus-4.6 (Claude Code), both executing on GPT-5.4 Nano as the backbone LLM. The generated agents are functional (they build, execute, and consume LLM tokens) but most fail to solve their benchmark tasks, with successful resolutions concentrated on a small subset of framework--benchmark pairs.

Resolution varies sharply by benchmark. $\tau^2$-bench is by far the most tractable: among agents that actually run, the median resolves 64\% and the best 80\%, because its conversational, tool-calling format is comparatively forgiving and leans on the shared backbone rather than on long, self-directed trajectories. The coding and terminal benchmarks are much harder, with medians of just 18\% (SWE-bench) and 14\% (TerminalBench): both demand long-horizon, iterative editing and command execution, where a single early misstep cascades into failure and the lightweight generated agents rarely sustain the loop. MCP-Atlas sits in between (median 39\% among the frameworks with execution data), rewarding agents that correctly discover and route requests through the MCP tool server. The spread within a single benchmark is also wide, from 0\% to 80\% on SWE-bench, confirming that the framework and its generated agent, not the task suite alone, drive the outcome.

The developer model that \emph{writes} the agent matters more than the model that runs it. Although both variants execute on the same GPT-5.4 Nano backbone, agents authored by Opus-4.6 resolve roughly twice as many tasks as those authored by GPT-5.4 (mean 41\% vs.\ 22\%, median 46\% vs.\ 8\%). The difference is visible in the code: GPT-5.4 frequently emits a minimal agent that issues a single model call and returns, whereas Opus-4.6 more often wires up the framework's full loop, registering tools and iterating over multiple turns. Agent \emph{code quality}, fixed at generation time, thus largely determines execution success, reinforcing that passing validation is necessary but far from sufficient.

\finding{ADK-generated agents are functional (they execute and consume tokens) but most fail to solve benchmark tasks: generation-time validation ensures syntactic correctness, not task-solving capability. Outcomes diverge sharply even among validated agents, and the developer model is decisive: Opus-authored agents resolve roughly twice as many tasks as GPT-authored ones (41\% vs.\ 22\%), all on the same execution backbone.}

\subsection{Comparison with Frontier Coding Agents}
\label{sec:reference}

\begin{table*}[t]
\centering
\caption{Production coding agents on four benchmarks. All agents use GPT-5.4 Nano. Token, cost, time, and call columns are per-task averages over the 50-task subset.}
\label{tab:reference}
\vspace{2pt}
\resizebox{\linewidth}{!}{%
\renewcommand{\arraystretch}{1.15}
\begin{tabular}{@{\hskip 2pt}lll crrrrrc@{\hskip 2pt}}
\toprule
\textbf{System} & \textbf{Version} & \textbf{Benchmark} & \textbf{Rate (\%)} & \textbf{In Tok} & \textbf{Out Tok} & \textbf{Cache Tok} & \textbf{Cost (USD)} & \textbf{Avg Time (s)} & \textbf{Avg Calls} \\
\midrule
  \multirow{4}{*}{\raisebox{-0.4ex}{\includegraphics[height=1.1em]{figures/logos/anthropic-sdk.png}}\,Claude Code} & \multirow{4}{*}{v2.1.150} & SWE-bench & 48.0 & 179.0K & 1.5K & 164.9K & 0.07 & 104.7 & 25.6 \\
    &  & $\tau^2$-bench & 46.0 & 15.5K & 226 & 7.1K & 0.00 & 65.1 & 3.0 \\
    &  & MCP-Atlas & 38.0 & 26.6K & 663 & 25.1K & 0.01 & 38.9 & 9.2 \\
    &  & TerminalBench & 30.0 & 254.1K & 4.2K & 243.7K & 0.10 & 207.7 & 27.8 \\
\midrule
  \multirow{4}{*}{\raisebox{-0.4ex}{\includegraphics[height=1.1em]{figures/logos/openai-agents.png}}\,Codex CLI} & \multirow{4}{*}{v0.95.0} & SWE-bench & 60.0 & 673.7K & 2.9K & 651.6K & 0.02 & 80.5 & 26.8 \\
    &  & $\tau^2$-bench & 46.0 & 233.4K & 1.1K & 219.7K & 0.01 & 178.3 & 3.6 \\
    &  & MCP-Atlas & 36.0 & 167.2K & 1.0K & 153.1K & 0.01 & 32.5 & 3.3 \\
    &  & TerminalBench & 16.0 & 372.9K & 3.2K & 353.4K & 0.01 & 112.9 & 12.7 \\
\midrule
  \multirow{4}{*}{\raisebox{-0.4ex}{\includegraphics[height=1.1em]{figures/logos/microsoft.png}}\,Copilot} & \multirow{4}{*}{v1.0.39} & SWE-bench & 74.0 & 2{,}578.3K & 41.1K & 2{,}094.4K & 0.57 & 592.5 & 64.8 \\
    &  & $\tau^2$-bench & 46.0 & 554.2K & 40.3K & 418.8K & 0.16 & 394.8 & 52.3 \\
    &  & MCP-Atlas & 58.0 & 327.1K & 16.4K & 198.9K & 0.09 & 180.0 & 14.5 \\
    &  & TerminalBench & 46.0 & 425.5K & 59.2K & 327.9K & 0.16 & 860.8 & 36.6 \\
\midrule
  \multirow{4}{*}{\raisebox{-0.4ex}{\includegraphics[height=1.1em]{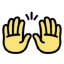}}\,OpenHands} & \multirow{4}{*}{v1.6.0} & SWE-bench & 60.0 & 2{,}357.8K & 47.6K & 1{,}550.0K & 0.29 & 617.4 & 44.6 \\
    &  & $\tau^2$-bench & 46.0 & 660.6K & 26.1K & 572.3K & 0.06 & 806.0 & 28.5 \\
    &  & MCP-Atlas & 52.0 & 498.7K & 12.6K & 269.4K & 0.07 & 248.2 & 16.7 \\
    &  & TerminalBench & 36.0 & 1{,}190.1K & 44.2K & 552.4K & 0.19 & 830.5 & 20.6 \\
\midrule
  \multirow{4}{*}{\raisebox{-0.4ex}{\includegraphics[height=1.1em]{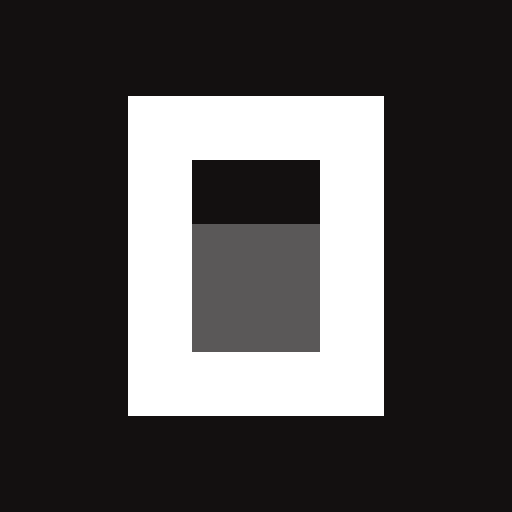}}\,OpenCode} & \multirow{4}{*}{v1.15.10} & SWE-bench & 76.0 & 885.9K & 7.4K & 851.1K & 0.10 & 271.9 & 31.2 \\
    &  & $\tau^2$-bench & 48.0 & 28.6K & 1.1K & 14.7K & 0.01 & 308.3 & 2.2 \\
    &  & MCP-Atlas & 18.0 & 92.4K & 1.9K & 84.4K & 0.01 & 66.2 & 5.0 \\
    &  & TerminalBench & 38.0 & 655.8K & 12.3K & 631.9K & 0.08 & 298.0 & 22.7 \\
\bottomrule
\end{tabular}%
}
\end{table*}

To contextualize ADK-generated agents, we compare against five frontier coding agents (\autoref{tab:reference}): Claude Code~\citep{claudecode}, GitHub Copilot~\citep{copilot}, Codex~\citep{codex}, OpenHands~\citep{openhands}, and OpenCode~\citep{opencode}. All use GPT-5.4 Nano as the backbone LLM to control for model effects.
Among the frontier agents themselves, Copilot achieves the highest average resolution across the four benchmarks (56\%: 74\% SWE-bench, 58\% MCP-Atlas, 46\% TerminalBench, 46\% $\tau^2$-bench), followed by OpenHands (48.5\%: 60\% SWE-bench, 52\% MCP-Atlas, 46\% $\tau^2$-bench, 36\% TerminalBench), OpenCode (45\%: 76\% SWE-bench, 48\% $\tau^2$-bench, 38\% TerminalBench, 18\% MCP-Atlas), Claude Code (40.5\%: 48\% SWE-bench, 46\% $\tau^2$-bench, 38\% MCP-Atlas, 30\% TerminalBench), and Codex (39.5\%: 60\% SWE-bench, 46\% $\tau^2$-bench, 36\% MCP-Atlas, 16\% TerminalBench). OpenCode posts the single highest scores on SWE-bench (76\%) and $\tau^2$-bench (48\%), while Copilot leads on MCP-Atlas (58\%). Crucially, the top of the accuracy ranking is also the least efficient: the two highest-resolution agents (Copilot 56\%, OpenHands 48.5\%) are by far the most token-hungry, consuming 3.9M and 4.7M input tokens across the four benchmarks. Claude Code and Codex are the standouts on efficiency: Claude Code resolves a competitive 40.5\% on just 0.48M tokens (8--10$\times$ fewer than Copilot or OpenHands) at roughly \$0.05 per task, and Codex is cheapest of all at \$0.01--0.02 per task, both reaching near the resolution of the leaders at a small fraction of their cost. The most accurate agent is thus not the most efficient, and a sharp accuracy--efficiency tradeoff separates the field.

Set against the ADK-generated agents, these frontier systems are strong but not dominant. The best ADK-generated agents (\autoref{tab:exec_full}) are competitive on individual benchmarks (resolving up to 80\% of tasks), and because each is a lightweight program generated for a single benchmark, the best ones can even exceed the general-purpose frontier agents at a fraction of the cost: on MCP-Atlas, \textit{Agno} resolves 68\% of tasks at 38K input tokens per task, beating Copilot's 58\% at 327K (roughly 9$\times$ fewer). The generated agents reach comparably low per-task cost in part because they terminate early on unsolved tasks, so the comparison is not purely an apples-to-apples one, but the top performers genuinely match or beat the frontier on their target benchmark. Performance is nonetheless highly uneven across the ecosystem, with the median framework--benchmark pair resolving only 32\%, so a well-chosen framework can match hand-built agents on some tasks while most frameworks fall well short.

\finding{Lightweight, single-benchmark ADK agents can beat general-purpose frontier agents far more cheaply. Yet this is a top-of-distribution effect: the median framework resolves only 32\%, so framework choice alone does not guarantee strong performance.}

\subsection{Stress Testing: Ablating Information Sources}
\label{sec:ablation}

\begin{table*}[t]
\caption{Generation effort under three ablation conditions (D\,/\,S\,/\,N = Docs only\,/\,Source only\,/\,No reference). Values averaged across benchmarks (K\,=\,thousands). \textbf{Pass}: number of benchmarks (of 4) that pass validation \emph{and} use the framework natively (thin-wrapper agents excluded). First row: GPT-5.4 (Codex). Second row: Opus-4.6 (Claude Code).}
\label{tab:gen_effort_ablation}
\centering
\resizebox{\linewidth}{!}{%
\tiny
\setlength{\tabcolsep}{2pt}
\begin{tabular}{@{}l c c c c c @{\hskip 6pt} | l c c c c c@{}}
\toprule
\textbf{Framework} & \textbf{Pass} & \textbf{In} & \textbf{Out} & \textbf{Cache} & \textbf{\$} &
\textbf{Framework} & \textbf{Pass} & \textbf{In} & \textbf{Out} & \textbf{Cache} & \textbf{\$} \\
\midrule
    \multirow{2}{*}{\raisebox{-0.4ex}{\includegraphics[height=1.1em]{figures/logos/langchain.png}}\,LangChain} & 1{\tiny/}1{\tiny/}1 & 1719{\tiny/}1651{\tiny/}1342 & 15{\tiny/}15{\tiny/}13 & 1084{\tiny/}1072{\tiny/}742 & 2.4{\tiny/}2.3{\tiny/}2.1 & \multirow{2}{*}{\raisebox{-0.4ex}{\includegraphics[height=1.1em]{figures/logos/swarms.png}}\,Swarms} & 2{\tiny/}0{\tiny/}1 & 2722{\tiny/}2092{\tiny/}1957 & 22{\tiny/}15{\tiny/}17 & 2113{\tiny/}1389{\tiny/}1371 & 3.1{\tiny/}2.8{\tiny/}2.5 \\
     & 3{\tiny/}3{\tiny/}2 & 1921{\tiny/}1331{\tiny/}1247 & 13{\tiny/}17{\tiny/}16 & 1798{\tiny/}1240{\tiny/}1172 & 1.5{\tiny/}1.5{\tiny/}1.3 &  & 2{\tiny/}3{\tiny/}1 & 1911{\tiny/}2010{\tiny/}1301 & 15{\tiny/}13{\tiny/}18 & 1815{\tiny/}1932{\tiny/}1241 & 1.8{\tiny/}1.7{\tiny/}1.4 \\
\addlinespace[0.25em]
    \multirow{2}{*}{\raisebox{-0.4ex}{\includegraphics[height=1.1em]{figures/logos/anthropic-sdk.png}}\,Anthropic SDK} & 0{\tiny/}0{\tiny/}0 & 1962{\tiny/}1824{\tiny/}1505 & 15{\tiny/}14{\tiny/}14 & 1366{\tiny/}1188{\tiny/}833 & 2.5{\tiny/}2.5{\tiny/}2.3 & \multirow{2}{*}{\raisebox{-0.4ex}{\includegraphics[height=1.1em]{figures/logos/mcp-agent.png}}\,MCP Agent} & 2{\tiny/}1{\tiny/}2 & 2130{\tiny/}2391{\tiny/}1988 & 14{\tiny/}18{\tiny/}16 & 1510{\tiny/}1680{\tiny/}1310 & 2.6{\tiny/}3.0{\tiny/}2.7 \\
     & 0{\tiny/}0{\tiny/}0 & 1287{\tiny/}1751{\tiny/}1591 & 14{\tiny/}14{\tiny/}16 & 1236{\tiny/}1694{\tiny/}1514 & 1.2{\tiny/}1.5{\tiny/}1.5 &  & 2{\tiny/}3{\tiny/}3 & 2107{\tiny/}1753{\tiny/}2220 & 15{\tiny/}14{\tiny/}14 & 2032{\tiny/}1665{\tiny/}2138 & 1.8{\tiny/}1.6{\tiny/}1.8 \\
\addlinespace[0.25em]
    \multirow{2}{*}{\raisebox{-0.4ex}{\includegraphics[height=1.1em]{figures/logos/langgraph.png}}\,LangGraph} & 1{\tiny/}2{\tiny/}3 & 2178{\tiny/}1738{\tiny/}1223 & 15{\tiny/}14{\tiny/}15 & 1458{\tiny/}1097{\tiny/}693 & 2.9{\tiny/}2.4{\tiny/}1.9 & \multirow{2}{*}{\raisebox{-0.4ex}{\includegraphics[height=1.1em]{figures/logos/guidance.png}}\,Guidance} & 1{\tiny/}1{\tiny/}1 & 2420{\tiny/}1846{\tiny/}1603 & 13{\tiny/}15{\tiny/}17 & 1593{\tiny/}1240{\tiny/}1024 & 3.2{\tiny/}2.4{\tiny/}2.3 \\
     & 3{\tiny/}3{\tiny/}3 & 1353{\tiny/}887{\tiny/}1217 & 16{\tiny/}15{\tiny/}17 & 1270{\tiny/}822{\tiny/}1138 & 1.5{\tiny/}1.1{\tiny/}1.3 &  & 1{\tiny/}3{\tiny/}0 & 2442{\tiny/}2140{\tiny/}1555 & 12{\tiny/}13{\tiny/}13 & 2388{\tiny/}2071{\tiny/}1513 & 1.8{\tiny/}1.7{\tiny/}1.3 \\
\addlinespace[0.25em]
    \multirow{2}{*}{\raisebox{-0.4ex}{\includegraphics[height=1.1em]{figures/logos/openai-agents.png}}\,OpenAI Agents} & 2{\tiny/}3{\tiny/}2 & 1932{\tiny/}1592{\tiny/}1650 & 14{\tiny/}15{\tiny/}16 & 1365{\tiny/}992{\tiny/}998 & 2.4{\tiny/}2.3{\tiny/}2.4 & \multirow{2}{*}{\raisebox{-0.4ex}{\includegraphics[height=1.1em]{figures/logos/atomic-agents.png}}\,AtomicAgents} & 1{\tiny/}2{\tiny/}1 & 2085{\tiny/}2139{\tiny/}1609 & 22{\tiny/}14{\tiny/}17 & 1451{\tiny/}1429{\tiny/}950 & 2.7{\tiny/}2.8{\tiny/}2.4 \\
     & 3{\tiny/}3{\tiny/}4 & 1411{\tiny/}1462{\tiny/}1364 & 15{\tiny/}14{\tiny/}14 & 1325{\tiny/}1391{\tiny/}1257 & 1.5{\tiny/}1.4{\tiny/}1.5 &  & 0{\tiny/}0{\tiny/}0 & 1096{\tiny/}1032{\tiny/}1446 & 18{\tiny/}21{\tiny/}15 & 1034{\tiny/}962{\tiny/}1393 & 1.3{\tiny/}1.3{\tiny/}1.3 \\
\addlinespace[0.25em]
    \multirow{2}{*}{\raisebox{-0.4ex}{\includegraphics[height=1.1em]{figures/logos/pydantic-ai.png}}\,PydanticAI} & 2{\tiny/}3{\tiny/}3 & 2183{\tiny/}1736{\tiny/}1517 & 14{\tiny/}15{\tiny/}17 & 1464{\tiny/}1087{\tiny/}977 & 2.9{\tiny/}2.5{\tiny/}2.1 & \multirow{2}{*}{\raisebox{-0.4ex}{\includegraphics[height=1.1em]{figures/logos/griptape.png}}\,Griptape} & 0{\tiny/}3{\tiny/}0 & 2522{\tiny/}2849{\tiny/}2010 & 15{\tiny/}16{\tiny/}16 & 1666{\tiny/}2057{\tiny/}1278 & 3.3{\tiny/}3.4{\tiny/}2.8 \\
     & 2{\tiny/}3{\tiny/}3 & 1262{\tiny/}1221{\tiny/}1196 & 12{\tiny/}13{\tiny/}13 & 1207{\tiny/}1163{\tiny/}1120 & 1.2{\tiny/}1.2{\tiny/}1.3 &  & 1{\tiny/}3{\tiny/}2 & 1618{\tiny/}1496{\tiny/}1517 & 15{\tiny/}16{\tiny/}17 & 1548{\tiny/}1428{\tiny/}1456 & 1.5{\tiny/}1.5{\tiny/}1.4 \\
\addlinespace[0.25em]
    \multirow{2}{*}{\raisebox{-0.4ex}{\includegraphics[height=1.1em]{figures/logos/llamaindex.png}}\,LlamaIndex} & 0{\tiny/}1{\tiny/}0 & 2583{\tiny/}2144{\tiny/}1654 & 13{\tiny/}16{\tiny/}13 & 1911{\tiny/}1400{\tiny/}963 & 3.0{\tiny/}2.9{\tiny/}2.5 & \multirow{2}{*}{\raisebox{-0.4ex}{\includegraphics[height=1.1em]{figures/logos/upsonic.png}}\,Upsonic} & 4{\tiny/}2{\tiny/}2 & 1989{\tiny/}2053{\tiny/}1792 & 14{\tiny/}13{\tiny/}13 & 1340{\tiny/}1368{\tiny/}1302 & 2.6{\tiny/}2.7{\tiny/}2.2 \\
     & 1{\tiny/}3{\tiny/}3 & 1307{\tiny/}1244{\tiny/}1092 & 14{\tiny/}15{\tiny/}15 & 1256{\tiny/}1172{\tiny/}1032 & 1.2{\tiny/}1.3{\tiny/}1.2 &  & 3{\tiny/}3{\tiny/}2 & 2340{\tiny/}1966{\tiny/}1937 & 14{\tiny/}14{\tiny/}15 & 2221{\tiny/}1898{\tiny/}1875 & 2.1{\tiny/}1.6{\tiny/}1.6 \\
\addlinespace[0.25em]
    \multirow{2}{*}{\raisebox{-0.4ex}{\includegraphics[height=1.1em]{figures/logos/crewai.png}}\,CrewAI} & 1{\tiny/}3{\tiny/}2 & 2550{\tiny/}1993{\tiny/}1302 & 14{\tiny/}14{\tiny/}15 & 1738{\tiny/}1127{\tiny/}754 & 3.3{\tiny/}3.0{\tiny/}2.0 & \multirow{2}{*}{\raisebox{-0.4ex}{\includegraphics[height=1.1em]{figures/logos/controlflow.png}}\,ControlFlow} & 0{\tiny/}0{\tiny/}0 & 2273{\tiny/}1783{\tiny/}1663 & 10{\tiny/}9{\tiny/}8 & 1538{\tiny/}1168{\tiny/}1020 & 2.9{\tiny/}2.4{\tiny/}2.3 \\
     & 2{\tiny/}3{\tiny/}3 & 1229{\tiny/}876{\tiny/}822 & 12{\tiny/}12{\tiny/}11 & 1170{\tiny/}817{\tiny/}772 & 1.2{\tiny/}1.0{\tiny/}0.9 &  & 0{\tiny/}0{\tiny/}0 & 1469{\tiny/}1239{\tiny/}976 & 9{\tiny/}9{\tiny/}9 & 1417{\tiny/}1201{\tiny/}939 & 1.2{\tiny/}1.0{\tiny/}0.9 \\
\addlinespace[0.25em]
    \multirow{2}{*}{\raisebox{-0.4ex}{\includegraphics[height=1.1em]{figures/logos/google-adk.png}}\,Google ADK} & 2{\tiny/}3{\tiny/}3 & 1872{\tiny/}2251{\tiny/}2217 & 15{\tiny/}14{\tiny/}13 & 1226{\tiny/}1560{\tiny/}1401 & 2.5{\tiny/}2.8{\tiny/}3.0 & \multirow{2}{*}{\raisebox{-0.4ex}{\includegraphics[height=1.1em]{figures/logos/agency-swarm.png}}\,AgencySwarm} & 0{\tiny/}2{\tiny/}3 & 534{\tiny/}1737{\tiny/}1677 & 3{\tiny/}15{\tiny/}12 & 164{\tiny/}1148{\tiny/}1029 & 1.1{\tiny/}2.3{\tiny/}2.4 \\
     & 2{\tiny/}3{\tiny/}1 & 1860{\tiny/}1410{\tiny/}1332 & 12{\tiny/}13{\tiny/}12 & 1797{\tiny/}1356{\tiny/}1286 & 1.5{\tiny/}1.3{\tiny/}1.2 &  & 3{\tiny/}3{\tiny/}3 & 1232{\tiny/}1144{\tiny/}1137 & 10{\tiny/}9{\tiny/}8 & 1146{\tiny/}1095{\tiny/}1080 & 1.3{\tiny/}1.0{\tiny/}1.0 \\
\addlinespace[0.25em]
    \multirow{2}{*}{\raisebox{-0.4ex}{\includegraphics[height=1.1em]{figures/logos/strands-agents.png}}\,Strands} & 3{\tiny/}3{\tiny/}3 & 2346{\tiny/}1787{\tiny/}1987 & 14{\tiny/}16{\tiny/}13 & 1665{\tiny/}1073{\tiny/}1215 & 2.9{\tiny/}2.6{\tiny/}2.8 & \multirow{2}{*}{\raisebox{-0.4ex}{\includegraphics[height=1.1em]{figures/logos/agently.png}}\,Agently} & 0{\tiny/}1{\tiny/}0 & 2592{\tiny/}2977{\tiny/}1767 & 12{\tiny/}12{\tiny/}14 & 1750{\tiny/}2070{\tiny/}1113 & 3.3{\tiny/}3.7{\tiny/}2.5 \\
     & 3{\tiny/}3{\tiny/}2 & 1398{\tiny/}1634{\tiny/}1566 & 14{\tiny/}14{\tiny/}18 & 1335{\tiny/}1566{\tiny/}1507 & 1.3{\tiny/}1.5{\tiny/}1.5 &  & 1{\tiny/}2{\tiny/}0 & 1224{\tiny/}1727{\tiny/}1926 & 22{\tiny/}19{\tiny/}18 & 1166{\tiny/}1662{\tiny/}1876 & 1.4{\tiny/}1.6{\tiny/}1.7 \\
\addlinespace[0.25em]
    \multirow{2}{*}{\raisebox{-0.4ex}{\includegraphics[height=1.1em]{figures/logos/semantic-kernel.png}}\,SemanticKernel} & 3{\tiny/}2{\tiny/}2 & 1981{\tiny/}1737{\tiny/}1758 & 15{\tiny/}20{\tiny/}17 & 1362{\tiny/}1085{\tiny/}1114 & 2.6{\tiny/}2.5{\tiny/}2.5 & \multirow{2}{*}{\raisebox{-0.4ex}{\includegraphics[height=1.1em]{figures/logos/lagent.png}}\,Lagent} & 0{\tiny/}0{\tiny/}0 & 2275{\tiny/}2171{\tiny/}1956 & 17{\tiny/}14{\tiny/}14 & 1517{\tiny/}1426{\tiny/}1211 & 3.0{\tiny/}2.9{\tiny/}2.8 \\
     & 2{\tiny/}3{\tiny/}3 & 1562{\tiny/}1362{\tiny/}1198 & 15{\tiny/}14{\tiny/}14 & 1510{\tiny/}1306{\tiny/}1149 & 1.4{\tiny/}1.3{\tiny/}1.2 &  & 0{\tiny/}0{\tiny/}0 & 1121{\tiny/}1801{\tiny/}1654 & 18{\tiny/}17{\tiny/}13 & 1061{\tiny/}1722{\tiny/}1603 & 1.3{\tiny/}1.7{\tiny/}1.4 \\
\addlinespace[0.25em]
    \multirow{2}{*}{\raisebox{-0.4ex}{\includegraphics[height=1.1em]{figures/logos/agno.png}}\,Agno} & 3{\tiny/}1{\tiny/}2 & 2317{\tiny/}2143{\tiny/}1587 & 16{\tiny/}14{\tiny/}15 & 1545{\tiny/}1469{\tiny/}902 & 3.1{\tiny/}2.7{\tiny/}2.4 & \multirow{2}{*}{\raisebox{-0.4ex}{\includegraphics[height=1.1em]{figures/logos/agentuniverse.png}}\,AgentUniverse} & 0{\tiny/}0{\tiny/}0 & 2470{\tiny/}2638{\tiny/}2696 & 15{\tiny/}19{\tiny/}16 & 1717{\tiny/}1768{\tiny/}1853 & 3.1{\tiny/}3.5{\tiny/}3.4 \\
     & 1{\tiny/}2{\tiny/}3 & 1602{\tiny/}1590{\tiny/}1283 & 14{\tiny/}15{\tiny/}16 & 1539{\tiny/}1527{\tiny/}1218 & 1.4{\tiny/}1.4{\tiny/}1.2 &  & 0{\tiny/}0{\tiny/}0 & 1820{\tiny/}1699{\tiny/}1852 & 18{\tiny/}15{\tiny/}16 & 1731{\tiny/}1632{\tiny/}1791 & 1.8{\tiny/}1.5{\tiny/}1.6 \\
\addlinespace[0.25em]
    \multirow{2}{*}{\raisebox{-0.4ex}{\includegraphics[height=1.1em]{figures/logos/ag2.png}}\,AG2} & 1{\tiny/}4{\tiny/}3 & 479{\tiny/}2091{\tiny/}1799 & 4{\tiny/}13{\tiny/}17 & 216{\tiny/}1319{\tiny/}1220 & 0.8{\tiny/}2.9{\tiny/}2.4 & \multirow{2}{*}{\raisebox{-0.4ex}{\includegraphics[height=1.1em]{figures/logos/evoagentx.png}}\,EvoAgentX\textsuperscript{\dag}} & 2{\tiny/}0{\tiny/}0 & 1449{\tiny/}2102{\tiny/}1761 & 9{\tiny/}14{\tiny/}10 & 860{\tiny/}1347{\tiny/}1101 & 2.1{\tiny/}2.9{\tiny/}2.4 \\
     & 3{\tiny/}3{\tiny/}3 & 1332{\tiny/}1197{\tiny/}991 & 16{\tiny/}18{\tiny/}23 & 1223{\tiny/}1045{\tiny/}921 & 1.5{\tiny/}1.2{\tiny/}1.4 &  & 0{\tiny/}1{\tiny/}0 & 1426{\tiny/}1815{\tiny/}1087 & 9{\tiny/}14{\tiny/}7 & 1365{\tiny/}1736{\tiny/}1047 & 1.2{\tiny/}1.6{\tiny/}0.9 \\
\addlinespace[0.25em]
    \multirow{2}{*}{\raisebox{-0.4ex}{\includegraphics[height=1.1em]{figures/logos/microsoft.png}}\,AutoGen} & 1{\tiny/}3{\tiny/}1 & 1827{\tiny/}1973{\tiny/}1889 & 14{\tiny/}14{\tiny/}15 & 1129{\tiny/}1296{\tiny/}1103 & 2.6{\tiny/}2.6{\tiny/}2.8 & \multirow{2}{*}{\raisebox{-0.4ex}{\includegraphics[height=1.1em]{figures/logos/nerve.png}}\,Nerve} & 1{\tiny/}3{\tiny/}3 & 2366{\tiny/}2184{\tiny/}3013 & 14{\tiny/}15{\tiny/}16 & 1778{\tiny/}1516{\tiny/}2119 & 2.7{\tiny/}2.8{\tiny/}3.7 \\
     & 2{\tiny/}3{\tiny/}3 & 1375{\tiny/}995{\tiny/}952 & 15{\tiny/}17{\tiny/}16 & 1313{\tiny/}926{\tiny/}895 & 1.3{\tiny/}1.2{\tiny/}1.1 &  & 2{\tiny/}3{\tiny/}2 & 1845{\tiny/}1728{\tiny/}1377 & 15{\tiny/}16{\tiny/}14 & 1775{\tiny/}1652{\tiny/}1314 & 1.6{\tiny/}1.6{\tiny/}1.3 \\
\addlinespace[0.25em]
    \multirow{2}{*}{\raisebox{-0.4ex}{\includegraphics[height=1.1em]{figures/logos/composio.png}}\,Composio} & 0{\tiny/}0{\tiny/}0 & 2109{\tiny/}1567{\tiny/}1827 & 19{\tiny/}14{\tiny/}16 & 1479{\tiny/}887{\tiny/}1205 & 2.7{\tiny/}2.4{\tiny/}2.5 & \multirow{2}{*}{\raisebox{-0.4ex}{\includegraphics[height=1.1em]{figures/logos/agent-squad.png}}\,AgentSquad\textsuperscript{\dag}} & 0{\tiny/}0{\tiny/}0 & 698{\tiny/}1993{\tiny/}2315 & 5{\tiny/}16{\tiny/}18 & 259{\tiny/}1282{\tiny/}1544 & 1.3{\tiny/}2.7{\tiny/}3.1 \\
     & 0{\tiny/}0{\tiny/}0 & 1591{\tiny/}1804{\tiny/}1658 & 14{\tiny/}14{\tiny/}14 & 1516{\tiny/}1730{\tiny/}1602 & 1.5{\tiny/}1.6{\tiny/}1.4 &  & 0{\tiny/}0{\tiny/}0 & 1675{\tiny/}1361{\tiny/}1210 & 21{\tiny/}19{\tiny/}18 & 1574{\tiny/}1275{\tiny/}1122 & 1.8{\tiny/}1.5{\tiny/}1.4 \\
\addlinespace[0.25em]
    \multirow{2}{*}{\raisebox{-0.4ex}{\includegraphics[height=1.1em]{figures/logos/haystack.png}}\,Haystack} & 2{\tiny/}3{\tiny/}3 & 2349{\tiny/}1973{\tiny/}1789 & 16{\tiny/}14{\tiny/}15 & 1575{\tiny/}1289{\tiny/}1075 & 3.1{\tiny/}2.7{\tiny/}2.6 & \multirow{2}{*}{\raisebox{-0.4ex}{\includegraphics[height=1.1em]{figures/logos/motleycrew.png}}\,MotleyCrew} & 3{\tiny/}4{\tiny/}3 & 2171{\tiny/}1673{\tiny/}1885 & 15{\tiny/}15{\tiny/}16 & 1549{\tiny/}1012{\tiny/}1233 & 2.7{\tiny/}2.4{\tiny/}2.6 \\
     & 2{\tiny/}3{\tiny/}3 & 1513{\tiny/}1460{\tiny/}971 & 13{\tiny/}17{\tiny/}17 & 1418{\tiny/}1375{\tiny/}906 & 1.4{\tiny/}1.5{\tiny/}1.2 &  & 2{\tiny/}3{\tiny/}1 & 1875{\tiny/}1447{\tiny/}1892 & 14{\tiny/}17{\tiny/}13 & 1797{\tiny/}1363{\tiny/}1844 & 1.6{\tiny/}1.5{\tiny/}1.5 \\
\addlinespace[0.25em]
    \multirow{2}{*}{\raisebox{-0.4ex}{\includegraphics[height=1.1em]{figures/logos/smolagents.png}}\,SmoLAgents\textsuperscript{\dag}} & 3{\tiny/}3{\tiny/}3 & 1769{\tiny/}1907{\tiny/}2112 & 15{\tiny/}16{\tiny/}19 & 1202{\tiny/}1284{\tiny/}1385 & 2.3{\tiny/}2.5{\tiny/}2.9 & \multirow{2}{*}{\raisebox{-0.4ex}{\includegraphics[height=1.1em]{figures/logos/taskflowai.png}}\,TaskflowAI} & 2{\tiny/}3{\tiny/}2 & 1686{\tiny/}1722{\tiny/}1860 & 16{\tiny/}14{\tiny/}17 & 1104{\tiny/}1132{\tiny/}1281 & 2.3{\tiny/}2.3{\tiny/}2.4 \\
     & 1{\tiny/}3{\tiny/}1 & 1317{\tiny/}1318{\tiny/}1405 & 15{\tiny/}14{\tiny/}12 & 1255{\tiny/}1265{\tiny/}1348 & 1.3{\tiny/}1.3{\tiny/}1.3 &  & 0{\tiny/}1{\tiny/}0 & 1122{\tiny/}1028{\tiny/}1202 & 18{\tiny/}21{\tiny/}17 & 1060{\tiny/}965{\tiny/}1145 & 1.3{\tiny/}1.3{\tiny/}1.3 \\
\addlinespace[0.25em]
    \multirow{2}{*}{\raisebox{-0.4ex}{\includegraphics[height=1.1em]{figures/logos/agent-framework.png}}\,AgentFramework} & 0{\tiny/}2{\tiny/}1 & 618{\tiny/}2550{\tiny/}2023 & 4{\tiny/}16{\tiny/}18 & 218{\tiny/}1870{\tiny/}1312 & 1.2{\tiny/}3.0{\tiny/}2.8 & \multirow{2}{*}{\raisebox{-0.4ex}{\includegraphics[height=1.1em]{figures/logos/council-ai.png}}\,CouncilAI} & 0{\tiny/}0{\tiny/}1 & 2274{\tiny/}2271{\tiny/}2500 & 13{\tiny/}15{\tiny/}15 & 1517{\tiny/}1533{\tiny/}1876 & 3.0{\tiny/}3.0{\tiny/}2.9 \\
     & 3{\tiny/}3{\tiny/}3 & 1652{\tiny/}1721{\tiny/}1466 & 16{\tiny/}14{\tiny/}15 & 1503{\tiny/}1644{\tiny/}1393 & 1.9{\tiny/}1.6{\tiny/}1.4 &  & 0{\tiny/}0{\tiny/}0 & 1762{\tiny/}6211{\tiny/}6874 & 21{\tiny/}58{\tiny/}251 & 1693{\tiny/}5140{\tiny/}6081 & 1.7{\tiny/}9.4{\tiny/}13.3 \\
\addlinespace[0.25em]
    \multirow{2}{*}{\raisebox{-0.4ex}{\includegraphics[height=1.1em]{figures/logos/qwen-agent.png}}\,Qwen Agent} & 0{\tiny/}0{\tiny/}0 & 2374{\tiny/}2567{\tiny/}1599 & 16{\tiny/}18{\tiny/}15 & 1651{\tiny/}1751{\tiny/}946 & 3.0{\tiny/}3.3{\tiny/}2.4 & \multirow{2}{*}{\raisebox{-0.4ex}{\includegraphics[height=1.1em]{figures/logos/agentflow.png}}\,AgentFlow\textsuperscript{\dag}} & 0{\tiny/}0{\tiny/}1 & 2394{\tiny/}2641{\tiny/}2226 & 16{\tiny/}15{\tiny/}20 & 1573{\tiny/}1725{\tiny/}1537 & 3.2{\tiny/}3.5{\tiny/}2.9 \\
     & 0{\tiny/}0{\tiny/}0 & 1462{\tiny/}1737{\tiny/}1622 & 16{\tiny/}14{\tiny/}13 & 1400{\tiny/}1676{\tiny/}1574 & 1.4{\tiny/}1.5{\tiny/}1.4 &  & 0{\tiny/}0{\tiny/}0 & 1309{\tiny/}1719{\tiny/}1540 & 12{\tiny/}13{\tiny/}12 & 1248{\tiny/}1627{\tiny/}1481 & 1.2{\tiny/}1.6{\tiny/}1.3 \\
\addlinespace[0.25em]
    \multirow{2}{*}{\raisebox{-0.4ex}{\includegraphics[height=1.1em]{figures/logos/agentscope.png}}\,AgentScope\textsuperscript{\dag}} & 0{\tiny/}0{\tiny/}1 & 2256{\tiny/}1973{\tiny/}1676 & 15{\tiny/}16{\tiny/}16 & 1513{\tiny/}1285{\tiny/}1056 & 3.0{\tiny/}2.7{\tiny/}2.4 & \multirow{2}{*}{\raisebox{-0.4ex}{\includegraphics[height=1.1em]{figures/logos/autoagent.png}}\,AutoAgent\textsuperscript{\dag}} & 1{\tiny/}2{\tiny/}2 & 2120{\tiny/}1975{\tiny/}2376 & 12{\tiny/}10{\tiny/}19 & 1386{\tiny/}1412{\tiny/}1451 & 2.8{\tiny/}2.4{\tiny/}3.4 \\
     & 0{\tiny/}1{\tiny/}0 & 2059{\tiny/}2325{\tiny/}2169 & 12{\tiny/}14{\tiny/}14 & 2006{\tiny/}2253{\tiny/}2107 & 1.6{\tiny/}1.8{\tiny/}1.7 &  & 2{\tiny/}3{\tiny/}3 & 877{\tiny/}963{\tiny/}673 & 8{\tiny/}9{\tiny/}7 & 831{\tiny/}904{\tiny/}628 & 0.9{\tiny/}1.0{\tiny/}0.7 \\
\addlinespace[0.25em]
    \multirow{2}{*}{\raisebox{-0.4ex}{\includegraphics[height=1.1em]{figures/logos/praisonai.png}}\,PraisonAI} & 1{\tiny/}2{\tiny/}1 & 1832{\tiny/}2341{\tiny/}2298 & 16{\tiny/}18{\tiny/}15 & 1163{\tiny/}1690{\tiny/}1537 & 2.6{\tiny/}2.9{\tiny/}3.0 & \multirow{2}{*}{\raisebox{-0.4ex}{\includegraphics[height=1.1em]{figures/logos/agentlite.png}}\,AgentLite\textsuperscript{\dag}} & 2{\tiny/}2{\tiny/}3 & 2023{\tiny/}2123{\tiny/}1560 & 17{\tiny/}15{\tiny/}16 & 1247{\tiny/}1369{\tiny/}998 & 2.9{\tiny/}2.9{\tiny/}2.2 \\
     & 0{\tiny/}3{\tiny/}3 & 1922{\tiny/}1574{\tiny/}1367 & 14{\tiny/}16{\tiny/}15 & 1860{\tiny/}1506{\tiny/}1266 & 1.6{\tiny/}1.5{\tiny/}1.5 &  & 1{\tiny/}2{\tiny/}2 & 1202{\tiny/}846{\tiny/}1072 & 35{\tiny/}52{\tiny/}33 & 1150{\tiny/}791{\tiny/}1017 & 1.7{\tiny/}2.0{\tiny/}1.6 \\
\addlinespace[0.25em]
    \multirow{2}{*}{\raisebox{-0.4ex}{\includegraphics[height=1.1em]{figures/logos/camel.png}}\,CAMEL\textsuperscript{\dag}} & 2{\tiny/}1{\tiny/}2 & 2060{\tiny/}1734{\tiny/}1313 & 14{\tiny/}14{\tiny/}15 & 1372{\tiny/}1056{\tiny/}724 & 2.7{\tiny/}2.5{\tiny/}2.1 & \multirow{2}{*}{\raisebox{-0.4ex}{\includegraphics[height=1.1em]{figures/logos/octotools.png}}\,Octotools} & 1{\tiny/}1{\tiny/}1 & 2134{\tiny/}2155{\tiny/}2147 & 15{\tiny/}17{\tiny/}20 & 1552{\tiny/}1404{\tiny/}1322 & 2.6{\tiny/}2.9{\tiny/}3.1 \\
     & 2{\tiny/}3{\tiny/}3 & 1528{\tiny/}1391{\tiny/}1213 & 14{\tiny/}14{\tiny/}14 & 1470{\tiny/}1332{\tiny/}1159 & 1.4{\tiny/}1.3{\tiny/}1.2 &  & 1{\tiny/}1{\tiny/}2 & 1141{\tiny/}1048{\tiny/}930 & 17{\tiny/}23{\tiny/}22 & 1062{\tiny/}965{\tiny/}845 & 1.4{\tiny/}1.5{\tiny/}1.4 \\
\addlinespace[0.25em]
    \multirow{2}{*}{\raisebox{-0.4ex}{\includegraphics[height=1.1em]{figures/logos/fast-agent.png}}\,FastAgent} & 0{\tiny/}1{\tiny/}0 & 2447{\tiny/}2603{\tiny/}2142 & 13{\tiny/}11{\tiny/}12 & 1655{\tiny/}1573{\tiny/}1454 & 3.1{\tiny/}3.7{\tiny/}2.7 & \multirow{2}{*}{\raisebox{-0.4ex}{\includegraphics[height=1.1em]{figures/logos/gptswarm.png}}\,GPTSwarm\textsuperscript{\dag}} & 0{\tiny/}1{\tiny/}0 & 1242{\tiny/}1584{\tiny/}1601 & 10{\tiny/}11{\tiny/}23 & 668{\tiny/}949{\tiny/}981 & 2.0{\tiny/}2.3{\tiny/}2.4 \\
     & 2{\tiny/}3{\tiny/}2 & 2287{\tiny/}2282{\tiny/}1574 & 10{\tiny/}12{\tiny/}11 & 2225{\tiny/}2213{\tiny/}1526 & 1.7{\tiny/}1.8{\tiny/}1.3 &  & 0{\tiny/}0{\tiny/}1 & 1956{\tiny/}1797{\tiny/}1731 & 16{\tiny/}16{\tiny/}14 & 1901{\tiny/}1733{\tiny/}1672 & 1.6{\tiny/}1.6{\tiny/}1.5 \\
\addlinespace[0.25em]
    \multirow{2}{*}{\raisebox{-0.4ex}{\includegraphics[height=1.1em]{figures/logos/solace-agent-mesh.png}}\,Solace} & 0{\tiny/}0{\tiny/}0 & 3192{\tiny/}2567{\tiny/}2182 & 12{\tiny/}11{\tiny/}12 & 2367{\tiny/}1934{\tiny/}1473 & 3.7{\tiny/}2.9{\tiny/}2.8 & \multirow{2}{*}{\raisebox{-0.4ex}{\includegraphics[height=1.1em]{figures/logos/taskweaver.png}}\,Taskweaver\textsuperscript{\dag}} & 0{\tiny/}0{\tiny/}0 & 2301{\tiny/}2308{\tiny/}2604 & 15{\tiny/}19{\tiny/}17 & 1675{\tiny/}1731{\tiny/}1915 & 2.8{\tiny/}2.7{\tiny/}3.1 \\
     & 0{\tiny/}1{\tiny/}0 & 1161{\tiny/}1054{\tiny/}1069 & 8{\tiny/}9{\tiny/}9 & 1110{\tiny/}1011{\tiny/}1027 & 1.0{\tiny/}1.0{\tiny/}0.9 &  & 0{\tiny/}0{\tiny/}0 & 1516{\tiny/}1934{\tiny/}1957 & 17{\tiny/}16{\tiny/}16 & 1452{\tiny/}1852{\tiny/}1893 & 1.5{\tiny/}1.7{\tiny/}1.7 \\
\addlinespace[0.25em]
    \multirow{2}{*}{\raisebox{-0.4ex}{\includegraphics[height=1.1em]{figures/logos/langroid.png}}\,Langroid} & 0{\tiny/}0{\tiny/}0 & 812{\tiny/}739{\tiny/}1044 & 5{\tiny/}4{\tiny/}6 & 465{\tiny/}437{\tiny/}641 & 1.2{\tiny/}1.1{\tiny/}1.5 & \multirow{2}{*}{\raisebox{-0.4ex}{\includegraphics[height=1.1em]{figures/logos/autoagents.png}}\,AutoAgents\textsuperscript{\dag}} & 0{\tiny/}0{\tiny/}0 & 2570{\tiny/}2444{\tiny/}2243 & 15{\tiny/}15{\tiny/}14 & 1691{\tiny/}1772{\tiny/}1494 & 3.4{\tiny/}2.9{\tiny/}2.9 \\
     & 0{\tiny/}0{\tiny/}0 & 243{\tiny/}452{\tiny/}82 & 1{\tiny/}2{\tiny/}0 & 206{\tiny/}420{\tiny/}61 & 0.3{\tiny/}0.4{\tiny/}0.1 &  & 0{\tiny/}0{\tiny/}0 & 1128{\tiny/}1413{\tiny/}1396 & 16{\tiny/}17{\tiny/}16 & 1066{\tiny/}1352{\tiny/}1332 & 1.3{\tiny/}1.4{\tiny/}1.4 \\
\addlinespace[0.25em]
    \multirow{2}{*}{\raisebox{-0.4ex}{\includegraphics[height=1.1em]{figures/logos/beeai.png}}\,BeeAI} & 3{\tiny/}2{\tiny/}2 & 2388{\tiny/}1688{\tiny/}1545 & 14{\tiny/}21{\tiny/}27 & 1769{\tiny/}1064{\tiny/}962 & 2.8{\tiny/}2.4{\tiny/}2.3 & \multirow{2}{*}{\raisebox{-0.4ex}{\includegraphics[height=1.1em]{figures/logos/opensage.png}}\,OpenSage\textsuperscript{\dag}} & 2{\tiny/}2{\tiny/}2 & 2618{\tiny/}2395{\tiny/}2170 & 12{\tiny/}13{\tiny/}13 & 1927{\tiny/}1742{\tiny/}1466 & 3.1{\tiny/}2.8{\tiny/}2.8 \\
     & 0{\tiny/}1{\tiny/}0 & 1586{\tiny/}1814{\tiny/}2282 & 12{\tiny/}29{\tiny/}23 & 1530{\tiny/}1754{\tiny/}2227 & 1.4{\tiny/}1.9{\tiny/}2.0 &  & 1{\tiny/}1{\tiny/}0 & 1588{\tiny/}1675{\tiny/}1419 & 12{\tiny/}12{\tiny/}10 & 1529{\tiny/}1621{\tiny/}1374 & 1.4{\tiny/}1.4{\tiny/}1.2 \\
\addlinespace[0.25em]
    \multirow{2}{*}{\raisebox{-0.4ex}{\includegraphics[height=1.1em]{figures/logos/metagpt.png}}\,MetaGPT\textsuperscript{\dag}} & 0{\tiny/}0{\tiny/}0 & 2506{\tiny/}2465{\tiny/}1561 & 13{\tiny/}16{\tiny/}12 & 1748{\tiny/}1810{\tiny/}1037 & 3.1{\tiny/}2.9{\tiny/}2.1 &  &  &  &  &  &  \\
     & 0{\tiny/}0{\tiny/}0 & 1219{\tiny/}1554{\tiny/}1101 & 13{\tiny/}15{\tiny/}13 & 1163{\tiny/}1485{\tiny/}1044 & 1.2{\tiny/}1.5{\tiny/}1.1 &  &  &  &  &  &  \\
\bottomrule
\end{tabular}}
\end{table*}

We ablate information sources across all 51 frameworks to test how robust generation is under degraded conditions (\autoref{tab:gen_effort_ablation}). The \textsc{Docs only} condition provides curated documentation but disables source code exploration. The \textsc{Source only} condition provides raw source code access but removes curated documentation. The \textsc{None} condition removes all reference material, forcing the LLM to rely entirely on parametric knowledge from pre-training.

Here \textbf{Pass} counts agents that both pass validation and use the framework \emph{natively} (thin-wrapper agents excluded), measuring genuine framework usage under each condition. The striking pattern is that genuine usage does \emph{not} rise with access to more documentation: \textsc{Source only} (40\%, raw code but no curated prose) yields the highest rate, above both \textsc{None} (33\%) and \textsc{Docs only} (28\%), a first sign that more curated information is not always better.
\begin{itemize}[nosep,leftmargin=*]
  \item \textbf{Source only} yields the \emph{highest} rate of genuine framework usage (40\%, 164/408): seeing the actual API in source code leads the developer to native calls more often than curated prose does.
  \item \textbf{Docs only} is the lowest (28\%, 114/408): curated documentation alone leaves the developer more likely to hand-roll a thin wrapper than when it can read the source.
  \item \textbf{None} is surprisingly survivable (33\%, 136/408): even with no documentation or source, a third of agents still use the framework natively, drawing on parametric knowledge of popular frameworks (\textit{LangChain}, \textit{CrewAI}, \textit{OpenAI Agents}). Strikingly, \textsc{None} even edges out \textsc{Docs only} (33\% vs.\ 28\%): the documentation is complete, but its worked examples (often minimal quickstarts) anchor the developer, which tends to copy or lightly edit those snippets without adapting them to the benchmark, leaving a thin wrapper rather than an agent that exercises the framework's native API. Stripped of examples to imitate, the developer instead falls back on the broader native-API usage it memorized from real code during pre-training. The survivors skew popular (the median framework that still passes ranks 22nd by monthly downloads versus 34th for those that fail) because widely-used APIs are heavily represented in the pre-training corpus, so the developer model has effectively memorized them; niche frameworks, sparse in that corpus, collapse once all reference material is removed.
  \item Token usage is nearly flat across the three conditions, so the differences reflect what the developer can \emph{discover} about the API, not how much it explores.
\end{itemize}

\finding{Genuine framework usage stays within a narrow 28--40\% band across information conditions and, crucially, does not increase with more curated documentation: \textsc{Docs only} is the lowest (28\%), while withholding documentation in favor of raw source \emph{raises} native usage (\textsc{Source only} 40\%, \textsc{None} 33\%). Documentation's worked examples anchor the developer toward thin wrappers; source code and parametric knowledge are largely substitutable, and no single information source is a hard bottleneck.}

\section{Limitations}
\label{sec:discussion}

\paragraph{Single execution model.}
All generated agents run on GPT-5.4 Nano as the execution backbone. This is a controlled-variable design rather than a confound: fixing the execution model isolates framework effects from model--framework interactions, which is precisely the quantity we set out to measure. It also does not bias the cross-framework comparison, because every framework reaches the model through the same in-container proxy, which normalizes their heterogeneous request formats (OpenAI chat-completions, the Azure Responses API, Anthropic Messages) onto one backend. All 51 frameworks are therefore evaluated against an identical model on identical infrastructure, keeping the comparison apples-to-apples. We chose Nano for its cost--performance tradeoff: the full campaign would cost ${\sim}$40$\times$ more with GPT-5.4 or Claude Opus. The pipeline is model-agnostic, and every ADK runs through this same infrastructure (adding one requires a single configuration entry), so sweeping additional model families is straightforward future work.

\paragraph{Training-data bias.}
The generator LLM has more training data for popular frameworks (LangChain, CrewAI) than newer ones. We argue this reflects a real-world advantage: frameworks with richer community resources are genuinely easier for \emph{any} developer to use. Our ablation (\S\ref{sec:ablation}) shows that even with no documentation or source code, 33\% of agents (136/408) still use the framework natively, drawing on parametric knowledge of popular frameworks, while genuine usage stays within a narrow 28--40\% band across all information conditions. Parametric knowledge thus provides a floor without being the dominant factor.

\section{Conclusion}
\label{sec:conclusion}

We presented LLM-as-a-Developer, an evaluation methodology that uses an LLM as a proxy developer to assess ADK frameworks at scale, and ADK Arena, a fully automated pipeline implementing this methodology across 51 frameworks and four benchmarks. Generation cost varies 5.6$\times$ across frameworks and serves as a quantitative proxy for API complexity. The best ADK-generated agents match frontier coding agents on individual benchmarks (up to 80\%), but the typical framework resolves far fewer (median 32\%), and among frontier agents accuracy trades off against efficiency, where the top resolvers spend 8--10$\times$ the tokens of the leanest. No single framework dominates all benchmarks, and ablation shows that genuine framework usage stays within a narrow 28--40\% band whether the developer has raw source code (highest, 40\%), curated documentation (lowest, 28\%), or no reference material at all (33\%), indicating that documentation, source code, and parametric knowledge are largely substitutable rather than any one being a hard bottleneck.

\bibliographystyle{plainnat}
\bibliography{custom}

\begin{thebibliography}{55}
\providecommand{\natexlab}[1]{#1}
\providecommand{\url}[1]{\texttt{#1}}
\expandafter\ifx\csname urlstyle\endcsname\relax
  \providecommand{\doi}[1]{doi: #1}\else
  \providecommand{\doi}{doi: \begingroup \urlstyle{rm}\Url}\fi

\bibitem[ope(2025)]{opencode}
Opencode.
\newblock \url{https://opencode.ai}, 2025.

\bibitem[{Anthropic}(2024)]{anthropic2024claude}
{Anthropic}.
\newblock The {Claude} 3 model family: {Opus}, {Sonnet}, {Haiku}.
\newblock \emph{Technical Report}, 2024.

\bibitem[{Anthropic}(2025)]{claudecode}
{Anthropic}.
\newblock Claude code.
\newblock
  \url{https://docs.anthropic.com/en/docs/agents-and-tools/claude-code}, 2025.

\bibitem[Bandi et~al.(2026)Bandi, Hertzberg, Boo, Polakam, Da, Hassaan, Sharma,
  Park, Hernandez, Rambado, Salazar, Cruz, Rane, Levin, Kenstler, and
  Liu]{mcpatlas2025}
Chaithanya Bandi, Ben Hertzberg, Geobio Boo, Tejas Polakam, Jeff Da, Sami
  Hassaan, Manasi Sharma, Andrew Park, Ernesto Hernandez, Dan Rambado, Ivan
  Salazar, Rafael Cruz, Chetan Rane, Ben Levin, Brad Kenstler, and Bing Liu.
\newblock {MCP-Atlas}: A large-scale benchmark for tool-use competency with
  real {MCP} servers.
\newblock \emph{arXiv preprint arXiv:2602.00933}, 2026.

\bibitem[Brown et~al.(2020)Brown, Mann, Ryder, Subbiah, Kaplan, Dhariwal,
  Neelakantan, Shyam, Sastry, Askell, et~al.]{brown2020language}
Tom Brown, Benjamin Mann, Nick Ryder, Melanie Subbiah, Jared~D Kaplan, Prafulla
  Dhariwal, Arvind Neelakantan, Pranav Shyam, Girish Sastry, Amanda Askell,
  et~al.
\newblock Language models are few-shot learners.
\newblock \emph{Advances in Neural Information Processing Systems},
  33:\penalty0 1877--1901, 2020.

\bibitem[Chen et~al.(2021)Chen, Tworek, Jun, Yuan, Pinto, Kaplan, Edwards,
  Burda, Joseph, Brockman, et~al.]{chen2021evaluating}
Mark Chen, Jerry Tworek, Heewoo Jun, Qiming Yuan, Henrique Pond{\'e}
  de~Oliveira Pinto, Jared Kaplan, Harri Edwards, Yuri Burda, Nicholas Joseph,
  Greg Brockman, et~al.
\newblock Evaluating large language models trained on code.
\newblock \emph{arXiv preprint arXiv:2107.03374}, 2021.

\bibitem[Chen et~al.(2024)Chen, Su, Zuo, Yang, Yuan, Chan, Yu, Lu, Hung, Qian,
  et~al.]{chen2024agentverse}
Weize Chen, Yusheng Su, Jingwei Zuo, Cheng Yang, Chenfei Yuan, Chi-Min Chan,
  Heyang Yu, Yaxi Lu, Yi-Hsin Hung, Chen Qian, et~al.
\newblock {AgentVerse}: Facilitating multi-agent collaboration and exploring
  emergent behaviors.
\newblock \emph{arXiv preprint arXiv:2308.10848}, 2024.

\bibitem[Chiang et~al.(2024)Chiang, Zheng, Sheng, Angelopoulos, Li, Li, Zhang,
  Zhu, Jordan, Gonzalez, and Stoica]{chiang2024chatbot}
Wei-Lin Chiang, Lianmin Zheng, Ying Sheng, Anastasios~Nikolas Angelopoulos,
  Tianle Li, Dacheng Li, Hao Zhang, Banghua Zhu, Michael Jordan, Joseph~E
  Gonzalez, and Ion Stoica.
\newblock Chatbot arena: An open platform for evaluating {LLMs} by human
  preference.
\newblock In \emph{Proceedings of the 41st International Conference on Machine
  Learning (ICML)}, 2024.

\bibitem[Clarke(2004)]{clarke2004measuring}
Steven Clarke.
\newblock Measuring {API} usability.
\newblock \emph{Dr. Dobb's Journal}, 2004.

\bibitem[Decan et~al.(2019)Decan, Mens, and Grosjean]{decan2019empirical}
Alexandre Decan, Tom Mens, and Philippe Grosjean.
\newblock An empirical comparison of dependency network evolution in seven
  software packaging ecosystems.
\newblock \emph{Empirical Software Engineering}, 24\penalty0 (1):\penalty0
  381--416, 2019.

\bibitem[{GitHub}(2025)]{copilot}
{GitHub}.
\newblock Github copilot coding agent.
\newblock \url{https://github.com/features/copilot}, 2025.

\bibitem[Hasan et~al.(2025)Hasan, Li, Fallahzadeh, Rajbahadur, Adams, and
  Hassan]{hasan2025testing}
Mohammed~Mehedi Hasan, Hao Li, Emad Fallahzadeh, Gopi~Krishnan Rajbahadur, Bram
  Adams, and Ahmed~E. Hassan.
\newblock An empirical study of testing practices in open source {AI} agent
  frameworks and agentic applications.
\newblock \emph{arXiv preprint arXiv:2509.19185}, 2025.

\bibitem[Hong et~al.(2024)Hong, Zhuge, Chen, Zheng, Cheng, Zhang, Wang, Wang,
  Yau, Lin, et~al.]{hong2024metagpt}
Sirui Hong, Mingchen Zhuge, Jonathan Chen, Xiawu Zheng, Yuheng Cheng, Ceyao
  Zhang, Jinlin Wang, Zili Wang, Steven Ka~Shing Yau, Zijuan Lin, et~al.
\newblock {MetaGPT}: Meta programming for a multi-agent collaborative
  framework.
\newblock \emph{arXiv preprint arXiv:2308.00352}, 2024.

\bibitem[Jimenez et~al.(2024)Jimenez, Yang, Wettig, Yao, Pei, Press, and
  Narasimhan]{jimenez2024swebench}
Carlos~E Jimenez, John Yang, Alexander Wettig, Shunyu Yao, Kexin Pei, Ofir
  Press, and Karthik Narasimhan.
\newblock {SWE-bench}: Can language models resolve real-world {GitHub} issues?
\newblock In \emph{Proceedings of the 12th International Conference on Learning
  Representations (ICLR)}, 2024.

\bibitem[Ko et~al.(2004)Ko, Myers, and Aung]{ko2004six}
Amy~J. Ko, Brad~A. Myers, and Htet~Htet Aung.
\newblock Six learning barriers in end-user programming systems.
\newblock In \emph{IEEE Symposium on Visual Languages and Human-Centric
  Computing}, pages 199--206, 2004.

\bibitem[Lemieux et~al.(2023)Lemieux, Inala, Qi, Lahiri, Sen, and
  Gao]{lemieux2023codamosa}
Caroline Lemieux, Jeevana~Priya Inala, Shuvendu~K. Qi, Shuvendu~K. Lahiri,
  Siddhartha Sen, and Kexin Gao.
\newblock {CodaMosa}: Escaping coverage plateaus in test generation with
  pre-trained large language models.
\newblock In \emph{International Conference on Software Engineering (ICSE)},
  2023.

\bibitem[Li et~al.(2023{\natexlab{a}})Li, Hammoud, Itani, Khizbullin, and
  Ghanem]{li2023camel}
Guohao Li, Hasan Abed Al~Kader Hammoud, Hani Itani, Dmitrii Khizbullin, and
  Bernard Ghanem.
\newblock {CAMEL}: Communicative agents for ``mind'' exploration of large
  language model society.
\newblock \emph{Advances in Neural Information Processing Systems}, 36,
  2023{\natexlab{a}}.

\bibitem[Li et~al.(2023{\natexlab{b}})Li, Allal, Zi, Muennighoff, Kocetkov,
  Mou, Marone, Akiki, Li, Chim, et~al.]{li2023starcoder}
Raymond Li, Loubna~Ben Allal, Yangtian Zi, Niklas Muennighoff, Denis Kocetkov,
  Chenghao Mou, Marc Marone, Christopher Akiki, Jia Li, Jenny Chim, et~al.
\newblock {StarCoder}: May the source be with you!
\newblock \emph{arXiv preprint arXiv:2305.06161}, 2023{\natexlab{b}}.

\bibitem[Li et~al.(2024)Li, Chiang, Frick, Dunlap, Wu, Zhu, Gonzalez, and
  Stoica]{li2024crowdsourced}
Tianle Li, Wei-Lin Chiang, Evan Frick, Lisa Dunlap, Tianhao Wu, Banghua Zhu,
  Joseph~E Gonzalez, and Ion Stoica.
\newblock From crowdsourced data to high-quality benchmarks: Arena-hard and
  {BenchBuilder} pipeline.
\newblock \emph{arXiv preprint arXiv:2406.11939}, 2024.

\bibitem[Liu et~al.(2022)Liu, Chen, Fan, Chen, Liu,
  et~al.]{liu2022demystifying}
Chengwei Liu, Sen Chen, Lingling Fan, Bihuan Chen, Yang Liu, et~al.
\newblock Demystifying the vulnerability propagation and its evolution via
  dependency trees in the npm ecosystem.
\newblock In \emph{Proceedings of the 44th International Conference on Software
  Engineering (ICSE)}, 2022.

\bibitem[Liu et~al.(2026)Liu, Upadhyay, Chhetri, Siddique, and
  Farooq]{liu2026multiagent}
Daniel Liu, Krishna Upadhyay, Vinaik Chhetri, A.B. Siddique, and Umar Farooq.
\newblock A large-scale study on the development and issues of multi-agent {AI}
  systems.
\newblock \emph{arXiv preprint arXiv:2601.07136}, 2026.

\bibitem[Liu et~al.(2023)Liu, Yu, Zhang, Xu, Lei, Lai, Gu, Ding, Men, Yang,
  et~al.]{liu2023agentbench}
Xiao Liu, Hao Yu, Hanchen Zhang, Yifan Xu, Xuanyu Lei, Hanyu Lai, Yu~Gu,
  Hangliang Ding, Kaiwen Men, Kejuan Yang, et~al.
\newblock {AgentBench}: Evaluating {LLMs} as agents.
\newblock \emph{arXiv preprint arXiv:2308.03688}, 2023.

\bibitem[Merrill et~al.(2026)Merrill, Carlini, Shaw, Schmidt,
  et~al.]{terminalbench2025}
Mike~A. Merrill, Nicholas Carlini, Alexander~G. Shaw, Ludwig Schmidt, et~al.
\newblock {Terminal-Bench}: Benchmarking agents on hard, realistic tasks in
  command line interfaces.
\newblock \emph{arXiv preprint arXiv:2601.11868}, 2026.

\bibitem[Mialon et~al.(2023)Mialon, Fourrier, Swift, Wolf, LeCun, and
  Scialom]{mialon2023gaia}
Gr{\'e}goire Mialon, Cl{\'e}mentine Fourrier, Craig Swift, Thomas Wolf, Yann
  LeCun, and Thomas Scialom.
\newblock {GAIA}: A benchmark for general {AI} assistants.
\newblock \emph{arXiv preprint arXiv:2311.12983}, 2023.

\bibitem[{MLCommons}(2024)]{mlperf}
{MLCommons}.
\newblock {MLPerf}: A benchmark suite for machine learning.
\newblock \url{https://mlcommons.org/benchmarks/}, 2024.

\bibitem[Murphy et~al.(2006)Murphy, Kersten, and Findlater]{murphy2006java}
Gail~C. Murphy, Mik Kersten, and Leah Findlater.
\newblock How are {Java} software developers using the {Eclipse IDE}?
\newblock \emph{IEEE Software}, 23\penalty0 (4):\penalty0 76--83, 2006.

\bibitem[{OpenAI}(2023)]{openai2023gpt4}
{OpenAI}.
\newblock {GPT-4} technical report.
\newblock \emph{arXiv preprint arXiv:2303.08774}, 2023.

\bibitem[{OpenAI}(2025)]{codex}
{OpenAI}.
\newblock Codex cli.
\newblock \url{https://github.com/openai/codex}, 2025.

\bibitem[Orogat et~al.(2026)Orogat, Rostam, and Mansour]{orogat2026mafbench}
Abdelghny Orogat, Ana Rostam, and Essam Mansour.
\newblock Understanding multi-agent {LLM} frameworks: A unified benchmark and
  experimental analysis.
\newblock \emph{arXiv preprint arXiv:2602.03128}, 2026.

\bibitem[Patil et~al.(2023)Patil, Zhang, Wang, and Gonzalez]{patil2023gorilla}
Shishir~G Patil, Tianjun Zhang, Xin Wang, and Joseph~E Gonzalez.
\newblock Gorilla: Large language model connected with massive {APIs}.
\newblock \emph{arXiv preprint arXiv:2305.15334}, 2023.

\bibitem[Qian et~al.(2024)Qian, Liu, Liu, Chen, Dang, Li, Yang, Chen, Su, Cong,
  et~al.]{qian2024chatdev}
Chen Qian, Wei Liu, Hongzhang Liu, Nuo Chen, Yufan Dang, Jiahao Li, Cheng Yang,
  Weize Chen, Yusheng Su, Xin Cong, et~al.
\newblock {ChatDev}: Communicative agents for software development.
\newblock \emph{arXiv preprint arXiv:2307.07924}, 2024.

\bibitem[Qin et~al.(2024)Qin, Liang, Ye, Zhu, Yan, Lu, Lin, Cong, Tang, Qian,
  et~al.]{qin2024toolllm}
Yujia Qin, Shihao Liang, Yining Ye, Kunlun Zhu, Lan Yan, Yaxi Lu, Yankai Lin,
  Xin Cong, Xiangru Tang, Bill Qian, et~al.
\newblock {ToolLLM}: Facilitating large language models to master 16000+
  real-world {APIs}.
\newblock \emph{arXiv preprint arXiv:2307.16789}, 2024.

\bibitem[Rama and Kak(2015)]{rama2015api}
Girish~Maskeri Rama and Avinash Kak.
\newblock Some structural measures of {API} usability.
\newblock \emph{Software: Practice and Experience}, 45\penalty0 (1):\penalty0
  75--110, 2015.

\bibitem[Rozi{\`e}re et~al.(2024)Rozi{\`e}re, Gehring, Gloeckle, Sootla, Gat,
  Tan, Adi, Liu, Sauvestre, Remez, et~al.]{roziere2024code}
Baptiste Rozi{\`e}re, Jonas Gehring, Fabian Gloeckle, Sten Sootla, Itai Gat,
  Xiaoqing~Ellen Tan, Yossi Adi, Jingyu Liu, Romain Sauvestre, Tal Remez,
  et~al.
\newblock Code {Llama}: Open foundation models for code.
\newblock \emph{arXiv preprint arXiv:2308.12950}, 2024.

\bibitem[Schick et~al.(2024)Schick, Dwivedi-Yu, Dess{\`\i}, Raileanu, Lomeli,
  Hambro, Zettlemoyer, Cancedda, and Scialom]{schick2024toolformer}
Timo Schick, Jane Dwivedi-Yu, Roberto Dess{\`\i}, Roberta Raileanu, Maria
  Lomeli, Eric Hambro, Luke Zettlemoyer, Nicola Cancedda, and Thomas Scialom.
\newblock Toolformer: Language models can teach themselves to use tools.
\newblock \emph{Advances in Neural Information Processing Systems}, 36, 2024.

\bibitem[Shinn et~al.(2023)Shinn, Cassano, Gopinath, Narasimhan, and
  Yao]{shinn2023reflexion}
Noah Shinn, Federico Cassano, Ashwin Gopinath, Karthik Narasimhan, and Shunyu
  Yao.
\newblock Reflexion: Language agents with verbal reinforcement learning.
\newblock \emph{Advances in Neural Information Processing Systems}, 36, 2023.

\bibitem[Sim et~al.(2003)Sim, Easterbrook, and Holt]{sim2003using}
Susan~Elliott Sim, Steve Easterbrook, and Richard~C. Holt.
\newblock Using benchmarking to advance research: A challenge to software
  engineering.
\newblock In \emph{Proceedings of the 25th International Conference on Software
  Engineering (ICSE)}, pages 74--83, 2003.

\bibitem[Stylos and Myers(2007)]{stylos2007usability}
Jeffrey Stylos and Brad Myers.
\newblock Mapping the space of {API} design decisions.
\newblock In \emph{IEEE Symposium on Visual Languages and Human-Centric
  Computing}, pages 50--60, 2007.

\bibitem[Sumers et~al.(2024)Sumers, Yao, Narasimhan, and
  Griffiths]{sumers2024cognitive}
Theodore~R Sumers, Shunyu Yao, Karthik Narasimhan, and Thomas~L Griffiths.
\newblock Cognitive architectures for language agents.
\newblock \emph{Transactions on Machine Learning Research}, 2024.

\bibitem[{TechEmpower}(2024)]{techempower}
{TechEmpower}.
\newblock {TechEmpower} web framework benchmarks.
\newblock \url{https://www.techempower.com/benchmarks/}, 2024.

\bibitem[Tichy(1998)]{tichy1998should}
Walter~F. Tichy.
\newblock Should computer scientists experiment more?
\newblock \emph{IEEE Computer}, 31\penalty0 (5):\penalty0 32--40, 1998.

\bibitem[Touvron et~al.(2023)Touvron, Lavril, Izacard, Martinet, Lachaux,
  Lacroix, Rozi{\`e}re, Goyal, Hambro, Azhar, et~al.]{touvron2023llama}
Hugo Touvron, Thibaut Lavril, Gautier Izacard, Xavier Martinet, Marie-Anne
  Lachaux, Timoth{\'e}e Lacroix, Baptiste Rozi{\`e}re, Naman Goyal, Eric
  Hambro, Faisal Azhar, et~al.
\newblock {LLaMA}: Open and efficient foundation language models.
\newblock \emph{arXiv preprint arXiv:2302.13971}, 2023.

\bibitem[{Transaction Processing Performance Council}(2024)]{tpch}
{Transaction Processing Performance Council}.
\newblock {TPC-H}: Decision support benchmark.
\newblock \url{https://www.tpc.org/tpch/}, 2024.

\bibitem[Wang et~al.(2024{\natexlab{a}})Wang, Ma, Feng, Zhang, Yang, Zhang,
  Chen, Tang, Chen, Lin, et~al.]{wang2024survey}
Lei Wang, Chen Ma, Xueyang Feng, Zeyu Zhang, Hao Yang, Jingsen Zhang, Zhiyuan
  Chen, Jiakai Tang, Xu~Chen, Yankai Lin, et~al.
\newblock A survey on large language model based autonomous agents.
\newblock \emph{Frontiers of Computer Science}, 18\penalty0 (6),
  2024{\natexlab{a}}.

\bibitem[Wang et~al.(2024{\natexlab{b}})Wang, Chen, Yuan, Zhang, Li, Peng, and
  Ji]{wang2024executable}
Xingyao Wang, Yangyi Chen, Lifan Yuan, Yizhe Zhang, Yunzhu Li, Hao Peng, and
  Heng Ji.
\newblock Executable code actions elicit better {LLM} agents.
\newblock \emph{arXiv preprint arXiv:2402.01030}, 2024{\natexlab{b}}.

\bibitem[Wang et~al.(2024{\natexlab{c}})Wang, Li, Song, Xu, Tang,
  et~al.]{openhands}
Xingyao Wang, Boxuan Li, Yufan Song, Frank~F. Xu, Xiangru Tang, et~al.
\newblock {OpenHands}: An open platform for {AI} software developers as
  generalist agents.
\newblock \emph{arXiv preprint arXiv:2407.16741}, 2024{\natexlab{c}}.

\bibitem[Wang et~al.(2025)Wang, Xu, Chen, Bi, Gu, and Zheng]{wang2025developer}
Yanlin Wang, Xinyi Xu, Jiachi Chen, Tingting Bi, Wenchao Gu, and Zibin Zheng.
\newblock An empirical study of agent developer practices in {AI} agent
  frameworks.
\newblock \emph{arXiv preprint arXiv:2512.01939}, 2025.

\bibitem[Wu et~al.(2024)Wu, Bansal, Zhang, Wu, Li, Zhu, Jiang, Zhang, Zhang,
  Liu, et~al.]{wu2024autogen}
Qingyun Wu, Gagan Bansal, Jieyu Zhang, Yiran Wu, Beibin Li, Erkang Zhu,
  Li~Jiang, Xiaoyun Zhang, Shaokun Zhang, Jiale Liu, et~al.
\newblock {AutoGen}: Enabling next-gen {LLM} applications via multi-agent
  conversation.
\newblock \emph{arXiv preprint arXiv:2308.08155}, 2024.

\bibitem[Xi et~al.(2025)Xi, Chen, Guo, He, Ding, Hong, Zhang, Wang, Jin, Zhou,
  et~al.]{xi2025rise}
Zhiheng Xi, Wenxiang Chen, Xin Guo, Wei He, Yiwen Ding, Boyang Hong, Ming
  Zhang, Junzhe Wang, Senjie Jin, Enyu Zhou, et~al.
\newblock The rise and potential of large language model based agents: A
  survey.
\newblock \emph{Science China Information Sciences}, 2025.

\bibitem[Xia et~al.(2024)Xia, Paltenghi, Tian, Pradel, and
  Zhang]{xia2024fuzz4all}
Chunqiu~Steven Xia, Matteo Paltenghi, Jia~Le Tian, Michael Pradel, and Lingming
  Zhang.
\newblock Fuzz4all: Universal fuzzing with large language models.
\newblock In \emph{International Conference on Software Engineering (ICSE)},
  2024.

\bibitem[Yang et~al.(2024)Yang, Jimenez, Wettig, Lieret, Yao, Narasimhan, and
  Press]{yang2024sweagent}
John Yang, Carlos~E Jimenez, Alexander Wettig, Kilian Lieret, Shunyu Yao,
  Karthik Narasimhan, and Ofir Press.
\newblock {SWE-agent}: Agent-computer interfaces enable automated software
  engineering.
\newblock \emph{arXiv preprint arXiv:2405.15793}, 2024.

\bibitem[Yao et~al.(2023)Yao, Zhao, Yu, Du, Shafran, Narasimhan, and
  Cao]{yao2023react}
Shunyu Yao, Jeffrey Zhao, Dian Yu, Nan Du, Izhak Shafran, Karthik Narasimhan,
  and Yuan Cao.
\newblock {ReAct}: Synergizing reasoning and acting in language models.
\newblock \emph{arXiv preprint arXiv:2210.03629}, 2023.

\bibitem[Yao et~al.(2024)Yao, Shinn, Razavi, and Narasimhan]{tau2bench}
Shunyu Yao, Noah Shinn, Pedram Razavi, and Karthik Narasimhan.
\newblock $\tau$-bench: A benchmark for tool-agent-user interaction in
  real-world domains.
\newblock \emph{arXiv preprint arXiv:2406.12045}, 2024.

\bibitem[Zheng et~al.(2023)Zheng, Chiang, Sheng, Zhuang, Wu, Zhuang, Lin, Li,
  Li, Xing, Zhang, Gonzalez, and Stoica]{zheng2023judging}
Lianmin Zheng, Wei-Lin Chiang, Ying Sheng, Siyuan Zhuang, Zhanghao Wu, Yonghao
  Zhuang, Zi~Lin, Zhuohan Li, Dacheng Li, Eric~P Xing, Hao Zhang, Joseph~E
  Gonzalez, and Ion Stoica.
\newblock Judging {LLM}-as-a-judge with {MT-Bench} and chatbot arena.
\newblock In \emph{Advances in Neural Information Processing Systems}, 2023.

\bibitem[Zimmermann et~al.(2019)Zimmermann, Staicu, Tenny, and
  Pradel]{zimmermann2019small}
Markus Zimmermann, Cristian-Alexandru Staicu, Cam Tenny, and Michael Pradel.
\newblock Small world with high risks: A study of security threats in the npm
  ecosystem.
\newblock In \emph{Proceedings of the 28th USENIX Security Symposium}, 2019.

\end{thebibliography}

\newpage
\appendix

\section{Ecosystem Analysis}
\label{sec:ecosystem}

Beyond evaluating what ADK frameworks \emph{do} on benchmarks, we examine how they \emph{relate} to each other and to the broader developer community. We analyze inter-framework dependencies (\S\ref{sec:dependencies}), which reveal supply-chain risks and hidden coupling, and downstream adoption (\S\ref{sec:downstream}), which quantifies real-world usage and market concentration.

\subsection{Inter-ADK Dependencies}
\label{sec:dependencies}

Many ADK frameworks reuse components from other ADKs, creating supply-chain concentration risks~\citep{zimmermann2019small,decan2019empirical} and hidden coupling invisible to package managers~\citep{liu2022demystifying}.

\begin{figure}[t]
\centering
\includegraphics[width=\linewidth]{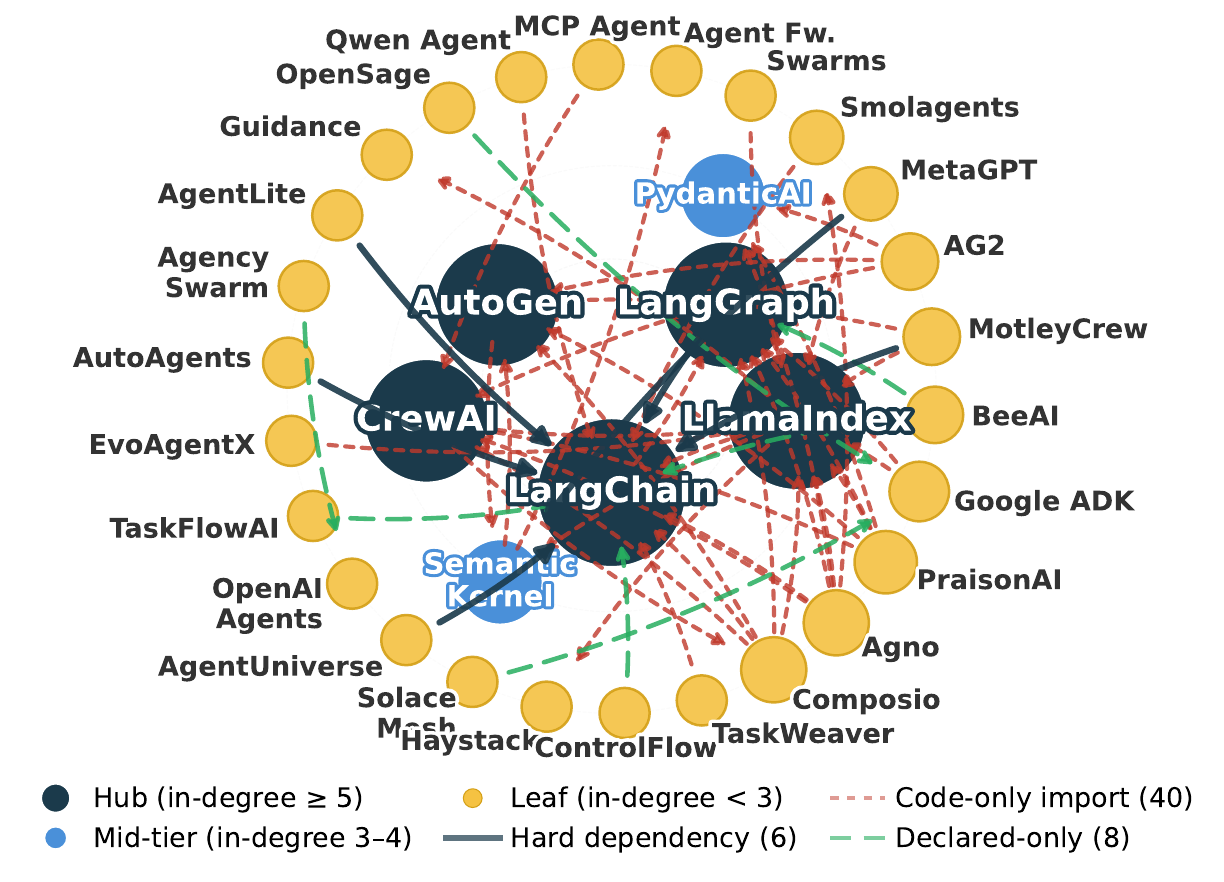}
\caption{Inter-ADK dependency network.}
\label{fig:inter_adk}
\end{figure}
We extract dependency relationships from both package metadata (\texttt{pyproject.toml}, \texttt{setup.py}, \texttt{requirements.txt}) and source code imports across all 51 repositories. An edge from framework $A$ to framework $B$ is classified into three categories (test and example imports are marked \emph{optional} and excluded from the core graph):

\begin{itemize}[nosep,leftmargin=*]
  \item \textbf{Hard}. Declared in package metadata \emph{and} imported in source code. The dependency is fully transparent; \texttt{pip install} automatically resolves it. We find 6 such edges.
  \item \textbf{Code-only}. Imported in source code but \emph{not} declared in metadata. Users who \texttt{pip install} the framework will encounter \texttt{ImportError} at runtime when a specific code path triggers the undeclared import. This is the dominant category, accounting for 40 of 54 non-optional edges (74\%).
  \item \textbf{Declared-only}. Declared in metadata but \emph{not} imported in source code. The package is installed but unused, likely a historical artifact or planned integration. We find 8 such edges.
\end{itemize}

\paragraph{Hub-and-spoke topology.}
The graph exhibits a hub-and-spoke structure. LangChain is the dominant hub (in-degree 14), with nearly half of all participating frameworks depending on it, five via hard dependencies (AgentLite, AgentUniverse, AutoAgents, LangGraph, MotleyCrew). LlamaIndex is the second hub (in-degree 10), but all its dependents use code-only imports, making the coupling less visible but equally fragile. On the consumer side, three frameworks have high out-degree: Agno (6), Composio (5), and PraisonAI (5), acting as integration layers that bridge multiple platforms. A total of 33 frameworks participate in the graph; the remaining 18 (35\%) are fully independent.

\paragraph{Shared infrastructure.}
Beyond inter-ADK edges, frameworks converge on a small set of shared infrastructure packages. \texttt{pydantic} is required by 38/51 frameworks (74.5\%), \texttt{openai} by 24 (47.1\%), \texttt{requests} by 23 (45.1\%), and \texttt{pyyaml} by 22 (43.1\%). A breaking change in \texttt{pydantic} alone propagates to three-quarters of the ecosystem. Across 1,223 dependency specifications, only 35.0\% use exact version pinning; 25.5\% are completely unbounded. This loose pinning, combined with the high rate of undeclared code-only imports, creates a fragile dependency surface where upstream changes can silently break downstream frameworks.

\paragraph{Mutual dependencies.}
Three framework pairs have bidirectional edges: LangChain $\leftrightarrow$ LangGraph (LangGraph hard-depends on \texttt{langchain\_core}, while LangChain code-imports LangGraph), AutoGen $\leftrightarrow$ Semantic Kernel (both code-import each other), and Composio $\leftrightarrow$ CrewAI. These mutual dependencies create tightly coupled clusters where a breaking change in either framework can cascade to the other.

\finding{The ecosystem has fragile supply-chain concentration: 74\% of inter-framework dependencies are undeclared, LangChain and LlamaIndex are hubs with in-degree 14 and 10 respectively, and \texttt{pydantic} breaking changes propagate to 75\% of the ecosystem.}

\begin{table*}[t]
\caption{Downstream adoption metrics for all 51 ADK frameworks, sorted by monthly downloads. \textbf{Ver.}: version evaluated in this study. $\bigstar$ = GitHub stars; DL/mo = PyPI monthly downloads; Repos = GitHub dependent repositories; CS = GitHub code-search hits in \texttt{requirements.txt} and \texttt{pyproject.toml}. ``NA'' indicates the framework is not published on PyPI or not yet indexed by GitHub. Frameworks marked with \dag{} originate from academic publications. Data collected April 2026.}
\label{tab:pypi_downstream}
\centering
\resizebox{\linewidth}{!}{%
\renewcommand{\arraystretch}{1.15}
\setlength{\tabcolsep}{4pt}
\begin{tabular}{@{\hskip 2pt}l l cccc @{\hskip 8pt} | @{\hskip 8pt} l l cccc@{\hskip 2pt}}
\toprule
\textbf{Framework} & \textbf{Ver.} & \textbf{$\bigstar$} & \textbf{DL/mo} & \textbf{Repos} & \textbf{CS} &
\textbf{Framework} & \textbf{Ver.} & \textbf{$\bigstar$} & \textbf{DL/mo} & \textbf{Repos} & \textbf{CS} \\
\midrule
\addlinespace[0.3em]
    \raisebox{-0.4ex}{\includegraphics[height=1.1em]{figures/logos/langchain.png}}\,Langchain & 1.2.17 & 135.2K & 233.1M & 279.7K & 32.0K & \raisebox{-0.4ex}{\includegraphics[height=1.1em]{figures/logos/swarms.png}}\,Swarms & 11.0.1 & 6.6K & 40.0K & 429 & 1.1K \\
    \raisebox{-0.4ex}{\includegraphics[height=1.1em]{figures/logos/anthropic-sdk.png}}\,Anthropic Agent Sdk & 0.98.0 & 3.3K & 103.8M & 50.0K & 22.9K & \raisebox{-0.4ex}{\includegraphics[height=1.1em]{figures/logos/mcp-agent.png}}\,Mcp Agent & 0.2.6 & 8.3K & 31.8K & 243 & 540 \\
    \raisebox{-0.4ex}{\includegraphics[height=1.1em]{figures/logos/langgraph.png}}\,Langgraph & 1.1.10 & 30.6K & 44.9M & 38.5K & 15.2K & \raisebox{-0.4ex}{\includegraphics[height=1.1em]{figures/logos/guidance.png}}\,Guidance & 0.3.1 & 21.4K & 28.3K & 1.1K & 2.3K \\
    \raisebox{-0.4ex}{\includegraphics[height=1.1em]{figures/logos/openai-agents.png}}\,Openai Agents & 0.15.1 & 25.3K & 24.2M & NA & 4.2K & \raisebox{-0.4ex}{\includegraphics[height=1.1em]{figures/logos/atomic-agents.png}}\,Atomic Agents & 2.7.5 & 5.8K & 20.1K & 91 & 159 \\
\addlinespace[0.4em]
    \raisebox{-0.4ex}{\includegraphics[height=1.1em]{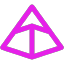}}\,Pydantic Ai & 1.89.1 & 16.7K & 23.4M & 3.9K & 4.6K & \raisebox{-0.4ex}{\includegraphics[height=1.1em]{figures/logos/griptape.png}}\,Griptape & 1.10.0 & 2.5K & 19.0K & 198 & 148 \\
    \raisebox{-0.4ex}{\includegraphics[height=1.1em]{figures/logos/llamaindex.png}}\,Llama Index & 0.12.52 & 49.0K & 10.1M & 24.0K & 17.9K & \raisebox{-0.4ex}{\includegraphics[height=1.1em]{figures/logos/upsonic.png}}\,Upsonic & 0.76.3 & 7.8K & 9.1K & 65 & 36 \\
    \raisebox{-0.4ex}{\includegraphics[height=1.1em]{figures/logos/crewai.png}}\,Crewai & 1.14.4 & 50.0K & 6.7M & 18.3K & 7.2K & \raisebox{-0.4ex}{\includegraphics[height=1.1em]{figures/logos/controlflow.png}}\,Controlflow & 0.12.1 & 1.4K & 7.0K & NA & 42 \\
    \raisebox{-0.4ex}{\includegraphics[height=1.1em]{figures/logos/google-adk.png}}\,Google Adk & 1.32.0 & 19.3K & 6.4M & NA & 4.9K & \raisebox{-0.4ex}{\includegraphics[height=1.1em]{figures/logos/agency-swarm.png}}\,Agency Swarm & 1.9.6 & 4.2K & 6.8K & 353 & 212 \\
\addlinespace[0.4em]
    \raisebox{-0.4ex}{\includegraphics[height=1.1em]{figures/logos/strands-agents.png}}\,Strands Agents & 1.38.0 & 5.7K & 5.1M & NA & 1.8K & \raisebox{-0.4ex}{\includegraphics[height=1.1em]{figures/logos/agently.png}}\,Agently & 4.1.0.2 & 1.6K & 3.5K & 1 & 36 \\
    \raisebox{-0.4ex}{\includegraphics[height=1.1em]{figures/logos/semantic-kernel.png}}\,Semantic Kernel & 1.41.3 & 27.8K & 2.7M & 2.6K & 1.9K & \raisebox{-0.4ex}{\includegraphics[height=1.1em]{figures/logos/lagent.png}}\,Lagent & 0.5.0rc3 & 2.2K & 3.2K & 174 & 2.7K \\
    \raisebox{-0.4ex}{\includegraphics[height=1.1em]{figures/logos/agno.png}}\,Agno & 2.6.4 & 39.7K & 1.7M & 2.5K & 3.5K & \raisebox{-0.4ex}{\includegraphics[height=1.1em]{figures/logos/agentuniverse.png}}\,Agentuniverse & 0.0.5 & 2.2K & 2.3K & 18 & 22 \\
    \raisebox{-0.4ex}{\includegraphics[height=1.1em]{figures/logos/ag2.png}}\,Ag2 & 0.12.2 & 4.5K & 1.5M & 774 & 1.6K & \raisebox{-0.4ex}{\includegraphics[height=1.1em]{figures/logos/evoagentx.png}}\,Evoagentx\textsuperscript{\dag} & 0.1.0 & 2.9K & 1.7K & NA & 9 \\
\addlinespace[0.4em]
    \raisebox{-0.4ex}{\includegraphics[height=1.1em]{figures/logos/microsoft.png}}\,Autogen & 0.7.5 & 57.5K & 1.4M & 4.1K & 2.0K & \raisebox{-0.4ex}{\includegraphics[height=1.1em]{figures/logos/nerve.png}}\,Nerve & 1.8.0 & 1.3K & 957 & NA & NA \\
    \raisebox{-0.4ex}{\includegraphics[height=1.1em]{figures/logos/composio.png}}\,Composio & 0.12.0 & 27.9K & 1.0M & 321 & 714 & \raisebox{-0.4ex}{\includegraphics[height=1.1em]{figures/logos/agent-squad.png}}\,Agent Squad & 1.0.2 & 7.6K & 651 & 10 & 17 \\
    \raisebox{-0.4ex}{\includegraphics[height=1.1em]{figures/logos/haystack.png}}\,Haystack & 2.28.0 & 25.0K & 712.4K & 1.4K & 1.1K & \raisebox{-0.4ex}{\includegraphics[height=1.1em]{figures/logos/motleycrew.png}}\,Motleycrew & 0.3.7 & 0.4K & 626 & NA & 4 \\
    \raisebox{-0.4ex}{\includegraphics[height=1.1em]{figures/logos/smolagents.png}}\,Smolagents\textsuperscript{\dag} & 1.24.0 & 26.9K & 559.4K & NA & 2.0K & \raisebox{-0.4ex}{\includegraphics[height=1.1em]{figures/logos/taskflowai.png}}\,Taskflowai & 0.5.13 & 0.1K & 191 & 6 & 7 \\
\addlinespace[0.4em]
    \raisebox{-0.4ex}{\includegraphics[height=1.1em]{figures/logos/agent-framework.png}}\,Agent Framework & 1.2.2 & 9.9K & 436.7K & NA & 1.6K & \raisebox{-0.4ex}{\includegraphics[height=1.1em]{figures/logos/council-ai.png}}\,Council Ai & 0.0.29 & 0.8K & 181 & 12 & 13 \\
    \raisebox{-0.4ex}{\includegraphics[height=1.1em]{figures/logos/qwen-agent.png}}\,Qwen Agent & 0.0.34 & 16.2K & 359.7K & 163 & 270 & \raisebox{-0.4ex}{\includegraphics[height=1.1em]{figures/logos/agentflow.png}}\,Agentflow\textsuperscript{\dag} & 0.1.2 & 1.8K & 146 & NA & 89 \\
    \raisebox{-0.4ex}{\includegraphics[height=1.1em]{figures/logos/agentscope.png}}\,Agentscope\textsuperscript{\dag} & 1.0.19 & 24.4K & 217.9K & 64 & 253 & \raisebox{-0.4ex}{\includegraphics[height=1.1em]{figures/logos/autoagent.png}}\,Autoagent\textsuperscript{\dag} & 0.1.0 & 9.2K & 123 & NA & 38 \\
    \raisebox{-0.4ex}{\includegraphics[height=1.1em]{figures/logos/praisonai.png}}\,Praisonai & 4.6.37 & 7.0K & 183.7K & 2 & 137 & \raisebox{-0.4ex}{\includegraphics[height=1.1em]{figures/logos/agentlite.png}}\,Agentlite\textsuperscript{\dag} & 0.1.12 & 0.6K & 86 & 3 & 2 \\
\addlinespace[0.4em]
    \raisebox{-0.4ex}{\includegraphics[height=1.1em]{figures/logos/camel.png}}\,Camel Ai\textsuperscript{\dag} & 0.2.90 & 16.8K & 134.9K & 380 & 959 & \raisebox{-0.4ex}{\includegraphics[height=1.1em]{figures/logos/octotools.png}}\,Octotools & 0.2.0 & 1.4K & 60 & NA & 3 \\
    \raisebox{-0.4ex}{\includegraphics[height=1.1em]{figures/logos/fast-agent.png}}\,Fast Agent & 0.6.26 & 3.8K & 105.3K & NA & 139 & \raisebox{-0.4ex}{\includegraphics[height=1.1em]{figures/logos/gptswarm.png}}\,Gptswarm\textsuperscript{\dag} & 0.1.0 & 1.0K & 28 & 2 & 5 \\
    \raisebox{-0.4ex}{\includegraphics[height=1.1em]{figures/logos/solace-agent-mesh.png}}\,Solace Agent Mesh & 1.23.1 & 3.3K & 83.1K & 8 & 76 & \raisebox{-0.4ex}{\includegraphics[height=1.1em]{figures/logos/taskweaver.png}}\,Taskweaver\textsuperscript{\dag} & 0.0.12 & 6.2K & NA & NA & 16 \\
    \raisebox{-0.4ex}{\includegraphics[height=1.1em]{figures/logos/langroid.png}}\,Langroid & 0.61.1 & 4.0K & 69.5K & 68 & 514 & Autoagents\textsuperscript{\dag} & 0.2 & 1.5K & NA & NA & 12 \\
\addlinespace[0.4em]
    \raisebox{-0.4ex}{\includegraphics[height=1.1em]{figures/logos/beeai.png}}\,Beeai & 0.1.79 & 3.2K & 49.5K & 43 & 157 & \raisebox{-0.4ex}{\includegraphics[height=1.1em]{figures/logos/opensage.png}}\,Opensage\textsuperscript{\dag} & NA & 75 & NA & NA & 5 \\
    \raisebox{-0.4ex}{\includegraphics[height=1.1em]{figures/logos/metagpt.png}}\,Metagpt\textsuperscript{\dag} & 0.8.1 & 67.5K & 42.0K & 122 & 150 & & & & & & \\
\bottomrule
\end{tabular}%
}
\end{table*}

\subsection{Downstream Adoption}
\label{sec:downstream}

We measure downstream adoption using four signals: GitHub stars, PyPI monthly downloads, GitHub dependent repositories, and GitHub code-search hits in dependency files (\texttt{requirements.txt}/\texttt{pyproject.toml}). \autoref{tab:pypi_downstream} reports all four metrics for all 51 frameworks. Three patterns stand out:

\paragraph{Adoption concentration.}
Monthly downloads follow a steep power law. The top 5 frameworks by downloads (LangChain, Anthropic SDK, LangGraph, OpenAI Agents, PydanticAI) account for over 93\% of all ADK downloads. LangChain alone receives 233M downloads/month, more than the next four combined. The top 14 frameworks each exceed 1M downloads/month; below that threshold, adoption drops sharply. At the tail, 3 frameworks are not published on PyPI at all, and 20 receive fewer than 100K downloads/month. By GitHub dependent repositories, concentration is equally pronounced: LangChain dominates (279.7K repos), followed by Anthropic SDK (50.0K) and LangGraph (38.5K), while the median framework has fewer than 200 dependents. This concentration implies that most developers building agents converge on a small set of frameworks, despite the ecosystem offering 51 alternatives.

\paragraph{Stars and downloads diverge.}
GitHub stars and PyPI downloads measure different things (community interest vs.\ production adoption), and the two metrics diverge for several frameworks. MetaGPT ranks 2nd by stars (67.5K) but receives only 42K downloads/month, suggesting research interest without corresponding production use; it is widely forked for academic experiments but rarely \texttt{pip install}ed as a dependency. Conversely, Anthropic SDK has just 3.3K stars but ranks 2nd by downloads (103.8M/month), reflecting its role as invisible infrastructure that other packages depend on rather than a project developers star for reference. AutoGen shows a similar pattern: 57.5K stars (3rd overall) but only 1.4M downloads/month (13th), likely because its original Microsoft repo accumulated stars before the AG2 fork split the user base. Newer vendor SDKs (Strands Agents, Google ADK) show the opposite: high download counts from CI pipelines and cloud deployments but few stars, as enterprise users rarely star vendor dependencies.

\paragraph{Code search reveals hidden adoption.}
GitHub's dependency graph has indexing lag for newer packages, making dependent-repository counts unreliable for frameworks released in the past 12 months. We complement this with GitHub code search over \texttt{requirements.txt} and \texttt{pyproject.toml} files across all public repositories. This reveals adoption invisible to traditional metrics: Google ADK shows 4.9K code-search hits despite zero indexed dependents, OpenAI Agents shows 4.2K, and Strands Agents shows 1.8K. Smolagents (2.0K hits, zero dependents) is another example: developers include it in their requirements files, but GitHub has not yet indexed these repositories as dependents. This multi-source approach provides complementary coverage, as no single metric captures all frameworks.

\finding{Adoption follows extreme power-law concentration (top 5 = 93\% of downloads). Stars and downloads diverge sharply, and no single metric captures all frameworks.}

\section{Details of Iteration Loop}
\label{sec:validation-patterns}

This appendix details the three validation levels of the Validate-and-Repair pipeline (\S\ref{par:validate-and-repair}). Each level is progressively more expensive but catches deeper failures; the pipeline short-circuits on the first error and feeds a structured diagnostic hint back to the generator LLM before the next iteration.

\subsection{Step 1: Static Analysis}
\label{sec:phase1-import}

Step~1 catches obvious errors without executing the agent or calling any LLM, making it both fast and free. It proceeds in three phases of increasing semantic depth.

\paragraph{Phase 1a: Compile and import.}
The validator dynamically loads \texttt{agent.py} via \texttt{importlib} and verifies that (1) the file parses without syntax errors, (2) a top-level \texttt{solve()} function exists, and (3) \texttt{solve()} accepts exactly two positional arguments (\texttt{problem\_statement}, \texttt{workdir}). Before loading, environment variables (\texttt{ADK\_BASE\_URL}, \texttt{ADK\_API\_KEY}, \texttt{ADK\_MODEL}, \texttt{ADK\_TEMPERATURE}) are injected to mimic the execution environment. Agents declaring \texttt{async def solve()} are immediately rejected since the harness calls \texttt{solve()} synchronously. This phase catches the most basic failures (syntax errors, missing entry points, wrong signatures) before more expensive analysis begins.

\paragraph{Phase 1b: Framework usage.}
\label{sec:phase2-usage}
A design requirement is that each generated agent \emph{genuinely uses} the target framework's orchestration API rather than bypassing it. The validator maintains a per-framework mapping from slug to expected import names (e.g., \texttt{crewai}: \{\texttt{crewai}, \texttt{crewai\_tools}\}; \texttt{google-adk}: \{\texttt{google.adk}, \texttt{google.genai}\}) and confirms at least one appears in source. It then rejects two common LLM workarounds: \emph{dummy references} (\texttt{\_ = framework.\_\_name\_\_}) that technically import the framework without calling any API, and \emph{raw API fallbacks} where \texttt{solve()} is implemented entirely via \texttt{httpx.Client()}, \texttt{requests.post()}, or \texttt{openai.OpenAI()} without any framework wrapping. These patterns are common in practice: LLMs often import the framework to satisfy the stated constraint while implementing the actual logic through familiar raw HTTP calls.

\paragraph{Phase 1c: Pattern-based static analysis.}
\label{sec:phase3-static}
An AST and regex scanner checks 40+ patterns organized into seven categories:
\begin{itemize}[nosep,leftmargin=*]
  \item \emph{URL double-path}: appending \texttt{"/v1"} to env var already containing \texttt{/v1} (causes 404).
  \item \emph{SDK mismatch}: Anthropic SDK pointed at OpenAI-format proxy.
  \item \emph{Missing timeouts}: HTTP calls without \texttt{timeout} (causes indefinite hang).
  \item \emph{Process-killing calls}: \texttt{sys.exit()} terminates container before output.
  \item \emph{Hardcoded URLs/keys}: secrets that should come from environment variables.
  \item \emph{Bare \texttt{Any} type hints}: tool schemas lacking JSON Schema \texttt{type} (causes 400).
  \item \emph{Broad exception handlers}: \texttt{except Exception:} without \texttt{raise} swallows diagnostics.
\end{itemize}
Additionally, a forbidden-dependency deny-list prevents importing other frameworks' packages, and benchmark-specific AST checks verify correct subclassing (e.g., $\tau^2$-bench \texttt{HalfDuplexAgent} signatures).

\paragraph{Runtime diagnostics and repair.}
\label{sec:runtime-diagnostics}
When any level fails, the validator pattern-matches the traceback against 70+ error signatures across 9 categories:
\begin{itemize}[nosep,leftmargin=*]
  \item \emph{Module/import}: missing packages, renamed APIs between versions.
  \item \emph{API configuration}: hardcoded URLs, wrong auth sources, SSL mismatches.
  \item \emph{Async/event loop}: nested \texttt{asyncio.run()}, unclosed loops, unawaited coroutines.
  \item \emph{Type/signature}: wrong argument counts, unexpected kwargs, removed attributes.
  \item \emph{Container/filesystem}: writes outside \texttt{/tmp} or output directory.
  \item \emph{Output parsing}: framework expects structured format but LLM returns plain text.
  \item \emph{Response handling}: None values, exhausted retries, Pydantic validation failures.
  \item \emph{Framework-specific}: renamed params, required base classes, breaking API changes.
  \item \emph{Abstract method}: missing implementations of required interfaces.
\end{itemize}

Beyond error matching, the proxy records comprehensive telemetry throughout execution: total LLM calls, tool execution count, whether tools were included in requests and whether tool calls appeared in responses, the names of all tools registered with the LLM, maximum conversation length observed, and whether a \texttt{/v1/v1} double-path was detected in any request URL. This telemetry enables precise failure diagnosis without requiring access to the agent's internal state. After \texttt{solve()} returns, a final output validation step checks for non-empty results and correct return types, catching common mistakes like returning a coroutine object (forgot \texttt{await}), a raw framework object (forgot to extract \texttt{.text} or \texttt{.content}), an unconsumed async generator, or framework metadata dicts instead of the expected string output.

\subsection{Step 2: Real LLM Smoke Test}
\label{sec:level2-real-llm}

While Step~1 catches structural problems, many failures only manifest when the agent interacts with a real LLM. Step~2 executes \texttt{solve()} with DeepSeek-V4-Flash through a dedicated token-recording proxy, catching model-specific incompatibilities: unsupported parameters (\texttt{reasoning\_content}, \texttt{developer} role), response format mismatches, authentication errors, and framework-internal routing issues that depend on actual model output.

\paragraph{Architecture.}
The test spawns an HTTP reverse proxy on a random port that transparently forwards requests to the DeepSeek API while recording per-call metrics (input/output/cached tokens, latency). The agent's \texttt{ADK\_BASE\_URL} and \texttt{ADK\_API\_KEY} point to this proxy. Runtime compatibility patches (identical to those applied during actual benchmark execution) are loaded before the agent module, ensuring full environment fidelity. This design means that if an agent passes Step~2, it will encounter the exact same runtime environment during the final benchmarking phase.

\paragraph{Benchmark-specific prompts and environments.}
Each benchmark receives a minimal test prompt designed to exercise core capabilities without expensive multi-step reasoning: MCP-Atlas forces tool calling (``What is the git status of the repository?''), SWE-bench forces code reasoning (``Fix the TypeError''), Terminal-Bench forces shell execution (``Create hello.txt''), and $\tau^2$-bench runs a 3-round multi-turn conversation with alternating user/assistant/tool messages. The validator also prepares a minimal environment matching what the real benchmark provides: a stub MCP server for MCP-Atlas, a domain configuration file for $\tau^2$-bench, and placeholder source files for SWE-bench.

\paragraph{Pass criteria and failure diagnostics.}
The test passes if the proxy records ${\geq}1$ LLM call (confirming the agent routes through the proxy), at least 2 calls for MCP-Atlas (tool discovery + usage), and the agent returns a non-empty string not matching known bad patterns (\texttt{<coroutine object}, \texttt{object at 0x}). A 45-second timeout with recorded proxy calls is also a pass, since iterative agent loops commonly exceed this budget. On failure, the validator distinguishes four scenarios based on proxy telemetry: 0 calls with timeout (agent bypasses proxy or hangs pre-API), ${\geq}1$ calls with timeout (agent loop lacks termination condition), 401 errors (wrong authentication source), and 400 errors (malformed request format). Each scenario produces a targeted diagnostic hint that tells the generator LLM exactly what to fix.

\subsection{Step 3: Real Benchmark Task}
\label{sec:level3-real-task}

Step~3 provides the strongest validation by running one task from the target benchmark's \emph{official} task set using the full benchmark adapter and execution environment. This catches failures invisible to Steps~1--2 that only emerge under realistic conditions: complex multi-step prompts with large context windows, tool schemas containing ${\sim}$20 tools with detailed JSON parameter definitions, multi-turn conversation management across dozens of messages, and real environment interactions with Docker containers, git repositories, and file systems.

\paragraph{Task selection and execution.}
Each benchmark has a designated smoke test task chosen for fast execution while exercising the full agent pipeline: a multi-tool git/file query for MCP-Atlas, \texttt{django\_\_django-10097} (focused bug fix) for SWE-bench, a git branch operation for Terminal-Bench, and a customer service interaction for $\tau^2$-bench. The validator launches the benchmark adapter in a daemon thread with the same configuration as the final benchmarking phase: identical runtime patches, proxy, and tool interfaces. Rather than blocking on completion, it polls the proxy's \texttt{/status} endpoint every 2 seconds for up to 60 seconds, enabling early termination once the agent demonstrates sufficient capability (e.g., reaching the minimum call count with tool executions).

\paragraph{Pass criteria.}
Success requires benchmark-specific minimum LLM calls (${\geq}2$ for MCP-Atlas, $\tau^2$-bench, and Terminal-Bench; ${\geq}1$ for SWE-bench), non-empty output where applicable (SWE-bench must produce a patch), and tool execution (MCP-Atlas must dispatch ${\geq}1$ tool call). When tools are registered but never executed, the validator performs root-cause analysis: tools never sent to the LLM indicates a configuration error, tool calls returned but not dispatched indicates a broken execution loop, and the LLM choosing not to call tools is treated as a prompt issue (warning only). Note that \emph{correctness} of the output is not evaluated at this stage; that responsibility belongs to the full benchmarking phase (\S\ref{sec:stage3}). Step~3 validates only that the agent can survive a complete execution cycle without crashing.

\end{document}